\documentclass[manuscript, screen, nonacm]{acmart}
\usepackage{hyperref}
\usepackage{hyperxmp}

\usepackage{packages}

\AtBeginDocument{%
  \providecommand\BibTeX{{%
    \normalfont B\kern-0.5em{\scshape i\kern-0.25em b}\kern-0.8em\TeX}}}

\setcopyright{rightsretained}
\acmJournal{CSUR}
\acmYear{2024} \acmVolume{1} \acmNumber{1} \acmArticle{1} \acmMonth{1}\acmDOI{10.1145/3704806}

\begin{document}

\title[Motivations, Challenges, Best Practices, and Benefits for Bots and Conversational Agents in Software Engineering]{Motivations, Challenges, Best Practices, and Benefits for Bots and Conversational Agents in Software Engineering: A Multivocal Literature Review}

\author{Stefano Lambiase}
\email{slambiase@unisa.it}
\affiliation{%
  \institution{SeSa Lab -- University of Salerno}
  \city{Salerno}
  \country{Italy}
}

\author{Gemma Catolino}
\email{gcatolino@unisa.it}
\affiliation{%
  \institution{SeSa Lab -- University of Salerno}
  \city{Salerno}
  \country{Italy}
}

\author{Fabio Palomba}
\email{fpalomba@unisa.it}
\affiliation{%
  \institution{SeSa Lab -- University of Salerno}
  \city{Salerno}
  \country{Italy}
}

\author{Filomena Ferrucci}
\email{fferrucci@unisa.it}
\affiliation{%
  \institution{SeSa Lab -- University of Salerno}
  \city{Salerno}
  \country{Italy}
}

\renewcommand{\shortauthors}{Lambiase et al.}

\begin{abstract}
\emph{\textbf{Bots}} are software systems designed to support users by automating specific processes, tasks, or activities.
When these systems implement a conversational component to interact with users, they are also known as \emph{\textbf{conversational agents}} or \emph{chatbots}. 
Bots—particularly in their conversation-oriented version and AI-powered—have seen increased adoption over time for software development and engineering purposes. 
Despite their exciting potential, which has been further enhanced by the advent of Generative AI and Large Language Models, bots still face challenges in terms of development and integration into the development cycle, as practitioners report that bots can add difficulties rather than provide improvements.
In this work, we aim to provide a taxonomy for characterizing bots, as well as a series of challenges for their adoption in software engineering, accompanied by potential mitigation strategies.
To achieve our objectives, we conducted a \emph{multivocal literature review}, examining both research and practitioner literature. 
Through such an approach, we hope to contribute to both researchers and practitioners by providing (i) a series of future research directions to pursue, (ii) a list of strategies to adopt for improving the use of bots for software engineering purposes, and (iii) fostering technology and knowledge transfer from the research field to practice—one of the primary goals of multivocal literature reviews.
\end{abstract}




\maketitle

\section{Introduction}
\label{sec:introduction}
In the last decade, continuous human activities increased—both in private and working life—thus raising the necessity for automation. 
Consequently, the adoption of \emph{\textbf{bots}}~\cite{storey2016_disrupting_developer_productivity_botDefinition, labeuf_2017_software_bots}—i.e., software systems designed to automate a specific function or set of activities—has grown and available for a plethora of purposes~\cite{storey_2016_bot_uses_taxonomy, motger2022_software_bot_SLR}. In particular, \emph{\textbf{conversational agents (CAs)}}~\cite{storey2016_disrupting_developer_productivity_botDefinition, motger2022_software_bot_SLR, shevat2017_designing_bots}—also known as \emph{chatbots}—are bots that communicate with users, by natural language or similar, using a communication channel. 
Recently, CAs—and, more generally, bots—have started to be adopted in the software development field~\cite{labeuf_2017_software_bots, perez2018_collaborative_modeling_with_chatbot}. For example, practitioners use bots to automate software maintenance and evolution activities. Similarly, they adopt bots for collaboration, speeding up communication and knowledge-sharing. Besides proposing them, the research community worked on frameworks to ease the development process of conversational systems.

%\revised{The exploration of bots and conversational agents across various systematic literature reviews reveals a rich tapestry of research focused on both the technological and strategic application of these tools within software engineering. Reviews such as those by Lewandowski et al.~\cite{lewandowski2021_bot_SLR} and Suhaili et al.~\cite{suhaili2021_bot_SLR} delve into the strategic implementation and technological underpinnings of conversational agents, highlighting a strong emphasis on the technical construction while advocating for a broader consideration of interaction design and natural language capabilities. Further analysis provided by Motger et al.~\cite{motger2022_software_bot_SLR} and Del Carpio and Angarita~\cite{del2023assistant} expands on how these agents are integrated into user interactions and the entire software development process. Moreover, Moguel-Sánchez et al.~\cite{moguel2023bots} focus on the practical applications of bots within software development, examining how these tools are employed to streamline tasks like project management and static code analysis. A common outcome across all the cited works and other authoritative studies in the field is that bots (especially given the recent emergence of Generative AI and Large Language Models) have incredible potential to enhance all aspects of development. Nevertheless, they are still problematic to develop and integrate into the development cycle, potentially introducing additional challenges that may worsen rather than improve it.}

The exploration of bots and conversational agents in systematic literature reviews reveals a focus on both technological and strategic applications within software engineering. Lewandowski et al.~\cite{lewandowski2021_bot_SLR} and Suhaili et al.~\cite{suhaili2021_bot_SLR} emphasize the technical construction and advocate for considering interaction design and natural language capabilities. Motger et al.~\cite{motger2022_software_bot_SLR} and Del Carpio and Angarita~\cite{del2023assistant} discuss integration into user interactions and software development. Moguel-Sánchez et al.~\cite{moguel2023bots} examine practical applications, such as project management and static code analysis. As a common finding, despite their potential, bots, especially with Generative AI and Large Language Models, present development and integration challenges.

To support both the research community—providing new research opportunities—and practitioners—delivering knowledge for adopting bots and CAs—we carried out a \emph{multivocal literature review (MLR)}~\cite{ogawa1991_MLR_definition, garousi2019_mlr_guidelines}, following the guidelines of Garousi et al.~\cite{garousi2019_mlr_guidelines}. Specifically, we aimed to understand the current role and impact of bots and CAs in the software engineering (SE) field, focusing on the (1) motivations for their adoption, (2) challenges, (3) best practices, and (4) benefits coming from their adoption.\footnote{From now and for the rest of the paper, we will use the term ``bot'' to refer to both bots without and with a conversational component (CAs) when it is not necessary to put in evidence the conversational component.}

Our work complements and diverges from existing literature by reviewing both academic and grey literature and identifying best practices for bot design and adoption. By focusing on engineering and interaction design challenges, we provide guidelines to enhance the practical application of bots in the industry. Additionally, this research aims to be useful to both researchers and practitioners. Moreover, the decision to conduct a Multivocal Literature Review (MLR) rather than an SLR for studying bots in software engineering is supported by Garousi et al.~\cite{garousi2019_mlr_guidelines}. The rapid technological evolution led by practitioners and the lack of extensive academic publications necessitates incorporating grey literature (GL) for current insights~\cite{park2022_bot_SLR, suhaili2021_bot_SLR}. The practical application of bots further underscores the need for GL to understand real-world impacts. Aligning academic research with practical experiences provides a comprehensive synthesis of knowledge~\cite{garousi2019_mlr_guidelines}, and including practitioner-driven insights is crucial to avoid publication bias~\cite{park2022_bot_SLR, suhaili2021_bot_SLR}.

We collected 107 literature items—79 formal studies and 28 grey ones. Results show how CAs are complex to design and deploy despite the diffusion of tools to support their development. Moreover, we noticed that these systems are prone to several problems that undermine their adoption. In particular, interaction—like interruption and noise—emerged as the most critical issue. Nevertheless, we discovered that bots and CAs benefit the practitioners' community considerably, justifying putting effort into increasing their adoption. In terms of contributions, we provide the following:
\begin{itemize}
    \item A taxonomy to categorize and describe bots and conversational agents based on the motivation of their use for software development purposes.
    \item A list of challenges arising from adopting bots and conversational agents.
    \item A set of best practices to deal with bots and conversational agents and an association between such practices and challenges (reported in the Discussion section).
    \item A list of benefits when adopting bots and conversational agents in the software development context.
    \item An \textbf{online appendix}~\cite{online_appendix} containing all our findings and all the data gathered during the steps of our multivocal literature review.
\end{itemize}
These findings can be helpful for researchers, proposing new research opportunities and standardizing the terminology for referring to bots and CAs, and practitioners, reporting best practices and knowledge to ease the adoption and development of conversational systems.

\section{Background and Related Work}
\label{sec:background_related}

According to Storey and Zagalsky~\cite{storey2016_disrupting_developer_productivity_botDefinition, storey_2016_bot_uses_taxonomy, storey2020_botse}, a \emph{software bot} is \emph{``a conduit or an interface between users and services, typically through a conversational user interface''}~\cite{storey2016_disrupting_developer_productivity_botDefinition}, that can be exploited for automating processes/activities, thus improving productivity~\cite{storey2016_disrupting_developer_productivity_botDefinition,storey_2016_bot_uses_taxonomy}.
For this reason, software bots are widely used for different purposes. For instance, in the context of software development, they have been adopted for automating application deployment—e.g., build, test, and report—or for supporting developers during communication activities—e.g., agenda and information retrieval~\cite{storey_2016_bot_uses_taxonomy, basu_2021_bot_for_distribution_of_service_requests, wessel_2019_bot_close_abbandoned_pull_request_and_issues}. 

\emph{Conversational agents (CAs)}—also known as chatbots— are particular types of software bots that use communication channels like \textsl{Slack}, \textsl{Teams}, or \textsl{Discord} to interact with users, generating human-like conversations~\cite{srivastava_2019_architecture_applications_with_conversational_components,wessel2021_dont_disturb_me_botChallenges}.

The properties of bots and conversational agents, other than their potential applicability to various fields, have attracted the interest of multiple research communities—e.g., Software Engineering and Human-Computer Interaction—over the last few years. Despite being a young research field, we have identified six systematic literature reviews that synthesize state of the art on the usage of bots and CAs in the software engineering context \cite{lewandowski2021_bot_SLR,park2022_bot_SLR,motger2022_software_bot_SLR,moguel2023bots} and in related fields~\cite{suhaili2021_bot_SLR}.

Lewandowski et al.~\cite{lewandowski2021_bot_SLR} conducted a systematic literature review aiming to provide a structured overview of how conversational agents might be used from a strategic standpoint, namely how they can be used within work and company processes, other than governance structures. The authors found 21 relevant scientific records from 2015 to 2020. The key findings of the systematic literature review suggested that most of the papers focused on the technical aspects behind conversational agents. At the same time, other perspectives should be considered when designing more practical and usable CAs. These aspects concern (1) natural language capabilities and training and (2) interaction design, namely, the way a conversational agent can interact with its stakeholders. According to Lewandowski et al.~\cite{lewandowski2021_bot_SLR}, the missing analysis of these aspects may prevent the broader adoption of conversational agents in the industry. Concerning this literature review, our work shares the goal of eliciting the best practices to design conversational agents.
Moreover, our analysis also considers themes connected to motivations, challenges, and benefits of conversational agents and bots in software engineering. Finally, our scope is larger since we also include resources from the grey literature, thus including complementary perspectives. As a result, we provide a more comprehensive analysis of the topic of interest.

Suhaili et al.~\cite{suhaili2021_bot_SLR} investigated the technology employed to develop conversational agents. Indeed, they primarily focus on the components and techniques used to design those tools, other than on the datasets for their training, evaluation metrics, and application domains. Their review covered the scientific literature from 2011 and 2020. The authors reported that deep and reinforcement learning represent popular methods for understanding users' intents and generating appropriate responses. In addition, they found that \textsc{Twitter}\footnote{The \textsc{Twitter} dataset: \url{https://www.kaggle.com/datasets/jp797498e/twitter-entity-sentiment-analysis}.} and \textsc{Airline Travel Information Systems}\footnote{The \textsc{Airline Travel Information Systems} dataset: \url{https://www.kaggle.com/datasets/hassanamin/atis-airlinetravelinformationsystem}.} represent the most widely used datasets to train and test the capabilities of conversational agents. Finally, metrics such as accuracy, F1-score, and BLUE index are more frequently used to evaluate the built agents. The work by Suhaili et al.~\cite{suhaili2021_bot_SLR} is complementary to ours; besides their goal, we aim to elicit the best practices for designing bots and conversational agents. 

Motger et al.~\cite{motger2022_software_bot_SLR} conducted a systematic literature review of secondary literature, reviewing 28 articles. Their goal was to analyze the anatomy of CAs by providing a taxonomy regarding (1) the human-computer interaction features more relevant for users when interacting with CAs, (2) which techniques are used for the design and implementation of CAs, and (3) which approaches are used for the training phase of the NLP and NLU modules of bots. Our work can be considered complementary to the one by Motger et al.~\cite{motger2022_software_bot_SLR}. First, we focus on the design of both bots and conversational systems, hence providing a more extensive overview than Motger et al., who only analyzed CAs. More importantly, we approach the literature review from the software engineering perspective rather than artificial intelligence. As such, we are interested in providing guidelines and recommendations on how to engineer bots and CAs rather than understanding which are the best practices to train their internal artificial intelligence components. Moreover, we elaborate on additional concerns that may inspire further research in software engineering, such as the definition of a taxonomy to characterize bots and a list of benefits from their adoption. Last but not least, our work is based on a multivocal review~\cite{ogawa1991_MLR_definition} and, therefore, we analyze both white and grey literature by following the guidelines by Kitchenham et al.~\cite{kitchenham2009_SLR_definition} and Garousi et al.~\cite{garousi2019_mlr_guidelines}.

Moguel-Sánchez et al.~\cite{moguel2023bots} conducted a systematic literature review to identify (1) the activities for which bots are used in software development, (2) the benefits, and (3) the problems. They reviewed 83 primary studies and conducted a thematic analysis to organize their quantitative and qualitative findings. They found that bots are mainly used for project management, to automate tasks with a low level of abstraction, such as tagging pull requests and commits, assigning team members to code reviews, performing static code analyses, and tracking changes in project repositories. Divergently from Moguel-Sánchez et al.~\cite{moguel2023bots}, we included in our research also the grey literature, resulting in a more comprehensive knowledge of the topic other than a larger quantity of literature. Moreover, we also investigated the best practices associated with using bots in SE and created a mapping between challenges and best practices as an ulterior contribution to our work. Finally, regarding the challenges, we rely on already published taxonomy to enhance it and continuously extend the state of the art.

Del Carpio and Angarita~\cite{del2023assistant} conducted an SLR studying the support given by software assistants (SAs)—including bots—to the entire software development process. Specifically, they studied (1) what processes are supported by SAs, (2) how practitioners interact with SAs, (3) what techniques are used to implement SAs, and (4) what challenges face SAs.
They reviewed 40 studies and found that most SAs are—actually—conversational agents, and their interaction with users is far from optimal. The work by Del Carpio and Angarita~\cite{del2023assistant} is complementary to ours; besides their goal, we aim to elicit the best practices for designing bots and conversational agents, and we focus on the benefits of using bots. Moreover, we included the grey literature, which helped us identify a set of practices to complement the findings of the white literature.

\section{Research Study Design}
\label{sec:methodology}

The \emph{goal} of the study is to understand the current role and impact of bots and conversational agents in software engineering, focusing on the (1) motivations for their adoption, (2) challenges, (3) best practices, and (4) benefits provided to software engineers. The purpose is to provide a comprehensive overview of the matter, which might be useful to learn about the current state of the art and of the practice. Moreover, we want to stimulate further research to be pursued. The \emph{perspective} is of both researchers and practitioners: the former are interested in having a unique resource to use as a starting point to deepen their knowledge of the research field; the latter are interested in understanding best practices, challenges, and benefits of using bots in practice.

To achieve our goal, we performed a \emph{Multivocal Literature Review (MLR)}~\cite{ogawa1991_MLR_definition, patton1991_MLR_definition_1}, namely, a research process where past published (formal) literature (e.g., journal and conference papers) and \emph{grey literature} (e.g., blog posts, videos, and white papers) on the topic investigated are systematically identified, selected, and critically assessed. An MLR builds on top of the concept of \textit{Systematic Literature Review (SLR)}~\cite{kitchenham2009_SLR_definition, keele2007_slr_guidelines_Kitchenham}—where only the formal literature is taken into account—to provide benefits not only to the research community but also to practitioners.

%\topar{generale descrizione delle linee guida usate e Reporting}
We relied on two guidelines when defining the protocol and conducting the review. In particular, concerning the formal literature analysis, we adopted the well-established approaches by Kitchenham et al.~\cite{kitchenham2009_SLR_definition, keele2007_slr_guidelines_Kitchenham}. In addition, we followed the guidelines defined by Garousi et al.~\cite{garousi2019_mlr_guidelines} when it comes to the grey literature.
Finally, when organizing and reporting the results, we followed the \textsl{``General Standard''} and \textsl{``Systematic Reviews''} guidelines provided by the \emph{ACM/SIGSOFT Empirical Standards}.\footnote{\emph{ACM/SIGSOFT Empirical Standards}: \url{https://github.com/acmsigsoft/EmpiricalStandards}.} 
In the following subsections, we describe the main steps of each phase.\footnote{The entire process is depicted in Appendix A—Fig. 1~\cite{online_appendix}.}

\begin{comment}
\begin{figure}
    \centering
    \includegraphics[width=\linewidth]{figures/methodology/methodology_review.pdf}
    \caption{Overview of the methods used for the study.}
    \label{fig:methodology}
\end{figure}
\end{comment}

%%%%%%%%%%%%%%%%%%%%%%%%%%%%%%%%%%%%%%%%%%%%%%%%%%%%%%%
%\todo[inline]{Magari ogni parte iniziale della fase potrebbe avere un paragrafo che descrive cosa si fa e un paragrafo che determina gli output della fase}

\subsection{Planning the Review}

\subsubsection{Initiation Phase}
\label{subsec:initial}

The Initiation phase consisted of three sub-phases.

\paragraph{\textbf{Initial Examination of Previous Studies.}} \label{par:initial_experimentation}
To collect insights and basic knowledge about the adoption of bots and conversational agents in software engineering, we conducted a preliminary investigation in which we reviewed a few existing studies on the matter, e.g., \cite{lewandowski2021_bot_SLR, suhaili2021_bot_SLR}. In particular, we used the \textsl{Repository for Bot-related Research}~\cite{bot_se_research_repository}---i.e., a publicly available repository of formal literature maintained by the software engineering research community. After reading the studies, we discarded them in a \textit{``throw-away prototyping''}---as suggested in the literature~\cite{kitchenham2009_SLR_definition,garousi2019_mlr_guidelines}---to start the main review process with basic knowledge, avoiding possible bias or influence from no systematic intervention.

\paragraph{\textbf{Motivation and Needs Identification.}} 
Before choosing to conduct an MLR, Garousi et al. ~\cite{garousi2019_mlr_guidelines} suggested investigating the current state of the art on the topic of interest to identify limitations that grey literature could address.
For this reason, Garousi et al.~\cite{garousi2019_mlr_guidelines} defined a checklist that consists of seven ``Yes'' /``No''  questions related to the complexity and interest of the topic. A high number of ``Yes'' justifies the inclusion of grey literature in the systematic analysis.
This step is crucial to prevent the inclusion of grey literature from being of no benefit to research and resulting in a futile effort.
As for our study, Table \ref{table:motivations_MLR} reports the filled checklist. To fill such checklist, we used the knowledge obtained during the initial phase (described in the previous paragraph) and the SLRs on the topic~\cite{lewandowski2021_bot_SLR,suhaili2021_bot_SLR,okonkwo2021_chatbots_in_education_SLR,park2022_bot_SLR}, as suggested by Garousi et al.~\cite{garousi2019_mlr_guidelines}.
Six out of seven responses were ``Yes,'' which allowed us to decide to proceed with including grey literature in our study.

\begin{itemize}
    \item The first two questions were about the topic itself rather than the work. We answered the first question positively since the use of bots in software engineering is intrinsically a complex topic from both a technical and social perspective~\citeA{wessel_2021_challenges_interacting_with_bot_on_OSS, zamanirad2017programming,srivastava_2019_architecture_applications_with_conversational_components}. Moreover, bot technology advances primarily on the practitioner side rather than the research side, thus leading to a fast evolution of technology that is difficult to capture by research studies alone, often requiring slow publication times to facilitate systematic review and process. Furthermore, the broad adoption of bots, coupled with the recent advent of generative AI, leads to a wide range of issues to analyze, many of them from both a research and practitioner perspective. Regarding the second question, since the topic of bots in software engineering is young, there is not enough literature to determine the presence of a lack of consensus. Nevertheless, the youthfulness of the topic further poses the need to conduct studies involving the gray literature as well.
    
    \item The third question is about the subject under study in the work. Literature has repeatedly shown that the study context is central to software engineering research~\cite{gargani2011works, petersen2009context}. This becomes even more true when it comes to bots and their impact on practitioners. The present work aims to elicit the motivations, challenges, strategies, and benefits of practitioners' use of bots for different purposes and in different conditions. Hence, the contribution of grey literature to characterize this context is essential. For such a reason, we answered the third question positively.

    \item Questions four, five, and six are about the work objectives. As mentioned above, the goal of the work is to provide knowledge suitable for (1) opening new avenues of research and (2) improving the use and adoption of technology by practitioners. Given this goal, corroborating academic research with practitioners' knowledge is essential. In fact, where formal research can provide a systematic and robust set of information, grey literature can augment that information with concrete and applicable strategies useful to both worlds~\cite{garousi2019_mlr_guidelines}. On the other hand, academic research may shine precisely in disproving false beliefs born in the practitioner world. In summary, the goal of this paper is certainly to provide knowledge that is useful to both worlds, and that is as true and reliable as possible.

    \item Regarding question seven, we found significant evidence about the extensive adoption of bots by practitioners, consequentially increasing the importance of the context on the matter~\cite{park2022_bot_SLR, suhaili2021_bot_SLR}. Beyond that, the professional world's commitment to process automation has led to the development of many frameworks to support the use, development, and adoption of bots.

    \item Last, including grey literature can provide current perspectives and address the gaps in formal academic literature. First, it helps to avoid publication bias, although it is noted that the grey literature accessed may not represent all unpublished studies. Moreover, excluding grey literature would risk losing critical insights and perspectives on the topic, a conclusion also observed in the case study presented in the guidelines papers~\cite{garousi2019_mlr_guidelines, mahood2014_searching_for_grey_literature}.
\end{itemize}

\begin{table}
    \centering
    \caption{Questions to decide whether to conduct an MLR~\cite{garousi2019_mlr_guidelines}.}
    
    \rowcolors{1}{graytable}{white}
    \resizebox{\linewidth}{!}{
    \begin{tabular}{p{0.025\linewidth} p{0.875\linewidth} P{0.1\linewidth}}
        \toprule
        \rowcolor{black}
        \textbf{\textcolor{white}{\#}} & \textbf{\textcolor{white}{Question}} & \textbf{\textcolor{white}{Answer}}\\
        \bottomrule
        1 & Is the subject ``complex'' and not solvable by considering only the formal literature? & Yes\\
        2 & Is there a lack of volume or quality of evidence, or a lack of consensus of outcome measurement in the formal literature? & No\\
        3 & Is the contextual information important to the subject under study? & Yes\\
        4 & Is it the goal to validate or corroborate scientific outcomes with practical experiences? & Yes\\
        5 & Is it the goal to challenge assumptions or falsify results from practice using academic research or vice versa? & Yes\\
        6 & Would a synthesis of insights and evidence from the industrial and academic community be useful to one or even both communities? & Yes\\
        7 & Is there a large volume of practitioner sources indicating high practitioner interest in a topic? & Yes\\
        \bottomrule
        \rowcolor{white}
        \multicolumn{3}{p{\linewidth}}{\textit{\textbf{Note}: The possible answers to each question are ``Yes'' or ``No''. One or more ``Yes'' responses suggest that it could be useful to conduct an MLR.}}\\
    \end{tabular}
    }
    
    \label{table:motivations_MLR}
\end{table}

\paragraph{\textbf{Goals and Research Questions Definition.}} Our main objective was to provide an overview of the use of bots and conversational agents for software engineering purposes. We investigated four aspects that led to multiple research questions to reach our goal.

First, we aimed to investigate the reason behind using bots in software engineering; our goal was to create a taxonomy to categorize bots based on their use. Hence, we formulated our first research question.

\sterqbox{RQ\textsubscript{1}—Motivations and Goals}{Which motivations and goals cause the use of bots and conversational agents in software engineering?}

While the prevalence of bots and conversational agents has been growing significantly in the recent past~\cite{park2022_bot_SLR, suhaili2021_bot_SLR}, developing, adopting, and interacting with them is still problematic~\citeA{wessel_2021_challenges_interacting_with_bot_on_OSS, wessel2021_dont_disturb_me_botChallenges}. For this reason, the second research question aimed at categorizing and mapping the challenges faced by practitioners, along with possible solutions.

\sterqbox{RQ\textsubscript{2}—Challenges}{What are the challenges related to using bots and conversational agents in software engineering?}

Based on the considerations above, we proceeded with our analysis by extracting a list of known best practices from the formal and grey literature, with the goal of helping practitioners adopt and develop bots for engineering purposes.
Moreover, we aimed to identify possible solutions to the limitations elicited while answering the second research question.

\sterqbox{RQ\textsubscript{3}—Best Practices}{What are the best practices for using bots and conversational agents in software engineering?}

Finally, the last research question aimed to investigate the impact of adopting bots, collecting and reporting the benefits of their usage. This led us to the definition of our last research question.

\sterqbox{RQ\textsubscript{4}—Benefits}{What are the benefits of using bots and conversational agents for software engineering?}

%%%%%%%%%%%%%%%%%%%%%%%%%%%%%%%%%%%%%%%%%%%%%%%%%%%%%%%
\subsubsection{Search Phase}

%\topar{Velocissima descrizione di cosa tratta questa fase citando \cite{keele2007_slr_guidelines_Kitchenham} e \cite{garousi2019_mlr_guidelines}}
The Search Phase consists of two sub-phases.

% : (1) Selecting the data sources from which to extract the literature and (2) Defining the search string to use on the data sources.
% In this section, these two steps are discussed.
%%%%%%%%%%%%%%%%%%%%%%%%%
\paragraph{\textbf{Data Sources Selection.}}
This phase aims to identify the most reliable databases to extract the literature and start our process.
This step differs between formal and grey literature.

\begin{itemize}
    \item \textit{Formal Literature.} To collect the formal literature, we selected \emph{Scopus},\footnote{\emph{Scopus} website: \url{https://www.scopus.com/home.uri}} \emph{IEEE Xplore},\footnote{\emph{IEEE Xplore} website: \url{https://ieeexplore.ieee.org/Xplore/home.jsp}} and \emph{ACM Digital Library}\footnote{\emph{ACM Digital Library} website: \url{https://dl.acm.org/}} as data sources. 
    These databases are considered the top three among the computer science formal literature sources,\footnote{The top list of computer science research databases: \url{https://paperpile.com/g/research-databases-computer-science/}} other than have been used in various other literature reviews~\cite{kitchenham2009_SLR_definition, islam2019_mlr_example_1, garousi2017_mlr_example_1, garousi2016_mlr_example_2}. Furthermore, they were recommended in the guidelines~\cite{keele2007_slr_guidelines_Kitchenham, garousi2019_mlr_guidelines} since they can provide a complete overview of the published research.
    As recommended by~\cite{keele2007_slr_guidelines_Kitchenham, garousi2019_mlr_guidelines}, we took all the results extracted from the first research and filtered them using eligibility criteria.
    \item \textit{Grey Literature.} To search the grey literature, we used the Google search engine.
    The reason behind our choice is that Google (1) has been used by other MLRs on Software engineering~\cite{islam2019_mlr_example_1, garousi2017_mlr_example_1, garousi2016_mlr_example_2}, and (2) is recommended in various MLR guidelines~\cite{garousi2019_mlr_guidelines, adams2017_grey_literature_inclusion_checklist}. 
    However, it is worth noting that deciding when to stop the search process is a complex problem---mainly caused by the high volume of available items~\cite{garousi2019_mlr_guidelines, ogawa1991_MLR_definition}. 
    For this reason, we adopted two different choices---as recommended by Garousi et al.~\cite{garousi2019_mlr_guidelines, garousi2017_mlr_example_1, garousi2016_mlr_example_2}:
    \begin{itemize}
    \item We stop the research when no new concepts emerge from the search results anymore, i.e., \emph{Theoretical saturation};
    \item We applied the \emph{Effort Bounded} that regards the inclusion of the top N search engine hits only. We analyzed the first ten pages produced by Google, observing that its algorithm retrieves and shows the most relevant results in the first few pages.
\end{itemize}
\end{itemize}

%%%%%%%%%%%%%%%%%%%%%%%%%
\paragraph{\textbf{Search String Design.}} In this phase, we defined different search strings to collect the formal and grey literature. In particular, we used the knowledge obtained during the initial examination to draft the search string. Its definition was based on (i) key terms on the topic of interest–gathered from relevant papers, (ii) synonyms and alternative terms, and (iii) logical connectors to combine different terms in different ways.

The initial version of the search string included basic terms such as “bots” and “software engineering.” During the preliminary exploration phase (Section \ref{par:initial_experimentation}), we began to use the string and improve it in accordance with the results obtained. We focused on the individuation of synonyms, which led first to the expansion of the search string for formal literature and then to the detailed definition of the different strings for grey literature.

As for the formal literature, the final search string is shown in the box below.
As it is possible to see, we did not include key terms of our research questions, e.g., \emph{``goals''}, since we aimed to collect as much formal literature as possible, even if the phase led to high effort.

\stesearchstringbox{Formal Literature Search String}{(``bot(s)'' $OR$ ``chatbot(s)'' $OR$ ``chat bot(s)'' $OR$ ``conversational agent(s)'') $AND$ (``software engineering'')}

As for the grey literature, following the guidelines provided by~\cite{kumara2021_sglr_example_1_infrastructure_code,garousi2016_mlr_example_2}, we started from a general search string to define a more specific one based on the goal of each research question. 
The reason behind our choice arose from the noise present in the output that characterizes informal databases, like the Google search engine. 
Table \ref{table:grey_search_string} reports the search strings for the grey literature based on the research questions of our study.

\begin{table}[b]
    \centering
    \caption{Search strings for grey literature.}
    
    \rowcolors{1}{graytable}{white}
    \resizebox{\linewidth}{!}{
    \begin{tabular}{p{0.025\linewidth} p{0.85\linewidth} P{0.125\linewidth}}
        \toprule
        \rowcolor{black}
        \textbf{\textcolor{white}{\#}} & \textbf{\textcolor{white}{Search String}} & \textbf{\textcolor{white}{RQ}}\\
        \bottomrule
        0 & {(``bot(s)'' $OR$ ``chatbot(s)'' $OR$ ``chat bot(s)'' $OR$ ``conversational agent(s)'') $AND$ (``software engineering'')} & General String\\
        1 & {(``bot(s)'' $OR$ ``chatbot(s)'' $OR$ ``chat bot(s)'' $OR$ ``conversational agent(s)'') $AND$ (``software engineering'') $AND$ (``goal(s)'' $OR$ ``motivation(s)'' $OR$ ``objective(s)'')} & RQ$_1$\\
        2 & {(``bot(s)'' $OR$ ``chatbot(s)'' $OR$ ``chat bot(s)'' $OR$ ``conversational agent(s)'') $AND$ (``software engineering'') $AND$ (``challenge(s)'' $OR$ ``limitation(s)'' $OR$ ``negative impact(s)'')} & RQ$_2$\\
        3 & {(``bot(s)'' $OR$ ``chatbot(s)'' $OR$ ``chat bot(s)'' $OR$ ``conversational agent(s)'') $AND$ (``software engineering'') $AND$ (``best practice(s)'')} & RQ$_3$\\
        4 & {(``bot(s)'' $OR$ ``chatbot(s)'' $OR$ ``chat bot(s)'' $OR$ ``conversational agent(s)'') $AND$ (``software engineering'') $AND$ (``benefit(s)'' $OR$ ``positive impact(s)'')} & RQ$_4$\\
        \bottomrule
        \rowcolor{white}
        \multicolumn{3}{p{\linewidth}}{\textit{\textbf{Note}: The search string \#0 was the same as the search string used for the formal literature and was used as a base for the search strings for the grey literature.}}\\
    \end{tabular}
    }
    
    \label{table:grey_search_string}
\end{table}

%%%%%%%%%%%%%%%%%%%%%%%%%%%%%%%%%%%%%%%%%%%%%%%%%%%%%%%
\subsubsection{Eligibility Criteria Phase}
In this phase, the aim is to define a series of criteria to apply after collecting the literature for filtering and identifying those that provide direct evidence based on our research questions~\cite{keele2007_slr_guidelines_Kitchenham, garousi2019_mlr_guidelines}. 
The phase consisted of two sub-phases: (1) Defining the formal literature eligibility criteria and (2) Defining the grey literature eligibility criteria. 

As for the \textit{formal literature}, the eligibility criteria are divided into three sets that had to be evaluated for each paper extracted from the search phase.

\begin{enumerate}
    \item \emph{Exclusion Criteria}, namely, criteria that, if met for at least one of them, decree the elimination of the study from the analysis set;
    \item \emph{Inclusion Criteria}, namely, criteria that, if met for at least one of them, decree the inclusion of the study into the analysis set;
    \item \emph{Quality Criteria}, namely, criteria that are used to grade studies and may lead to their elimination if the grade is below a certain threshold.
\end{enumerate}

Table \ref{table:exclusion_inclusion_criteria_formal} shows exclusion and inclusion criteria.
We defined such criteria according to the guidelines by Kitchenham et al.~\cite{keele2007_slr_guidelines_Kitchenham, garousi2019_mlr_guidelines} and past literature in the Software Engineering field~\cite{islam2019_mlr_example_1,kumara2021_sglr_example_1_infrastructure_code}. Each criterion could be only \textsl{True} or \textsl{False}. 
Using these filters, we could exclude all preliminary research results, e.g., workshops or posters, and avoid considering duplicated papers derived from the combination of three data sources. Nevertheless, we decided not to exclude workshop papers reported in the Repository for Bot-related Research~\cite{bot_se_research_repository} since these could be highly related to this work's main reference community.
In addition, we included EC6 in order to ensure that the collected papers were concerned with the topic of interest. 
Moreover, as for other LRs~\cite{park2022_bot_SLR,suhaili2021_bot_SLR,lewandowski2021_bot_SLR,islam2019_mlr_example_1,garousi2016_mlr_example_2}, we defined our inclusion criteria to match our objective and research questions.
Regarding our exclusion criteria, it is worth discussing the EC5—Studies written before 2014. We decided to apply such criteria based on the findings of previous LRs~\cite{park2022_bot_SLR, suhaili2021_bot_SLR}, which report that literature on software bots—particularly in their communication-oriented form—started growing after 2014.
The other exclusion criteria are widespread among other LRs, and some have been applied using the filters integrated into the selected databases for data source selection.

Regarding the quality criteria, we defined them as the following:
\begin{itemize}
    \item[\textbf{Q1}] Is the valuable information for the review clearly reported and indicated in the paper?
    \item[\textbf{Q2}] Are the findings clearly reported in the paper?
    \item[\textbf{Q3}] Are the RQs or objectives clearly defined?
    \item[\textbf{Q4}] How many pages does the paper consist of? (with respect to the top venue formats)
\end{itemize}
The first three questions could be answered as \textsl{``Yes''}, \textsl{``Partially''}, \textsl{``No''}. Each answer corresponded to a numerical value, i.e., \textsl{`1'}, \textsl{`0.5'}, \textsl{`0'}, respectively. The sum of these values reflects the quality score of the study. If the score was lower than 2, we excluded it from the review.
The last one could be answered as \textsl{``1 to 2''}, \textsl{``3 to 4''}, and \textsl{``4 and up''}.

Concerning the \emph{grey literature} eligibility criteria evaluation, we followed the guideline proposed by Garousi et al.~\cite{garousi2019_mlr_guidelines} combining inclusion and exclusion criteria with quality assessment criteria (Appendix A—Table 1).

\begin{table}
    \centering
    \caption{Exclusion and Inclusion Criteria for Formal Literature.}
    
    \rowcolors{1}{graytable}{white}
    \resizebox{\linewidth}{!}{
    \begin{tabular}{p{0.5\linewidth} p{0.5\linewidth}}
        \toprule
        \rowcolor{black}
        \multicolumn{1}{P{0.5\linewidth}}{\textbf{\textcolor{white}{Exclusion Criteria}}} & \multicolumn{1}{P{0.5\linewidth}}{\textbf{\textcolor{white}{Inclusion Criteria}}}\\
        \bottomrule
        \begin{itemize}[
            nosep,
            before={\begin{minipage}[t]{\hsize}},
            after ={\end{minipage}} 
            ]  
          \item[EC1] Studies not written in English.
          \item[EC2] Posters.
          \item[EC3] Workshop papers.
          \item[EC4] Study not entirely available for free.
          \item[EC5] Studies written before 2014.
          \item[EC6] Studies not related to the topic of interest.
          \item[EC7] Duplicated studies.
        \end{itemize} 
        &
        \cellcolor{graytable}
        \begin{itemize}[
            nosep,
            before={\begin{minipage}[t]{\hsize}},
            after ={\end{minipage}} 
            ] 
          \item[IC1] Studies that report about the uses of bots in Software Engineering.
          \item[IC2] Studies that report about the challenges of bots in Software Engineering.
          \item[IC3] Studies that report about the best practices in the use of bots in Software Engineering.
          \item[IC4] Studies that report about the benefits of using bots for Software Engineering.
        \end{itemize}
        \\
        \bottomrule
    \end{tabular}
    }
    
    \label{table:exclusion_inclusion_criteria_formal}
\end{table}

\subsection{Conducting Phase}

\subsubsection{Studies Selection Phase}
In the studies selection phase, the aim was to apply the planned strategy to collect the literature~\cite{keele2007_slr_guidelines_Kitchenham, garousi2019_mlr_guidelines}. The analysis included all studies published from 2014 to 4/4/2024 {(Appendix A—Fig. 2~\cite{online_appendix}).

As for the extraction and selection of the formal literature, we started by running the search string on the three selected databases, e.g., Scopus, obtaining a set of 2108 studies. We imported the results of the three datasets in Excel, and the first two authors of the paper started the inclusion and exclusion process. As a first filtering step, we started deleting duplicates by the three datasets merged. Then, the remaining set of papers was divided in two, and the first two authors started applying inclusion and exclusion criteria on the two separate sets; the authors determined if they include or exclude the paper starting by reading the title, abstract, and—if necessary—the entire body of the studies. After analyzing 20 papers each, the two authors reviewed corresponding work and discussed; no conflict arose, and then the authors continued applying the exclusion and inclusion criteria, resulting in 173 studies included and 1935 excluded. Next, an ulterior session of filtering was applied jointly by all the authors on the 173 studies that remained; this aimed at identifying potential errors in the inclusion process. After this step, we excluded other 64 studies, obtaining a final set of 109 studies. Finally, the first two authors of the paper applied the quality criteria, obtaining a final set of 79 studies from the white literature.

Regarding the grey literature, the first author started using the search strings on the Google Search Engine to collect the literature. For each search string, the first 10 pages were checked, and the items judged related to the work objectives were imported into an Excel file until theoretical saturation was reached~\cite{garousi2019_mlr_guidelines}. By doing so, we obtained a set of 52 items. Then, the first two authors of the paper started applying all the eligibility criteria simultaneously, as Garousi et al.~\cite{garousi2019_mlr_guidelines} suggested. Such a process led to identifying 28 items for the grey literature.

After completing the extraction and selection process for both the formal and grey literature, we combined the two sets. This final step resulted in a comprehensive collection of 107 studies, which formed the basis for the subsequent data extraction phase.

Regarding the distribution of literature for type—i.e., blogs (for grey literature), conferences, workshops, and journals—most of the works are conference papers (52), while 28 are blogs, 13 are workshop papers, and 14 are journal papers. This is not a surprise, seeing that the research topic specifically concerning conversational agents—is extremely recent and tends to be more relegated to conferences and live discussion settings.
Moreover, as expected, a large amount of the literature comes from the \emph{International Workshop on
Bots in Software Engineering (BotSE)}; it is the main research community on bots and CAs for Software Engineering purposes. In addition, there are also top venues on Software Engineering, e.g., \emph{International Conference on Software Engineering (ICSE)} and \emph{Transaction on Software Engineering (TSE)} (Appendix A—Fig.3). More details are in the online appendix~\cite{online_appendix}.

\begin{comment}
\begin{figure}
    \centering
    \includegraphics[width=0.75\linewidth]{figures/methodology/Event_Distribution_Pie_Chart.pdf}
    \caption{Selected formal literature's venues.}
    \label{fig:venues}
\end{figure}
\end{comment}

%%%%%%%%%%%%%%%%%%%%%%%%%%%%%%%%%%%%%%%%%%%%%%%%%%%%%%%
\subsubsection{Data Extraction Phase}

%\topar{Velocissima descrizione di cosa tratta questa fase citando \cite{keele2007_slr_guidelines_Kitchenham} e \cite{garousi2019_mlr_guidelines}}
In the data extraction phase, we started from our objectives and research questions to define a series of data extraction forms, i.e., sets of questions to identify useful elements to answer a research question in the studies~\cite{keele2007_slr_guidelines_Kitchenham, garousi2019_mlr_guidelines}.

Aligning with other similar works, we defined the extraction forms jointly—through brainstorming and discussion—and then the first author applied them.
We identified and extracted the relevant data using a pre-defined data extraction form from each selected source that we needed to answer the research questions. 
We also extracted some general information---e.g., authors names, published year, and keywords---as recommended by guidelines on literature review~\cite{keele2007_slr_guidelines_Kitchenham, garousi2019_mlr_guidelines}.
To validate and test our forms, we conducted a preliminary extraction on a set of 15 randomly selected sources---like in other literature reviews~\cite{garousi2016_mlr_example_2, garousi2017_mlr_example_1, mantyla2016_mlr_example_1_gamification_of_software_testing, islam2019_mlr_example_1}.

From a practical standpoint, we imported all the papers into a new Excel sheet and added a column for each data form identified. Then, we began extracting information from the papers in random order. If the extracted information was different from what had been extracted previously, we would create a new column in the file under the data form group (following a hierarchical structure). However, if the information extracted from a paper had already been indicated by another paper, meaning the column already existed, we would simply mark the corresponding cell with a sign and add any new information. For example, one of the data forms for RQ1 was “what are the reasons for using bots in SE?” Through the paper by Van Tonder and Le Goues~\citeA{van_2019_programming_repair_bots}, we extracted the information “repair bots, to identify issues in the code and automatically resolve them.”. Subsequently, the paper by Monperrus~\citeA{monperrus_2019_automated_bug_fixes} also reported the same information, so an "X" was added to the corresponding cell in the repair bots column.

All the extracted data are in our online appendix~\cite{online_appendix}. We maintained traceability between the forms and the research questions as recommended by MLR guidelines~\cite{garousi2019_mlr_guidelines}.

%%%%%%%%%%%%%%%%%%%%%%%%%%%%%%%%%%%%%%%%%%%%%%%%%%%%%%%
\subsubsection{Data Synthesis and Analysis Phase}

%\topar{Velocissima descrizione di cosa tratta questa fase citando \cite{keele2007_slr_guidelines_Kitchenham} e \cite{garousi2019_mlr_guidelines}}
In the data synthesis phase, we synthesized data previously collected to present them to the target audience in a readable and usable way. 

After extracting and dividing the data between the various form groups, the first two authors analyzed them. In this way, we avoided losing important information or misunderstanding data \cite{garousi2019_mlr_guidelines, keele2007_slr_guidelines_Kitchenham}. If the first two authors were unsure how to synthesize a study's results, the plan was to ask the remaining two authors for help; this case never occurred.

To conduct a correct data synthesis, we relied on two qualitative analysis methods, namely, \emph{narrative synthesis}—i.e., the description and ordering of primary evidence with commentary and interpretation of a set of data—and \emph{thematic analysis}—i.e., the summary of studies under a set of recurrent themes in literature~\cite{cruzes2010_synthesizing_evidence_in_SE}. 
First, we used narrative synthesis to perform an initial analysis of the gathered data from the previous phase. Then, we used thematic analysis to classify and categorize the data.

Concretely, we started with the data extracted in the previous phase and began annotating them with descriptions and labeled comments using keywords, such as “Repair bots” and “Maintenance bots”. Subsequently, we identified descriptions that closely resembled each other or appeared to be related and began aggregating them into individual categories. Each category became a new column in the Excel file, and through iterative repetition of this process, some columns were grouped into thematic similarity groups. For instance, “Repair bots” and “Bots used for identifying flaky tests” were grouped under the category test bots. To maintain traceability, within the file, all papers contributing to the generation of a category were marked with an “X” corresponding to the paper and the column dedicated to that category.

The first two authors jointly conducted the steps mentioned above—at least for a subset of the items—to identify possible discussion points to strengthen the final results. At the end of the process, we obtained a new set of insights and information categorized under four macro categories—one for each research question. 
We reported all our findings in the online appendix for our work~\cite{online_appendix}.

%%%%%%%%%%%%%%%%%%%%%%%%%%%%%%%%%%%%%%%%%%%%%%%%%%%%%%%
\subsection{Results Reporting Phase}

%\topar{Velocissima descrizione di cosa tratta questa fase citando \cite{keele2007_slr_guidelines_Kitchenham} e \cite{garousi2019_mlr_guidelines}}
Based on the target audience of this review, i.e., practitioners and researchers, we relied on the guidelines provided by Garousi et al.~\cite{garousi2019_mlr_guidelines} to enhance the readability---particularly from the practitioners' side. 
For such a purpose, we (1) decided to include a practical-oriented Discussion section (Section \ref{Sec_addressing_limitation}), (2) asked practitioners for feedback on our results, and (3) adopted various tools to enhance results summaries.
%Moreover, we provided all the material used in this study in our online appendix\todo{cita l'appendice online}.

\section{Analysis of the Results}
\label{sec:results}

In the following section, we describe the results achieved for each research question and report some statistics of our papers collection.

%%%%%%%%%%%%%%%%%%%%%%%%%%%%%%%%%%%%%%
%%%%%%%%%%%%%%%%%%%%%%%%%%%%%%%%%%%%%%
\subsection{RQ\textsubscript{1}: On the use of bots and conversational agents for software engineering purposes}
%\filomena{è epressa diversamente precedentemente}
%\topar{obiettivo della domanda di ricerca}
As a first goal, we aimed to identify the motivations and objectives behind the adoption of bots for software engineering purposes. Our purpose was to gather general and fundamental pieces of information to (1) define a taxonomy to categorize bots based on their use and (2) better perform the following research step.

%\topar{come si sono riportati i risultati?}
Our investigation into answering the first research question (described in Section \ref{sec:methodology}) led to the identification of 13 fundamental motivations for adopting bots and conversational agents in software engineering. In order to provide a more comprehensive and understandable taxonomy, we categorize each of the fundamental motivations into 4 macro categories of bots (Depicted in Appendix B—Fig. 1~\cite{online_appendix}). The identified motivations are not mutually exclusive; the same bot can be used for different motivations according to a plethora of reasons, e.g., working roles, needs, and types of projects. In the following section, we elaborated on each category and associated motivations, corroborating the discussion with the primary studies and grey literature supporting them.

\begin{comment}
\begin{figure}[b]
    \centering
    \includegraphics[width=1\linewidth]{figures/motivation_review.pdf}
    \caption{Motivation for adopting bots and CAs.}
    \label{fig:motivations}
\end{figure}
\end{comment}

%%%%%%%%%%%%%%%%%%%
\subsubsection{\underline{Software Product Bots}}
In this category, we group all the bots and associated motivations of use used to perform automatic operations on source code or related artifacts and activities. We identified the following sub-categories: 

\begin{description}[leftmargin=0.3cm]
    \item[Development.] This sub-category refers to bots that \textit{are used to generate source code instead of developers or other practitioners}. In our investigation, we found 11 studies, 9 from the white literature~\citeA{gilson_2019_NLP_and_collaborativeSE,ren_2020_collaborative_modelling_bot_vs_online_tools,gilson_2020_recording_design_decision_on_the_fly,purwoko2023analysis,saini2022automated,almonte2021automating,qasse2023chat2code,scoccia2023exploring} and 2 from the grey one~\citeB{new_grey_2, new_grey_6} ($\approx 10.28\%$ of the entire dataset).\\
    Part of the literature focuses on the use of bots (in most cases empowered by Generative AI) to automatically write source code~\citeA{purwoko2023analysis,scoccia2023exploring}\citeB{new_grey_2, new_grey_6} and code for machine learning purposes~\citeA{purwoko2023analysis}; in this way, practitioners can save time and increase productivity. Nevertheless, Generative AI is still raw in this task, and software engineers do not rely heavily on it~\citeA{purwoko2023analysis,scoccia2023exploring}\citeB{new_grey_2, new_grey_6}. Another part of the literature combined bots and models-based language to generate source code and entire applications~\citeA{gilson_2019_NLP_and_collaborativeSE,ren_2020_collaborative_modelling_bot_vs_online_tools,saini2022automated,almonte2021automating,qasse2023chat2code,perez_2019_CA_flexible_modelling}; in this context, bots act like facilitators in creating models representing complex systems and modules. Interestingly, bots to generate smart contracts~\citeA{ren_2020_collaborative_modelling_bot_vs_online_tools, qasse2023chat2code} and recommendation systems~\citeA{almonte2021automating} (thus, bots used to generate other bots) are being proposed.
    
    \item[Refactoring.] This sub-category refers to bots that are used to \textit{(1) improve source code} and (2) \textit{avoid and/or resolve technical debts by means of refactoring operations}, i.e., \textit{Refactoring Bots}. Our dataset has 11 studies, 8 from the white literature~\citeA{wyrich_2019_bot_for_code_refactoring,alizadeh_2019_refbot_software_refactoring_bot,hu_2019_improving_feedback_pull_request_with_bots,carvalho_2020_C_3PR_fix_static_analysis_violations,wyrich_2020_acceptance_of_refactoring_bots,wyrich_2021_bots_interaction_with_pull_requests,serban_2021_fixes_for_statyc_analysis_warning,purwoko2023analysis} and 3 from the grey one~\citeB{best_bots_to_improve_your_software_development_process,four_Things_You_Absolutely_Need_to_Know_About_Software_Bots,software_bot_explained} ($\approx 10.28\%$ of the entire dataset).\\
    Some of them are related to technical debt~\citeA{wyrich_2019_bot_for_code_refactoring, hu_2019_improving_feedback_pull_request_with_bots}; for example, bots in such a context are used to refactor code smells automatically. Moreover, bots can also be used to support quality aspects of source code, e.g., by means of quality criteria~\citeA{alizadeh_2019_refbot_software_refactoring_bot} or code readability~\citeB{best_bots_to_improve_your_software_development_process}, and provide fixes when static analysis tools detect violations~\citeA{carvalho_2020_C_3PR_fix_static_analysis_violations, wyrich_2020_acceptance_of_refactoring_bots}. In the recent past, refactoring has also been automated using Generative AI, even without satisfying results~\citeA{purwoko2023analysis}.
    
    \item[Testing.] This sub-category refers to bots that \textit{support testers and similar roles during the testing activity and for executing test code}. Our dataset has 11 studies, 10 from the white literature~\citeA{wyrich_2020_acceptance_of_refactoring_bots,storey_2016_bot_uses_taxonomy,urli_2018_repairnator_project,paikari_2018_framework_chatbot_and_their_future,van_2019_programming_repair_bots,hu_2019_improving_feedback_pull_request_with_bots,monperrus_2019_automated_bug_fixes,erlenhov_2020_bot_characteristics_and_challenges_from_practitioners,erlenhov_2020_challenges_and_guidelines_for_test_bot_testing,okanovi_2020_chatbot_to_support_load_testing} and 1 from the grey one ($\approx 10.28\%$ of the entire dataset).\\
    Most of the literature regards the use of bots for automation of test cases execution~\citeA{storey_2016_bot_uses_taxonomy,urli_2018_repairnator_project,van_2019_programming_repair_bots,hu_2019_improving_feedback_pull_request_with_bots,monperrus_2019_automated_bug_fixes,erlenhov_2020_challenges_and_guidelines_for_test_bot_testing,okanovi_2020_chatbot_to_support_load_testing}. Furthermore, we found that many bots are used to identify bugs and perform automatic fixes~\citeA{storey_2016_bot_uses_taxonomy,urli_2018_repairnator_project,van_2019_programming_repair_bots,hu_2019_improving_feedback_pull_request_with_bots,monperrus_2019_automated_bug_fixes,wyrich_2020_acceptance_of_refactoring_bots}\citeB{best_bots_to_improve_your_software_development_process}, i.e., \emph{Repair Bot}. Last, some works propose bots to (i) support developers in writing test code~\citeA{okanovi_2020_chatbot_to_support_load_testing} and (ii) implement heuristics to detect \emph{flaky tests}~\citeA{storey_2016_bot_uses_taxonomy}—i.e., tests exhibiting both a passing and failing behavior when run against the same code~\cite{luo2014_test_flakiness_definition}.
    
    \item[Configuration.] This sub-category concerns bots used to support practitioners during the setup and configuration processes of new software projects. Our dataset has 6 studies, 4 from the white literature~\citeA{storey_2016_bot_uses_taxonomy,dey_2020_bot_that_commit_code,dey_2020_bot_commits,wyrich_2020_acceptance_of_refactoring_bots,he2023automating} and 1 from the grey one~\citeB{software_bot_explained} ($\approx 5.61\%$ of the entire dataset).\\
    From the identified literature, two studies propose bots for automatically creating/modifying software configuration files or similar~\citeA{dey_2020_bot_that_commit_code,dey_2020_bot_commits}, and four studies~\citeA{storey_2016_bot_uses_taxonomy,wyrich_2020_acceptance_of_refactoring_bots,he2023automating}\citeB{software_bot_explained} report about bots used to manage and monitor project dependencies.

    \item[CI/CD.] This sub-category regards the usage of bots \textit{to support everyday activities in the context of CI/CD}. In our investigation, we found 8 studies, 4 from the white literature~\citeA{storey_2016_bot_uses_taxonomy,beschastnikh_2017_research_and_bot,wessel_2019_bot_close_abbandoned_pull_request_and_issues,erlenhov_2020_bot_characteristics_and_challenges_from_practitioners} and 4 from the grey one~\citeB{best_bots_to_improve_your_software_development_process,storey2020_botse,software_bot_explained,new_grey_1} ($\approx 7.48\%$ of the entire dataset).\\
    Some studies analyze the usage of bots for monitoring software systems (for example, on \textsc{GitHub}) and performing automatic operations if certain conditions are met (e.g., automatically build the system on branch merge) ~\citeA{storey_2016_bot_uses_taxonomy,erlenhov_2020_bot_characteristics_and_challenges_from_practitioners}. Finally, five works study the use of bots to automatically trigger actions when specific criteria—related to no-code platforms—are met~\citeA{storey_2016_bot_uses_taxonomy,beschastnikh_2017_research_and_bot,wessel_2019_bot_close_abbandoned_pull_request_and_issues}\citeB{best_bots_to_improve_your_software_development_process,storey2020_botse}. For example, automatically creating an issue on GitHub when a bug is reported, or a task is assigned in the collaboration platform of the team (e.g., \textsl{Trello} and \textsl{Jira}). Moreover, some bots to support DevOps-related activities started to emerge~\citeB{new_grey_1}.
\end{description}

%%%%%%%%%%%%%%%%%%%
\subsubsection{\underline{Software Process Bots}}

Under this category, we group all the bots used to improve and enhance communication, collaboration, and management activities. We identified the following three sub-categories:

\begin{description}[leftmargin=0.3cm]
    \item[Team.] This sub-category refers to bots used \textit{to support practitioners in handling team quality attributes like turnover and socio-technical aspects}. We found 6 studies, 3 from the white literature~\citeA{mirsaeedi_2020_mitigating_turnover_with_bot, voria2022cadocs, wang2023optimizingEliteDev} and 3 from the grey one~\citeB{best_bots_to_improve_your_software_development_process, storey2020_botse, bots_and_AI_in_project_management} ($\approx 5.61\%$ of the entire dataset).\\
    Regarding the motivations, some bots can be used to support management activities such as risk management~\citeB{bots_and_AI_in_project_management} and \textit{community smells} management~\citeA{voria2022cadocs}, i.e., anti-patterns precursors of social debt~\cite{tamburri2019software}. Moreover, a study proposes to use bots specifically to reduce \emph{turnover}—the phenomenon of continuous influx and retreat of human resources in a team~\cite{foucault2015_impact_of_turnover}—in an open-source software community by improving the reviewer selection process in code review~\citeA{mirsaeedi_2020_mitigating_turnover_with_bot}. More related to the open-source communities, bots are used to help onboard new contributors~\citeB{best_bots_to_improve_your_software_development_process} and alleviate burnout and stress~\citeA{wang2023optimizingEliteDev}\citeB{storey2020_botse} by—for example—supporting core contributors' activities.

    \item[Communication and Collaboration.] This sub-category concerns bots used \textit{to help communication and collaboration between practitioners}. During our investigation, we found 8 studies: 7 from the white literature~\citeA{hefny2021proactive,kim_2020_bot_to_facilitate_group_discussion, gilson_2019_NLP_and_collaborativeSE,gilson_2020_recording_design_decision_on_the_fly, paikari_2019_chatbot_for_conflict_detection_and_resolution,mirsaeedi_2020_mitigating_turnover_with_bot, cerezo_2019_builing_an_expert_recommender_bot} and 1 from the grey one~\citeB{storey2020_botse} ($\approx 7.48\%$ of the entire dataset).\\
    Two studies reported that practitioners use bots to facilitate meetings and communication in community chats, encouraging team participation~\citeA{kim_2020_bot_to_facilitate_group_discussion}\citeB{storey2020_botse}. Bots can also assist managers with scheduling and planning meetings~\citeA{hefny2021proactive}. Additionally, bots help detect when two individuals work on the same artifact, preventing code conflicts~\citeA{paikari_2019_chatbot_for_conflict_detection_and_resolution}. Bots can record past decisions to avoid information loss and support future decisions~\citeA{gilson_2019_NLP_and_collaborativeSE,gilson_2020_recording_design_decision_on_the_fly}. They also identify experts to reduce turnover and assist newcomers~\citeA{mirsaeedi_2020_mitigating_turnover_with_bot, cerezo_2019_builing_an_expert_recommender_bot}.
    
    %Two studies reported that practitioners use bots to facilitate meetings and communication in community chats—e.g., Slack—by nudging the participation of all team members~\citeA{kim_2020_bot_to_facilitate_group_discussion}\citeB{storey2020_botse}. Moreover, and more related to the management side, a bot can also be used as a manager's companion for scheduling and planning meetings~\citeA{hefny2021proactive}. Regarding the contribution part, practitioners can insert bots in their work process to detect when two individuals are working on the same artifact~\citeA{paikari_2019_chatbot_for_conflict_detection_and_resolution}, thus avoiding code conflicts and supporting merge operations. More related to the collaboration side, bots can also be integrated into the workflow to record past decisions and rationale to avoid loss of information and support future decisions~\citeA{gilson_2019_NLP_and_collaborativeSE,gilson_2020_recording_design_decision_on_the_fly}. Furthermore, bots are also adopted to record the experts of a component and support their retrieval to reduce turnover and support newcomers~\citeA{mirsaeedi_2020_mitigating_turnover_with_bot, cerezo_2019_builing_an_expert_recommender_bot}.

    \item[Task.] Practitioners adopt bots \textit{to support activities related to task management}, e.g., evaluation, assignation, and planning. During our investigation, we found 5 studies: 3 from the white literature~\citeA{hefny2021proactive, basu_2021_bot_for_distribution_of_service_requests, ni_2019_CrowDevBot_CA_for_software_crowdsourcing} and 2 from the grey one~\citeB{bots_and_AI_in_project_management, bots_and_AI_in_project_management_2} ($\approx 4.67\%$ of the entire dataset).\\
    Some of the bots are adopted to find the right person for the right task by using metrics and developers profiling~\citeA{ni_2019_CrowDevBot_CA_for_software_crowdsourcing, basu_2021_bot_for_distribution_of_service_requests}\citeB{bots_and_AI_in_project_management}. More related to the managers, a work proposed an all-round bot for task management assistance; the focus of such a bot is task monitoring, issues management, planning, and also discussion of task characteristics to assigning it to the right person~\citeA{hefny2021proactive}.

    \item[Code Review.] The review revealed that practitioners use bots to support the code review process, often associating it with pull requests. Our dataset has 6 studies on this topic, 5 from the white literature~\citeA{wessel_2020_code_review_bot_OSS, wessel2022bots_orchestratore, wessel2022qualityBotCodeReview, ribeiro2022together,mirsaeedi_2020_mitigating_turnover_with_bot} and 1 from the grey one~\citeB{storey2020_botse} ($\approx 5.61\%$ of the entire dataset).\\
    In general, researchers report that the adoption of such bots could effectively benefit software projects in different ways.~\citeA{wessel_2020_code_review_bot_OSS, wessel2022qualityBotCodeReview,mirsaeedi_2020_mitigating_turnover_with_bot} For example, turnover decreases, and communication becomes more effective; moreover, pull requests increase in quality, and contributions to open-source projects are more rapid than without the bots. Nevertheless, the risk of introducing noise exists; some studies proposed the adoption of bots to orchestrate code review bots and summarize their activity and outputs more concretely and cleanly~\citeA{wessel2022bots_orchestratore, ribeiro2022together}.

\end{description}

%%%%%%%%%%%%%%%%%%%

\subsubsection{\underline{Knowledge Bots}}

Under the category Knowledge Bots, we group all the bots used to store, share, and manage information and knowledge about the software project. We identified the following sub-categories:

\begin{description}[leftmargin=0.3cm]
    \item[Documentation.] Practitioners use bots \textit{to support documentation writing}. Our dataset has 10 studies regarding this type of bot, 9 from the white literature~\citeA{storey_2016_bot_uses_taxonomy,paikari_2018_framework_chatbot_and_their_future,matthies_2019_bot_and_agile_retrospectives,gilson_2019_NLP_and_collaborativeSE,dey_2020_bot_that_commit_code,dey_2020_bot_commits,zhang2024automatic,ahmad2023towardsLLMArchitecture,ren_2020_collaborative_modelling_bot_vs_online_tools} and 1 from the grey one~\citeB{storey2020_botse} ($\approx 9.35\%$ of the entire dataset).\\
    Several works focus their attention on bots used to automatically create and update documenting artifacts~\citeA{gilson_2019_NLP_and_collaborativeSE,dey_2020_bot_that_commit_code,dey_2020_bot_commits}\citeB{storey2020_botse} (e.g., code comments, documents, and reports). The state of the art~\citeA{storey_2016_bot_uses_taxonomy,ren_2020_collaborative_modelling_bot_vs_online_tools} has proposed bots to automatize the writing of notes for new software product releases, using NLP to synthesize developers' commit messages and comments. Moreover, practitioners started to use Generative AI to write code commits messages~\citeA{zhang2024automatic,ahmad2023towardsLLMArchitecture}. Other studies propose bots to improve and speed up agile processes by automatizing the agile retrospective step by collecting metrics from different sources~\citeA{matthies_2019_bot_and_agile_retrospectives,ren_2020_collaborative_modelling_bot_vs_online_tools}. Last, bots are also used to record and document design decisions and, for some of them, generate static or behavioral models on-the-fly~\citeA{gilson_2019_NLP_and_collaborativeSE,ren_2020_collaborative_modelling_bot_vs_online_tools} (e.g., UML graphs).

    \item[Metrics.] Bots in the Metrics category are used \textit{to monitor the team environment (both product and process) and collect metrics}. We found 7 studies related to this category, 4 from the white literature~\citeA{storey_2016_bot_uses_taxonomy,kumar_2019_sankie,erlenhov_2020_bot_characteristics_and_challenges_from_practitioners,lin_2020_bot_framework_for_service_development_assistance} and 3 from the grey one~\citeB{bots_and_AI_in_project_management,bots_and_AI_in_project_management_2,new_grey_1} ($\approx 6.54\%$ of the entire dataset).\\
    Some studies are related to collecting and analyzing product and process metrics, allowing practitioners to make informed decisions~\citeA{kumar_2019_sankie,lin_2020_bot_framework_for_service_development_assistance}\citeB{bots_and_AI_in_project_management,bots_and_AI_in_project_management_2}; this is particularly important when dealing with services and micro-services applications~\citeA{lin_2020_bot_framework_for_service_development_assistance}. Moreover, also in this case, bots seem to be used as managers' peers to improve the overall management process~\citeB{bots_and_AI_in_project_management,bots_and_AI_in_project_management_2}. Furthermore, some bots have also been adopted to support DevOps activities by computing metrics and suggesting ways to improve the pipeline~\citeA{kumar_2019_sankie}\citeB{new_grey_1}.
\end{description}

%%%%%%%%%%%%%%%%%%%

\subsubsection{\underline{Emergent Bots}}

In this last category we have included three reasons for adopting bots that have been emerging in recent times and that we believe are worthy of mention although still in an embryonic stage. 

\begin{description}[leftmargin=0.3cm]

    \item[Digital Twins Bots.] Bots in this category are integrated into the development workflow \textit{to support specific developers and act as companions}. We identified 19 studies about this type of bot, 16 from the white literature~\citeA{storey_2016_bot_uses_taxonomy,carr2016automatic,tian2017APIbot,paikari_2018_framework_chatbot_and_their_future,hu_2019_improving_feedback_pull_request_with_bots,fukui_2019_bot_suggesting_questions,brown_2019_bots_for_effective_recommendations,khanan_2020_JITBot_just_in_time_defect_prediction_bot,erlenhov_2020_bot_characteristics_and_challenges_from_practitioners,romero_2020_experiences_building_an_answer_bot,dominic_2020_bot_for_SE_newcomers,chatterjee_2021_automatic_extraction_of_opinion,sadi_2021_RAPID_bot_for_designing_web_APIs,robe2022designing,wang2023optimizingEliteDev,robe2022pairProgramming_dataset} and 3 from the grey one~\citeB{best_bots_to_improve_your_software_development_process,storey2020_botse,bots_and_AI_in_project_management_2} ($\approx 17.76\%$ of the entire dataset).\\
    The majority of works report that bots are used to support the retrieval of information regarding general implementation questions and, particularly, API documentation and guidelines~\citeA{storey_2016_bot_uses_taxonomy,tian2017APIbot,fukui_2019_bot_suggesting_questions,romero_2020_experiences_building_an_answer_bot,dominic_2020_bot_for_SE_newcomers,chatterjee_2021_automatic_extraction_of_opinion}\citeB{storey2020_botse}. Then, some studies focus on using bots to enforce the application of formatting guidelines or requirements~\citeA{hu_2019_improving_feedback_pull_request_with_bots,sadi_2021_RAPID_bot_for_designing_web_APIs}\citeB{best_bots_to_improve_your_software_development_process} and good programming practices~\citeA{carr2016automatic}\citeB{best_bots_to_improve_your_software_development_process}. Furthermore, some bots are also used like actual peers during the programming activity (pair-programming bots)~\citeA{robe2022designing,robe2022pairProgramming_dataset}; among these, some are used to detect possible implementation choices that could lead to defects in the code and propose solutions~\citeA{khanan_2020_JITBot_just_in_time_defect_prediction_bot}. Related to the open-source field, some bots act like peers to recommend useful tools to solve a problem~\citeA{brown_2019_bots_for_effective_recommendations}, support elite developers in general tasks, and help them stay focused on the work by eliminating noise~\citeA{wang2023optimizingEliteDev}.

    \item[Orchestration Bots.] Despite being useful for practitioners, a well-known reported problem with bots is the \textit{noise} arising from their use. In short, since bots are still unable to behave according to the operational context, if their interaction is not well designed, it could result in them providing too much information to practitioners, often unhelpful, causing distraction and discomfort~\citeA{labeuf_2017_software_bots,erlenhov_2020_bot_characteristics_and_challenges_from_practitioners,wessel_2021_challenges_interacting_with_bot_on_OSS,saadat_2021_bots_modify_workflow_of_github_teams}\citeB{storey2020_botse}. To solve this problem, some practitioners operationalize bots specifically designed \textit{to orchestrate and summarize the results of other bots}. In such a sense, these Orchestration Bots act like bot managers. We identified 2 studies about this type of bot, both in the white literature~\citeA{wessel2022bots_orchestratore, ribeiro2022together} ($\approx 1.87\%$ of the entire dataset).\\
    Wessel et al.~\citeA{wessel2022bots_orchestratore} proposed a series of guidelines for developing these types of bots, while Ribeiro et al.~\citeA{ribeiro2022together} created \textsc{FunnelBot}, a bot designed using the previously mentioned guidelines to orchestrate pull-request bots.

    \item[Technology Transfer Bots.] Last but not least, and more attractive for researchers, some bots are starting to be used \textit{to support the technology transfer between research and the practitioner field}. Indeed, the limit of research-developed tools is often the poor effort invested in developing their non-functional requirements, resulting in poor quality. Some bots are starting to be developed specifically to surpass such a limitation. We identified 3 studies about this type of bot, 2 in the white literature~\citeA{beschastnikh_2017_research_and_bot, voria2022cadocs} and 1 in the grey one~\citeB{storey2020_botse} ($\approx 2.80\%$ of the entire dataset).\\
    In this sense, Beschastnikh et al.~\citeA{beschastnikh_2017_research_and_bot} and Voria et al.~\citeA{voria2022cadocs} proposed an ecosystem of bots specifically designed to integrate other research-originated bots and make them more easily available to and usable by practitioners.
     
\end{description}

\color{black}

%%%%%%%%%%%%%%%%%%%%%%%%%%%%%%%%%%%%%%
%%%%%%%%%%%%%%%%%%%%%%%%%%%%%%%%%%%%%%
\subsection{RQ\textsubscript{2}: On the challenges of bots and conversational agents for software engineering purposes}

The diffusion of bots and conversational agents for supporting software development processes has also led to the emergence of new challenges for computer scientists. 
For this reason, we investigated the challenges related to the interaction with, adoption, and development of bots for software engineering purposes.

Regarding the existing literature, Wessel et al.~\cite{wessel2021_dont_disturb_me_botChallenges} identified a series of challenges in the interactions with software bots in the open-source development context, with a particular focus on the concept of \emph{``noise''}---namely, annoying bot behaviors, e.g., verbosity and unsolicited actions, which lead to deterioration of communication in teams. 
In terms of contributions, the authors provided a series of challenges experimented by practitioners, categorized under three main categories, i.e., \textit{Interaction}, \textit{Adoption}, and \textit{Development Challenges}.

On the one side, we found challenges already covered by Wessel et al.~\cite{wessel2021_dont_disturb_me_botChallenges}. On the other side, we contributed state of the art with new challenges—indicated in Figure \ref{fig:challenges} in bold and red text color. 
Specifically, we decided to maintain the high-level separation proposed by Wessel et al.~\cite{wessel2021_dont_disturb_me_botChallenges}—the left side of Figure \ref{fig:challenges}—and to re-design the low-level categorization—the right side of the figure—based on the findings of our MLR.
The following subsections report each identified challenge in detail.

\begin{figure}
    \centering
    \includegraphics[width=1\linewidth]{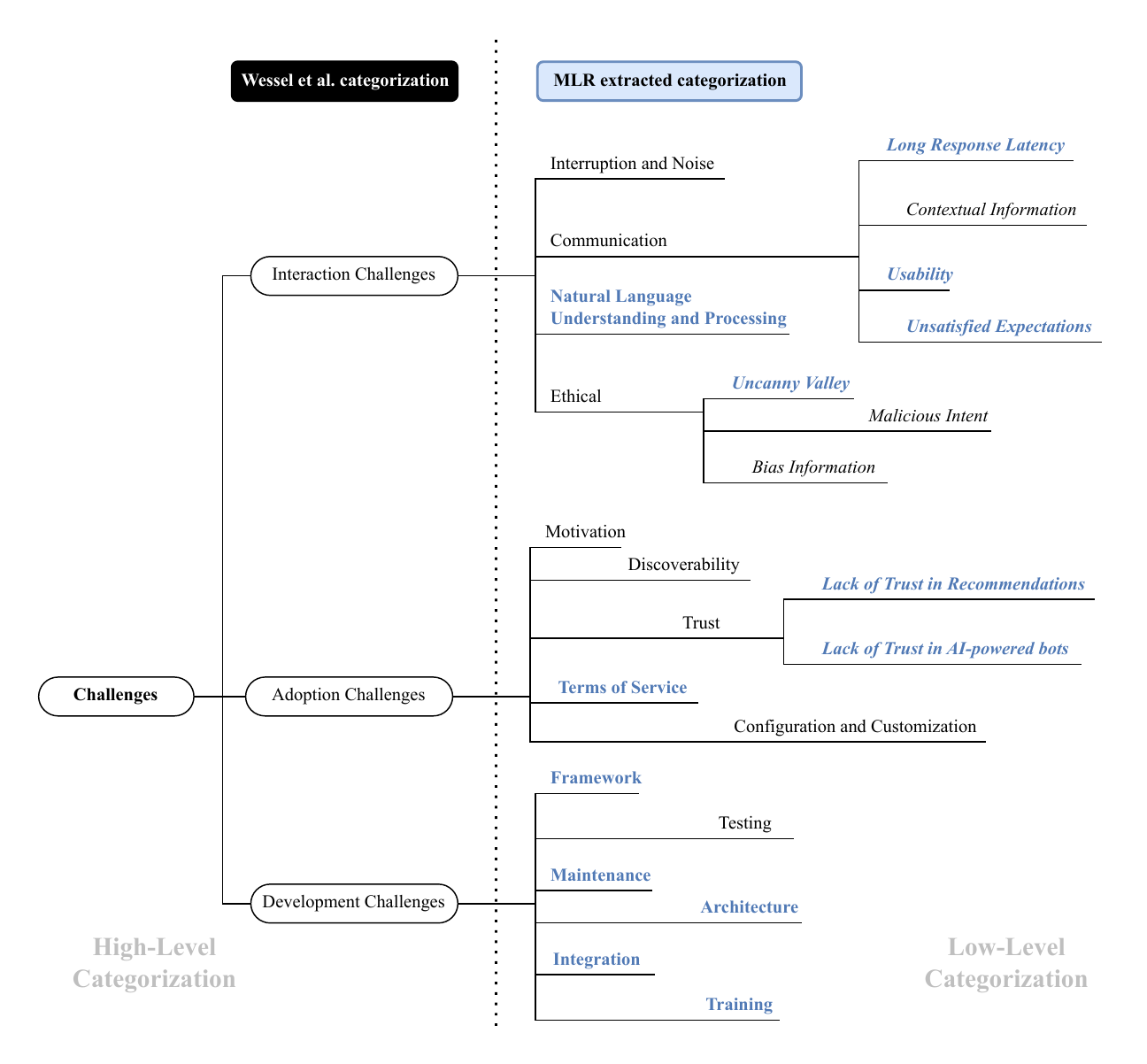}
    \caption{Categorization of challenges associated with bots in the context of software engineering. This categorization is based on the structure proposed by Wessel et al. ~\cite{wessel2021_dont_disturb_me_botChallenges}. New challenges originated from our work are shown in blue.}
    \label{fig:challenges}
\end{figure}

%%%%%%%%%%%%
\subsubsection{\underline{Interaction Challenges}}
\label{sec:interaction_challenges}

Most of the challenges identified by the literature regard the interaction between practitioners and bots. 
Such interaction can be intended as unidirectional—the user receives a message from the bot and reacts to it—and bidirectional—the user and the bot actively exchange messages.
%\topar{un poco di dati statistici, quanti paper parlano di questa categoria di challenge? 10 white e 4 grey}
During our investigation we found 27 studies; 17 from the white literature~\citeA{labeuf_2017_software_bots,srivastava_2019_architecture_applications_with_conversational_components,cerezo_2019_builing_an_expert_recommender_bot,pinheiro_2019_bot_development_challenges_motivations,lee_2019_accelerating_bots_development,castro_2019_bot_usability,melo_2020_what_do_devs_expect_from_CAs,erlenhov_2020_bot_characteristics_and_challenges_from_practitioners,abdellatif_2020_challenges_in_chatbot_development_from_stackoverflow,wyrich_2020_acceptance_of_refactoring_bots,wessel_2021_challenges_interacting_with_bot_on_OSS,saadat_2021_bots_modify_workflow_of_github_teams, abdellatif2021comparison, wessel2022bots_orchestratore, erlenhov2022dependency, de2023meet, robe2022pairProgramming_dataset} and 10 from the grey one~\citeB{storey2020_botse,bots_challenges_for_development,bots_challenges_for_development_2,bots_NLP_or_NLU,7_chatbot_development_challenges,new_grey_1,new_grey_3,new_grey_4,new_grey_5,new_grey_6} that report on this challenge ($\approx 25.23\%$ of the entire dataset).

\begin{description}[leftmargin=0.3cm]
    \item[Interruption and Noise.] One of the most frequently noted challenges involves certain bots exhibiting a tendency towards verbosity or producing irrelevant output~\citeA{labeuf_2017_software_bots,erlenhov_2020_bot_characteristics_and_challenges_from_practitioners,wessel_2021_challenges_interacting_with_bot_on_OSS,saadat_2021_bots_modify_workflow_of_github_teams,wessel2022bots_orchestratore,erlenhov2022dependency}\citeB{storey2020_botse}. This issue, commonly referred to as “noise” in the literature, can significantly affect bot usability as well as team communication and collaboration. The origin of this behavior often lies in the bot's design—sometimes, bots are intentionally programmed to “harass” developers. On the other side, errors in the bot's operation can also result in unexpected and disruptive noise.
    
    \item[Communication.] Another challenge refers to the communication between the user and the bot~\citeA{srivastava_2019_architecture_applications_with_conversational_components,cerezo_2019_builing_an_expert_recommender_bot,pinheiro_2019_bot_development_challenges_motivations,lee_2019_accelerating_bots_development,castro_2019_bot_usability,melo_2020_what_do_devs_expect_from_CAs,erlenhov_2020_bot_characteristics_and_challenges_from_practitioners,abdellatif_2020_challenges_in_chatbot_development_from_stackoverflow,wyrich_2020_acceptance_of_refactoring_bots,wessel_2021_challenges_interacting_with_bot_on_OSS,erlenhov2022dependency,de2023meet}\citeB{storey2020_botse,bots_challenges_for_development,bots_challenges_for_development_2,7_chatbot_development_challenges,new_grey_1,new_grey_3,new_grey_6}. For example, \textbf{ensuring a short response time} comes to be mandatory for bot use: various bots—especially those based on API calls or off-the-shelf modules—are characterized by long responses to users' inputs and consequent frustration in them~\citeA{lee_2019_accelerating_bots_development}\citeB{bots_challenges_for_development_2}. Nevertheless, architectures of this type are often recommended for these software systems, and developers are faced with a trade-off between high modularity and maintainability—given by using such architectures—and fast response time.
    To continue, several works report the difficulty of bots in \textbf{providing practical and contextualized information}: indeed, bots should have information about the environment in which they are operating (e.g., repositories address, commit template, and team composition), but this knowledge is difficult to collect automatically~\citeA{melo_2020_what_do_devs_expect_from_CAs,wyrich_2020_acceptance_of_refactoring_bots,wessel_2021_challenges_interacting_with_bot_on_OSS,erlenhov2022dependency}\citeB{storey2020_botse,bots_challenges_for_development_2} and is often limited by technology limitations of Generative AI tools~\citeA{de2023meet}\citeB{new_grey_3}. Another primary concern regards how bots communicate with the user, specifically in the conversation design, generation of reply messages, and design of the chatbot's graphical user interface~\citeA{srivastava_2019_architecture_applications_with_conversational_components,cerezo_2019_builing_an_expert_recommender_bot,pinheiro_2019_bot_development_challenges_motivations,castro_2019_bot_usability,melo_2020_what_do_devs_expect_from_CAs,erlenhov_2020_bot_characteristics_and_challenges_from_practitioners,abdellatif_2020_challenges_in_chatbot_development_from_stackoverflow,wessel_2021_challenges_interacting_with_bot_on_OSS}\citeB{storey2020_botse,bots_challenges_for_development,7_chatbot_development_challenges}. Moreover, most practitioners struggle to \textbf{provide straightforward and satisfying user interfaces}, and researchers underline the \textbf{lack of guidelines and standards for bots usability}~\citeA{srivastava_2019_architecture_applications_with_conversational_components,cerezo_2019_builing_an_expert_recommender_bot,pinheiro_2019_bot_development_challenges_motivations,castro_2019_bot_usability,melo_2020_what_do_devs_expect_from_CAs,erlenhov_2020_bot_characteristics_and_challenges_from_practitioners,abdellatif_2020_challenges_in_chatbot_development_from_stackoverflow,wessel_2021_challenges_interacting_with_bot_on_OSS}\citeB{storey2020_botse,bots_challenges_for_development,7_chatbot_development_challenges,new_grey_6}. The last challenge is related to the use of Generative AI-powered chatbots; this technology is relatively recent and, like many other disruptive technologies, is at the point of maximum expectation and illusion about its actual use by practitioners based on the Gartner hype cycle. However, the technology still needs to be updated to meet user demands, specifically regarding software engineering professionals. This aspect causes \textbf{disillusionment and frustration because of unsatisfied expectations} when interacting with such types of chatbots~\citeB{new_grey_1}.

    \item[Natural Language Understanding and Processing.] As already described in Section \ref{sec:background_related}, conversational agents' interactions are based and modeled on the concept of \emph{intent}, i.e., a conceptual representation of a possible user request characterized by a set of mandatory information to fulfill the request (called \emph{entities}). Their correct interpretation is crucial for the bot operation and is often based on text interpretation using natural language processing (NLP) and understanding (NLU) algorithms. Therefore, \textbf{training NLP and NLU machine learning models} represented the most mentioned  challenge~\citeA{srivastava_2019_architecture_applications_with_conversational_components,cerezo_2019_builing_an_expert_recommender_bot,abdellatif_2020_challenges_in_chatbot_development_from_stackoverflow,abdellatif2021comparison,robe2022pairProgramming_dataset}\citeB{bots_NLP_or_NLU,new_grey_1,new_grey_4,new_grey_6}; the main problem lies on the fact that a conversational agent works in singular contexts, and it is difficult to (1) extract sentences to train models and (2) reuse sentences from other contexts.
 operation~\citeA{cerezo_2019_builing_an_expert_recommender_bot,wessel_2021_challenges_interacting_with_bot_on_OSS}\citeB{storey2020_botse,new_grey_4,new_grey_5,new_grey_6}. \textbf{ skeptical of cause of past experiences with bots exposing malicious intents}
    For example, practitioners tend to be \textbf{skeptical because of past experiences with bots exposing malicious intents}—e.g., theft of sensitive information or insertion of malicious code. So, interacting with these bots can be challenging because their identification is not always an easy task~\citeA{wessel_2021_challenges_interacting_with_bot_on_OSS}.
    In addition, another challenge is related to a well-known topic such \emph{fairness}~\cite{verma2018_fairness_definition}, intended as the equitable and just treatment—without discrimination—assumed by software systems—particularly in the context of artificial intelligence. Specifically, conversational agents are based on trained machine learning models using datasets containing sentences and assertions by potential users, thus possibly leading the bot to be biased given the different cultures and backgrounds of the individuals who built the dataset, which could \textbf{lead to unexpected and often discriminatory behaviors}~\citeA{wessel_2021_challenges_interacting_with_bot_on_OSS}\citeB{new_grey_4,new_grey_5,new_grey_6}.
    Last but not least, the literature highlighted the phenomena known as \emph{Uncanny Valley}—i.e., the phenomenon whereby a computer behaving like a human being evokes a sense of unease and revulsion in the person interacting with it~\cite{mori2012_uncanny_valley, kageki2012_uncanny_valley}. The phenomenon is extremely diffused, especially in the interaction with conversational agents. It is related to the humanization of robots: people tend to exhibit unpleasant feelings—such as revulsion and uneasiness comparable to perturbation—when interacting with machines with overly ``human'' attitudes. Managing this phenomenon so that \textbf{the bots turn out to be human enough to improve communication without being disturbing} is a difficult challenge~\citeA{cerezo_2019_builing_an_expert_recommender_bot,wessel_2021_challenges_interacting_with_bot_on_OSS}\citeB{storey2020_botse}.
\end{description}

%%%%%%%%%%%%
\subsubsection{\underline{Adoption Challenges}}
\label{sec:adoption_challenges}

%\todo[inline]{\stefano{Da riscrivere perché secondo me è assai bruttina}}

Several studies reported challenges in adopting bots and chatbots for development reasons. 
We found 19 studies: 12 from the white literature~\citeA{storey_2016_bot_uses_taxonomy,van_2019_programming_repair_bots,monperrus_2019_automated_bug_fixes,brown_2019_bots_for_effective_recommendations,erlenhov_2020_bot_characteristics_and_challenges_from_practitioners,brown_2020_recommendation_for_bot_architectures,wyrich_2021_bots_interaction_with_pull_requests,wessel_2021_challenges_interacting_with_bot_on_OSS,he2023automating,robe2022designing,gao2022does,gao2022understandingBotsSentiment} and 7 from the grey one~\citeB{storey2020_botse,software_bot_explained,7_chatbot_development_challenges,new_grey_1,new_grey_4,new_grey_5,new_grey_6} that report on such a category ($\approx 17.76\%$ of the entire dataset).

\begin{description}[leftmargin=0.3cm]
    \item[Motivation and Trust.] According to various studies, the first challenge involves \textbf{motivating practitioners to use bots}~\citeA{monperrus_2019_automated_bug_fixes,erlenhov_2020_bot_characteristics_and_challenges_from_practitioners,wessel_2021_challenges_interacting_with_bot_on_OSS,gao2022understandingBotsSentiment}\citeB{storey2020_botse,software_bot_explained,7_chatbot_development_challenges,new_grey_1}. In fact, various practitioners are skeptical of bots because of bias—they think they can implement malicious behaviors or are useless and noisy, underlining a \textbf{lack of trust}~\citeA{monperrus_2019_automated_bug_fixes,brown_2019_bots_for_effective_recommendations,erlenhov_2020_bot_characteristics_and_challenges_from_practitioners,brown_2020_recommendation_for_bot_architectures,wyrich_2021_bots_interaction_with_pull_requests,he2023automating,robe2022designing,gao2022does,gao2022understandingBotsSentiment}\citeB{storey2020_botse,software_bot_explained,7_chatbot_development_challenges,new_grey_1} of developers in bots' recommendations.Furthermore, \textbf{the problem of trust becomes harder when dealing with Generative AI-powered bots}. In fact, such AI is notoriously subject to hacking actions performed by malicious persons and aimed at influencing their behaviors~\citeB{new_grey_4,new_grey_5} by introducing bugs or defects in the source code~\citeB{new_grey_1}. Furthermore, the risk of unexpected privacy violations~\citeB{new_grey_1} and the lack of transparency in personal data management~\citeB{new_grey_4,new_grey_6} constitute a severe limit to the adoption of Generative AI-powered bots.

    \item[Discoverability and Customization.] Furthermore, \textbf{identifying the correct bot to solve a problem is challenging}~\citeA{storey_2016_bot_uses_taxonomy,wessel_2021_challenges_interacting_with_bot_on_OSS} for two reasons: first, there are limited search mechanisms to identify appropriate bots given a problem; second, the lack of a unified standard to characterize bots leads to difficulties in their description and categorization.
    This becomes even more problematic when practitioners identify the correct bot but have \textbf{difficulty configuring it}~\citeA{van_2019_programming_repair_bots,wessel_2021_challenges_interacting_with_bot_on_OSS,he2023automating}\citeB{new_grey_1}. In fact, such systems often have few configuration options, even if they have to be used in significantly different contexts.

    \item[Terms of Services.] The adoption of bots can be \textbf{hindered by the specific terms of service of the platforms they are intended to operate on}, which may restrict their functionalities~\citeB{software_bot_explained}. This legal and regulatory environment can severely limit the actions that bots are allowed to perform, thereby reducing their effectiveness and potential benefits. As a result, practitioners often hesitate to integrate bots into their workflows due to concerns over compliance and the potential legal ramifications of bot activities that violate these terms. This fear of inadvertently breaching platform policies can lead to a reluctance to adopt bot technology despite its potential to enhance productivity and efficiency.
\end{description}

%%%%%%%%%%%%
\subsubsection{\underline{Development Challenges}}
\label{sec:development_challenges}

The last category of identified challenges concerns with the development of bots. 
Specifically, we found 19 studies, 13 from the white literature~\citeA{zamanirad2017programming,srivastava_2019_architecture_applications_with_conversational_components,pinheiro_2019_bot_development_challenges_motivations,abdellatif_2020_challenges_in_chatbot_development_from_stackoverflow,brown_2020_recommendation_for_bot_architectures,erlenhov_2020_challenges_and_guidelines_for_test_bot_testing,perez_2021_creating_and_migrating_chatbot,cabot_2021_testing_challenges_for_NLP_intensive_bots,wessel_2021_challenges_interacting_with_bot_on_OSS,ouaddi2024architecture,robe2022pairProgramming_dataset,ahmad2023towardsLLMArchitecture,cerezo_2019_builing_an_expert_recommender_bot} and 6 from the grey one~\citeB{bots_challenges_for_development_2,10_best_chatbot_framework,top_framework_part_1,probot,8_chatbot_environments,understanding_CA_architecture} that report on such a category ($\approx 17.76\%$ of the entire dataset).

\begin{description}[leftmargin=0.3cm]
    \item[Framework, Architecture, and Integration.] First of all, various actors have proposed various frameworks to support developers in the development of bots and chatbots (e.g., Microsoft Azure Bot Framework, Google DialogFlow, Amazon Lex, and Rasa). Although such variety can provide practitioners with a large plethora of possibilities to reach their objectives, the literature identifies a \textbf{difficulty in identifying the correct framework to start development}~\citeA{pinheiro_2019_bot_development_challenges_motivations,abdellatif_2020_challenges_in_chatbot_development_from_stackoverflow,perez_2021_creating_and_migrating_chatbot,wessel_2021_challenges_interacting_with_bot_on_OSS,ouaddi2024architecture,ahmad2023towardsLLMArchitecture}\citeB{10_best_chatbot_framework,top_framework_part_1,probot,8_chatbot_environments}. Such a challenge arises for three main reasons. First, the enormous plethora of frameworks that implement the same functionalities makes it hard to select the better one. Second, the lack of readable documentation for each framework ulteriorly hardens the decision process. Third, \textbf{bots are complex systems to design from an architectural point of view}~\citeA{srivastava_2019_architecture_applications_with_conversational_components,brown_2020_recommendation_for_bot_architectures,ahmad2023towardsLLMArchitecture,ouaddi2024architecture}\citeB{understanding_CA_architecture} because, to improve modularity, engineers push to adopt client-server architectures in which the core functionalities are implemented on the server and the interaction with the user on the client side. This, in addition to adding complexity due to the management of communication protocols between the various components (e.g., HTTPS, API Call, and Service Calls), often makes it necessary to use different technologies for the same system, and \textbf{not all bot frameworks allow easy integration with them}~\citeA{pinheiro_2019_bot_development_challenges_motivations,abdellatif_2020_challenges_in_chatbot_development_from_stackoverflow,ahmad2023towardsLLMArchitecture}. Furthermore, all the previously mentioned problems become harder when practitioners deals with AI-powered bots; in this type of bot, in addition to the integration of the already numerous technologies involved, it is necessary to integrate artificial intelligence systems in a way that meets the required nonfunctional requirements~\citeA{ahmad2023towardsLLMArchitecture}.

    \item[Testing.] Furthermore, despite such a large amount of frameworks, \textbf{a significant minority support developers in testing activities of bots and CAs}~\citeA{srivastava_2019_architecture_applications_with_conversational_components,pinheiro_2019_bot_development_challenges_motivations,erlenhov_2020_challenges_and_guidelines_for_test_bot_testing,cabot_2021_testing_challenges_for_NLP_intensive_bots}\citeB{bots_challenges_for_development_2}. This is a significant problem because the lack of testing for such a system contributes to the mistrust of practitioners when they have to use them. Underlying this lack is the difficulty in automating the interactions between a human user and the bot. Furthermore, the large number of frameworks for bot development makes it difficult to provide a single tool capable of testing for all.

    \item[Maintenance.] Such a challenge in terms of testing leads to the emergence of another one: \textbf{the maintenance—and consequent evolution—of bots is challenging}~\citeA{zamanirad2017programming,erlenhov_2020_challenges_and_guidelines_for_test_bot_testing,wessel_2021_challenges_interacting_with_bot_on_OSS}. In fact, the lack of tests and the difficulty in monitoring such systems can impact maintenance activities, making it hard to identify and fix a bug.

    \item[Training.] Training AI-powered bots for natural language understanding and processing is a complex endeavor, especially within the specific context of software engineering. Obtaining suitable datasets for training these tools is a significant challenge~\citeA{srivastava_2019_architecture_applications_with_conversational_components,cerezo_2019_builing_an_expert_recommender_bot,abdellatif_2020_challenges_in_chatbot_development_from_stackoverflow,abdellatif2021comparison,robe2022pairProgramming_dataset}. Not only are these datasets scarce, but they are also particularly difficult to acquire when tailored to the nuanced requirements of software engineering~\citeA{robe2022pairProgramming_dataset,ahmad2023towardsLLMArchitecture}. This scarcity and specificity of data lead to \textbf{considerable difficulties in effectively training AI modules to perform their intended tasks}.
\end{description}

%%%%%%%%%%%%%%%%%%%%%%%%%%%%%%%%%%%%%%
%%%%%%%%%%%%%%%%%%%%%%%%%%%%%%%%%%%%%%
\subsection{RQ\textsubscript{3}: On the best practices to use bots and conversational agents for software engineering purposes}

%\topar{Obiettivo della RQ e riassunto delle motivazioni}
Besides collecting limitations and challenges for software bots, we were also interested in the possible best practices to overcome them. For such a purpose, we formulated our third research question to extract a list of best practices from the formal and grey literature to help practitioners adopt and develop bots for engineering purposes.

%\topar{Citare la tabella e iniziare tramite desciption a descrivere ogni punto di essa}
We decided to divide the identified best practices into two categories based on the approach's scope.
In the following, we elaborated on each best practice individually.

%%%%%%%%%%%%%%%%%%%%%%%%%%%%
\subsubsection{\underline{Development and Design Best Practices}}
%\stefano{Descrivere la categoria, citare la tabella e introdurre l'elenco}
As already reported in Section \ref{sec:development_challenges}, the large number of available technologies and the need for standardized methodologies make developing and designing bots challenging. In the following, we report the best practices identified by researchers and practitioners that can facilitate the implementation of bots and conversational systems (Appendix B—Table 1~\cite{online_appendix}). %Table \ref{table:best_practices_developmnent} summarizes each best practice.

\paragraph{\textbf{Follow a modular architecture}}
Despite their significant heterogeneity, bots—particularly conversational agents—are characterized by the need to work similarly on different platforms. For example, CAs should be used on different communication platforms—e.g., Slack and Discord—but still implement the same functionalities. In these situations, a client-service architecture—or derivated ones—represents the best choice~\citeA{zamanirad2017programming,lee_2019_accelerating_bots_development}.
Srivastava and Prabhakar~\citeA{srivastava_2019_architecture_applications_with_conversational_components} defined a reference architecture for conversational agents which specifies that these systems should be divided into two macro-components: (1) \emph{Conversational Subsystem}, which implements the input analysis, the extraction of user intent, and the response; (2) \emph{Business Component}, which contains the logic to fulfill user intents and is implemented like an API.
Another possible architecture~\citeB{understanding_CA_architecture} can organize chatbots into five modules:

\begin{enumerate}
    \item \emph{Environment} represents the core Natural Learning Process (NLP) engine and context interpretation.
    \item \emph{Question and Answer System} is where the system interprets the question and responds with relevant answers from the knowledge base.
    \item \emph{Plugins} is the module that uses API calls to interrogate other systems for executing bot operations.
    \item \emph{Node Server} is the module that handles the traffic requests from users and routes them to appropriate components.
    \item \emph{Front-end Systems} are the possible client-facing platforms users can use to interact with the system.
\end{enumerate}

%%%%
\paragraph{\textbf{Make bots adaptive and able to learn over time using AI}}
Software bots, specifically in the figure of CAs, need to communicate with a large plethora of users characterized by different cultural and language characteristics. Therefore, in most cases, such interaction and the bot's user understanding rely on an NLP module--that uses datasets composed of possible phrases and assertions in different languages--to interpret the user's intent. However, the dataset employed should increase over time based on the user's usage. For this reason, implementing a machine learning system able to learn over time and modify the dataset accordingly to new conditions is needed—namely, \emph{active learning systems}~\cite{settles2009_active_learning}—becomes necessary~\citeB{storey2020_botse,bots_challenges_for_development,bots_challenges_for_development_2,new_grey_4}.

%%%%
\paragraph{\textbf{Adopt a specific lifecycle process for bots}}
Bot are complex systems to design, both from a technical and logical side, especially if integrated with a conversational module, i.e., like for CAs. Consequently, practitioners focused their attention on defining rigorous steps to orchestrate these systems from a development and maintenance perspective~\citeB{implement_bot_12_steps,10_steps_for_chatbot_strategy,8_best_practices_for_bot_development}. In the following, we reported the elicited steps for conversational agents development:

\begin{multicols}{2}
\begin{enumerate}
    \item clearly defines the process to be automated or supported by the CA;
    \item catch the target audience's needs by engaging with software developers through interviews and usability testing sessions to identify common challenges they face during the development cycle;
    \item use existing sources to analyze competitors by conducting surveys of similar tools already existing or interviewing the potential audience;
    \item define audience intents using use cases and journey maps after conducting individual interviews with the audience;
    \item design the interaction workflow for each intent by using modeling languages, e.g., UML;
    \item choose technologies for the CA development and operation platforms according to company needs;
    \item design the CA architecture to be modular as much as possible and adherent to the guidelines proposed by researchers~\citeA{zamanirad2017programming,lee_2019_accelerating_bots_development,srivastava_2019_architecture_applications_with_conversational_components}\citeB{understanding_CA_architecture};
    \item develop a measurement framework to evaluate the CA by establishing metrics such as reduction in time to fix bugs, frequency of use, and developer satisfaction to assess the impact of the CA on the development process~\citeA{kumar_2019_sankie,hu2023bot,wang2023optimizingEliteDev}\citeB{storey2020_botse,understanding_CA_architecture};
    \item implement an artificial intelligence (NLP, NLU, and active learning) module capable of understanding technical language specific to software development, using, for example, libraries that can parse and make sense of code snippets and technical queries~\citeA{robe2022pairProgramming_dataset};
    \item implement the CA application and interaction logic in a way that can handle diverse inputs from developers, such as textual descriptions, code blocks, or even spoken commands, and provide appropriate responses or actions~\citeA{labeuf_2017_software_bots,brown_2019_bots_for_effective_recommendations,melo_2020_what_do_devs_expect_from_CAs,wang2023optimizingEliteDev,robe2022designing,arteaga2024support}\citeB{storey2020_botse,implement_bot_12_steps,10_steps_for_chatbot_strategy,8_best_practices_for_bot_development};
    \item deploy the CA and adopt a continuous improvement process.
\end{enumerate}
\end{multicols}

%%%%
\paragraph{\textbf{Carefully select the correct framework based on bot requirements}}
As mentioned in Section \ref{sec:development_challenges}, the increasing adoption of bots and conversational agents in software development leads diverse actors to develop frameworks for supporting their development. Some examples are \textls{Azure Bot} from Microsoft—able to provide a single environment to develop bots for different communication channels—or \textls{Rasa}—an open-source solution to develop bots and train NLP models all in the same place. Nevertheless, determining which framework best fits a project's requirements is far from easy~\citeA{pinheiro_2019_bot_development_challenges_motivations,abdellatif_2020_challenges_in_chatbot_development_from_stackoverflow,perez_2021_creating_and_migrating_chatbot,wessel_2021_challenges_interacting_with_bot_on_OSS,abdellatif2021comparison}\citeB{10_best_chatbot_framework,top_framework_part_1,probot,8_chatbot_environments}.
To address this challenge, practitioners identified criteria and key motivations that should guide while selecting the development framework, i.e., core technology, the target platform of operation, and the distinctive characteristics~\citeB{10_best_chatbot_framework,top_framework_part_1,probot,8_chatbot_environments}. We reported the main framework with associated motivations in Appendix B—Table 2~\cite{online_appendix}.

%%%%
\paragraph{\textbf{Adopt Domain Specific Language Models to develop bots}}
Developing bots requires expertise in several areas, such as software engineering and natural language processing. Although frameworks and platforms simplify the implementation of CA, there is still a significant obstacle caused by the inability of such frameworks to support all the developers' desires. Some studies suggest approaches based on domain-specific languages (DSLs) for model-driven development, reducing the workload of developers and designers~\citeA{ouaddi2024architecture}. Beyond that, such studies indicate that there are already bots developed to support the developer in building models that can then be used to develop systems—e.g., recommendation systems—in an automated manner~\citeA{almonte2021automating}.

%%%
\paragraph{\textbf{Make the bot able to save and use contextual information using RAG and Vector DBs.}}
Maintaining contextual information is critical for adopting bots in software engineering~\citeA{melo_2020_what_do_devs_expect_from_CAs,wyrich_2020_acceptance_of_refactoring_bots,wessel_2021_challenges_interacting_with_bot_on_OSS,erlenhov2022dependency}\citeB{storey2020_botse,bots_challenges_for_development_2}. Integrating Retrieval-Augmented Generation (RAG) models and Vector Databases can address this challenge~\citeA{arteaga2024support,de2023meet}\citeB{new_grey_3}. RAG models combine retrieval and generation capabilities to access and use context-specific information effectively. Vector Databases store and query contextual data embeddings, enabling precise and context-aware responses. This synergy ensures chatbots provide more meaningful and context-rich conversations, enhancing the overall user experience.

%%%
\paragraph{\textbf{Conduct Wizard of the Oz studies to create datasets for NL modules.}}
Wizard of Oz studies create datasets for training NLP and NLU models for developer-support bots~\citeA{robe2022pairProgramming_dataset}. Human operators simulate AI responses, allowing researchers to collect nuanced language data from developers. This method captures realistic insights into the support developers need and the queries they pose. Despite the complexity of these studies, collecting extensive data is essential for developing effective models~\citeB{new_grey_4}.

\subsubsection{\underline{Interaction and Adoption Best Practices}}
%\stefano{Descrivere la categoria, citare la tabella e introdurre l'elenco}
Interaction for bots is an important—if not the most important—aspect to take care of during their development and design. 
For such a reason, it is not surprising that the state of the art focused more on best practices to improve how bots communicate with users (Appendix B—Table 3~\cite{online_appendix}).

\paragraph{\textbf{Integrate bots actions in a way that does not interrupt developer workflow}}
Bots are naturally designed to automate some processes—or part of them—to allow users to waste less time and focus on other tasks~\cite{storey2016_disrupting_developer_productivity_botDefinition}. Such automation comes with no problems: first, bots' actions should be secure to allow users to trust them; second, bots should not interrupt users in their operation, or the advantage of the automation loses its value and usefulness~\citeA{monperrus_2019_automated_bug_fixes,erlenhov_2020_bot_characteristics_and_challenges_from_practitioners,wessel_2021_challenges_interacting_with_bot_on_OSS,wessel2022bots_orchestratore,ribeiro2022together}\citeB{storey2020_botse,software_bot_explained}.
For example, many users found tool-recommender-bot interrupted existing processes—most notably by breaking software builds—and were discouraged from merging pull requests from our system~\citeA{brown_2019_bots_for_effective_recommendations}.
For the reasons mentioned above, software bots should be designed so that the results of their actions should not slow down or hinder the user's actions, at least not unexpectedly and covertly~\citeA{brown_2019_bots_for_effective_recommendations,erlenhov_2020_bot_characteristics_and_challenges_from_practitioners,he2023automating,robe2022designing,wang2023optimizingEliteDev}\citeB{storey2020_botse,implement_bot_12_steps,10_steps_for_chatbot_strategy,8_best_practices_for_bot_development}. In recent years, orchestrator bots, i.e., bots designed to manage and guide the actions of other bots, have emerged as an efficient way to integrate a large number of bots easily without interrupting the practitioner's workflow~\citeA{wessel2022bots_orchestratore,ribeiro2022together}.

%%%%
\paragraph{\textbf{Allow developers to edit bot configuration easily}}
As mentioned in Section \ref{sec:adoption_challenges}, configuring bots to allow their operation in different contexts is a crucial challenge~\citeA{van_2019_programming_repair_bots,wessel_2021_challenges_interacting_with_bot_on_OSS}—both for their adoption and development.
Therefore, developing the system providing users with easy ways to configure it—e.g., editable configuration files, graphical interfaces for configuration, and exhaustive documentation—could be extremely useful in increasing bot adoption~\citeA{van_2019_programming_repair_bots,alizadeh_2019_refbot_software_refactoring_bot,erlenhov_2020_bot_characteristics_and_challenges_from_practitioners,he2023automating,wang2023optimizingEliteDev}. Moreover, Erlenhov et al.~\citeA{erlenhov_2020_bot_characteristics_and_challenges_from_practitioners} demonstrated that such a best practice could mitigate the \emph{Interruption and Noise problem}~\citeA{wessel_2021_challenges_interacting_with_bot_on_OSS}—elaborated in Section \ref{sec:interaction_challenges}—through a design or configuration that is mindful of whom to notify of which events.

%%%%
\paragraph{\textbf{Provide bots with personality}}
User perception is essential for bots. Various studies~\citeA{labeuf_2017_software_bots,monperrus_2019_automated_bug_fixes,robe2022designing} and grey literature~\citeB{implement_bot_12_steps,10_steps_for_chatbot_strategy,8_best_practices_for_bot_development} reported that this perception can be highly influenced by the bot's language register—as well as by its name and graphical elements. Specifically, ensuring that the bot is recognizable and distinguishable, i.e., providing it with a distinctive \emph{personality}, can help in its adoption and use~\citeA{labeuf_2017_software_bots,robe2022designing}.
Therefore, the bot's language should be casual, accessible, and fun; in a way that is not dull but friendly.
Nevertheless, it should be paid attention to not trigger the \emph{unchanny valley} effect, i.e., the phenomenon whereby a computer bears a near-identical resemblance to a human being and evokes a sense of unease or aversion in the person interacting with it~\cite{mori2012_uncanny_valley, kageki2012_uncanny_valley}.

%%%%
\paragraph{\textbf{Design a Smooth and frictionless interaction}}
Various works identified and analyzed the phenomena of ``noise''—i.e., annoying bot behaviors that lead to deterioration of communication in teams~\cite{wessel2021_dont_disturb_me_botChallenges}. Consequently, designing a smooth and frictionless interaction becomes mandatory, and this could be achieved by carefully planning conversational flows, especially for conversational bots. This can be achieved by (1) allowing bots to detect dead ends in conversations and prompt users by giving hints on how to continue the interaction and (2) implementing mechanisms to provide continuous feedback to the user that the bot is listening~\citeA{labeuf_2017_software_bots,brown_2019_bots_for_effective_recommendations,melo_2020_what_do_devs_expect_from_CAs,wang2023optimizingEliteDev,robe2022designing}\citeB{storey2020_botse,implement_bot_12_steps,10_steps_for_chatbot_strategy,8_best_practices_for_bot_development}. Moreover, designing bots to be proactive—maybe integrating Generative AI in them—is a successful way to please practitioners and make their interactions smooth~\citeA{arteaga2024support}. However, this can be achievable if the natural language comprehension of the bot is well enough designed and developed~\citeA{melo_2020_what_do_devs_expect_from_CAs}. Moreover, designers should evaluate the use of graphical user interface elements—e.g., cards, buttons, and boxes—to facilitate the interaction and make the conversation more efficient~\citeA{labeuf_2017_software_bots}.  

%%%%
%%%%
\paragraph{\textbf{Enforce transparency in bot actions and outputs}}
Trust is a vital aspect of adopting bots for software development since such systems are often designed to automate product development and release operations, which are considered vital to the success of a software project. 
The first way to improve it is related to the \emph{uncanny valley} effect: when users interact with a machine that seems too human to them, they begin feeling a sense of unease and revulsion~\cite{mori2012_uncanny_valley, kageki2012_uncanny_valley}. 
To avoid such a situation, developers should ensure that the bot's software nature is always apparent and that situations perceived as ambiguous by the user are never created~\citeA{labeuf_2017_software_bots,monperrus_2019_automated_bug_fixes,he2023automating,wang2023optimizingEliteDev}\citeB{storey2020_botse}. Moreover, it should always be evident when the user is asked to take action and why the request is made. For example, implementing graphical elements that are activated only at the moment when the user has to give input can be an excellent way to request an action overt~\citeA{labeuf_2017_software_bots,monperrus_2019_automated_bug_fixes}\citeB{storey2020_botse}.

%%%%
\paragraph{\textbf{Integrating concepts from behavioral science in interaction flow}}
As already said, \emph{recommendation bots}—namely, bots designed to provide user tips and recommendations—are largely adopted for software development purposes. Nevertheless, such bots' recommendations seem to be ineffective if the interaction with the users violates social contexts within software engineering and interrupts the development workflow of programmers. For This reason, Brown and Parning~\citeA{brown_2020_recommendation_for_bot_architectures} proposed adopting behavioral science concepts to design and guide the interaction flow of conversational systems—e.g., \emph{nudging}~\cite{thaler2009_nudge_theory_definition} and \emph{choice architecture}~\cite{thaler2013_choice_architecture}. Notably, elaborated in their work, they use 11 practical tools—provided by Johnson et al.~\cite{johnson2012_choice_architecture_tools}—to integrate \emph{choice architecture}~\cite{thaler2013_choice_architecture}—the way options are organized and presented to humans—in \emph{recommendation bots} and consequentially improve the recommendation process. 
From a practical point of view, bots should be designed according to three interaction principles:
\begin{itemize}
    \item \emph{Actionability}. It refers to the ease with which users can act on recommendations. To improve developers' perception of recommendations, better choices should be easier to take and implement.
    \item \emph{Feedback}. It refers to the way pieces of information are provided to the user. To improve developers' perception of recommendations, useful and essential information should be provided to the users.
    \item \emph{Locality}. It refers to the setting—both physical and temporal—surrounding the context of developer recommendations. To improve developers' perception of recommendations, they should be presented during their working hours and when they are in the workplace.
\end{itemize}
Moreover, a study performed by Robe et al.~\citeA{robe2022designing} demonstrated that practitioners are more inclined to adopt and use bots that demonstrate adaptability, motivation, and social presence skills; such characteristics support developers' trust.

%%%%
\paragraph{\textbf{Provide users with a high level of control over bot}}
Another way to improve interaction is through \textit{nudges}—i.e., any aspect of the choice architecture that alters people's behavior in a predictable way without forbidding any options or significantly changing their economic incentives~\cite{thaler2009_nudge_theory_definition}. As a way to use nudge for conversational systems, Brown et al.~\citeA{brown_2019_bots_for_effective_recommendations} reported that making easy and cheap recommendations—instead of forcing updates and changes to developers—could improve the users' perception of recommendation bots~\citeA{brown_2019_bots_for_effective_recommendations,alizadeh_2019_refbot_software_refactoring_bot,erlenhov_2020_bot_characteristics_and_challenges_from_practitioners,wyrich_2020_acceptance_of_refactoring_bots,melo_2020_what_do_devs_expect_from_CAs,wang2023optimizingEliteDev}\citeB{storey2020_botse}. For example, in the context of a bot that analyzes the source code to find quality problems, making comments above the incriminated line of code instead of automatically refactoring it could be a better choice.
Another way to provide a sense of control to the user—as proposed and reported by Erlenhov et al.~\citeA{erlenhov_2020_bot_characteristics_and_challenges_from_practitioners}—is to make the bot ask for confirmation before starting to work on a task or at the end of interactions—in a \emph{transaction-fashion}—to avoid misunderstanding.

%%%%
\paragraph{\textbf{Create a reliable test infrastructure}}
An alternative way to make developers trust a development bot is to provide it with a reliable and public test infrastructure~\citeA{erlenhov_2020_bot_characteristics_and_challenges_from_practitioners}\citeB{bots_challenges_for_development_2}. 
In such a way, the potential audience of the tool can evaluate it in advance and feel more comfortable with it. 
Nevertheless, testing bots can be challenging due to the need to simulate all possible user interactions~\citeA{srivastava_2019_architecture_applications_with_conversational_components,pinheiro_2019_bot_development_challenges_motivations,erlenhov_2020_challenges_and_guidelines_for_test_bot_testing,cabot_2021_testing_challenges_for_NLP_intensive_bots}\citeB{bots_challenges_for_development_2}. In order to facilitate such activity, practitioners propose to (1) use one of the provided frameworks for bots—and chatbots—testing—e.g., \textls{testYourBot}\footnote{\textls{testYourBot} site: \url{https://shankar96.github.io/testYourBot/}} and \textls{Dimon}\footnote{\textls{Dimon} site: \url{http://dimon.co/}}—and (2) develop and perform manual testing of bots based on their interaction flow~\citeB{bots_challenges_for_development_2}.

%%%%
\paragraph{\textbf{Integrate a feedback system allowing users to evaluate bot choices}}
Most problems that characterize software bots regard how systems interact with users. In most cases, these issues could be fixed by asking users what they want and what the specific problem is; therefore, implementing a system for practitioners to provide feedback and evaluate bots' actions is highly recommended. Although an automatic approach based on heuristics would be the most appropriate solution—but not the easiest one to implement—in which the user is directly asked to provide opinions can also reach the goal~\citeA{kumar_2019_sankie,hu2023bot,wang2023optimizingEliteDev}\citeB{storey2020_botse,understanding_CA_architecture}.

%%%
\paragraph{\textbf{Adapt the bots to the privacy policies of organizations}}
To enhance trust in bots using generative AI, developers should implement adapters at the bot interfaces. These adapters control the bot's access to information and responses, aligning operations with privacy policies~\citeB{new_grey_4} and ensuring confidentiality and user consent. Configured to limit sensitive data collection, adapters reduce data breach risks and support data minimization~\citeB{new_grey_4}. This integration enhances compliance with privacy regulations and bolsters trust in the bot's application.

%%%%%%%%%%%%%%%%%%%%%%%%%%%%%%%%%%%%%%
%%%%%%%%%%%%%%%%%%%%%%%%%%%%%%%%%%%%%%
\subsection{RQ\textsubscript{4}: On the benefits of bots and conversational agents for software engineering purposes}

%\topar{obiettivo della RQ}
In this RQ, we aimed to identify the benefits for development teams of adopting bots for software engineering purposes. Specifically, our goal was to determine which areas are most impacted by bots to identify possible fields yet to be explored and propose future steps for research to deepen (Appendix B—Table 4~\cite{online_appendix}).

\subsubsection{\underline{Productivity Benefits}}

%\topar{"molti dei benefici riguardano, ovviamente, la produttività"}
Most of the benefits provided by bots regard the productivity aspects of software development. This is because bots have been designed initially \textbf{to automate processes considered unimportant and tedious for practitioners}~\citeA{okanovi_2020_chatbot_to_support_load_testing,gilson_2020_recording_design_decision_on_the_fly}\citeB{bots_agile,5_reasons_for_codebots,microsoft_azure_to_DevSecOps}. For this reason, it does not look surprising that the most positive impact of their adoption—recorded by researchers and practitioners—is related to the productivity.

%\topar{"la produttività può essere impattata sia influenzando le azioni degli sviluppatori (da 1 a 3) ..."}
Furthermore, bots can \textbf{help developers complete tasks related to meaningful goals for the project}~\citeA{storey_2016_bot_uses_taxonomy,labeuf_2017_software_bots,gilson_2019_NLP_and_collaborativeSE,wessel_2020_code_review_bot_OSS,gilson_2020_recording_design_decision_on_the_fly,basu_2021_bot_for_distribution_of_service_requests,wang2023optimizingEliteDev,ribeiro2022together}\citeB{bots_and_AI_in_project_management,bots_and_AI_in_project_management_2}. Such a benefit is defined by Storey et al.~\citeA{storey_2016_bot_uses_taxonomy} under the title of \emph{effectiveness} and can be achieved in various ways. For example, by automatically capturing data and disseminating them to the entire team, bots can increase the developers' awareness of the project situation~\citeA{storey_2016_bot_uses_taxonomy,labeuf_2017_software_bots}. An ulterior way to improve developers' effectiveness is through conversation moderation—i.e., providing information during meetings or bringing the discussion back into focus when participants tend to digress~\citeA{gilson_2019_NLP_and_collaborativeSE}.

Bots can also impact productivity by \textbf{helping developers complete tasks more quickly}~\citeA{storey_2016_bot_uses_taxonomy,carr2016automatic,tian2017APIbot,wyrich_2019_bot_for_code_refactoring,gilson_2019_NLP_and_collaborativeSE,wessel_2020_code_review_bot_OSS,dominic_2020_bot_for_SE_newcomers,okanovi_2020_chatbot_to_support_load_testing,gilson_2020_recording_design_decision_on_the_fly,wyrich_2020_acceptance_of_refactoring_bots,zhang2024automatic,scoccia2023exploring,wang2023optimizingEliteDev,wessel2022qualityBotCodeReview,ribeiro2022together}\citeB{bots_agile,5_reasons_for_codebots,new_grey_1}—this is identified by \emph{efficiency}~\citeA{storey_2016_bot_uses_taxonomy}. Such a benefit can be reached in many ways: (1) bots can spare developers from reading large amounts of documentation by understanding what they search for and providing it immediately~\citeA{storey_2016_bot_uses_taxonomy,tian2017APIbot}; (2) bots can support during source code writing by providing code snippets or tips~\citeA{carr2016automatic,wyrich_2019_bot_for_code_refactoring}\citeB{5_reasons_for_codebots}; (3) bots can also help in documentation phases, by collecting pieces of information from different sources and summary them~\citeA{gilson_2019_NLP_and_collaborativeSE}\citeB{bots_agile}.

%\topar{"... sia andando ad automatizzare dei task (da 4 a 5)"}
In practice, bots are also used to perform simple tasks, e.g., building a new version of a software project, executing test suites, and preparing execution reports, thus implying an increase in workers' free time and optimizing the execution of these tasks. This is because bots \textbf{can work without interruption}~\citeA{erlenhov_2020_bot_characteristics_and_challenges_from_practitioners}\citeB{best_bots_to_improve_your_software_development_process,5_reasons_for_codebots,bots_and_AI_in_project_management,bots_and_AI_in_project_management_2} and \textbf{handle large amounts of tasks simultaneously}~\citeA{erlenhov_2020_bot_characteristics_and_challenges_from_practitioners}\citeB{bots_agile,5_reasons_for_codebots,bots_and_AI_in_project_management,bots_and_AI_in_project_management_2,new_grey_1}.

\subsubsection{\underline{Collaboration Benefits}}

%\topar{"i benefici dei bot sono anche dal punto di vista sociale e comunicativo (da 6 a 8)"}
Bots also provide benefits when comes to knowledge sharing and collaboration in development teams~\citeA{labeuf_2017_software_bots,matthies_2019_bot_and_agile_retrospectives,gilson_2019_NLP_and_collaborativeSE,kim_2020_bot_to_facilitate_group_discussion,gilson_2020_recording_design_decision_on_the_fly,wessel2022qualityBotCodeReview}\citeB{bots_agile,bots_and_AI_in_project_management,new_grey_1}. 
For example, if correctly used, bots can distribute to the team the status of other tasks, consequently \textbf{building team trust and cooperation}~\citeA{labeuf_2017_software_bots}. 
Furthermore, bots can enhance the recording of decisions, performing: (1) automatic fulfillment of artifacts templates (e.g., user stories), (2) summary build from discussions, (3) automatic generation of diagrams, and (4) monitoring of conversations to automatically refer previous decisions. Moreover, bots can enhance developer collaboration by performing (1) recording and retrieving technical discussions, (2) fetching and updating models based on conversation, and (3) parsing committed messages. By doing all of these, bots can \textbf{support the decision-making process} of different figures involved in software engineering~\citeA{wessel2022bots_orchestratore}\citeB{new_grey_1}.

Furthermore, by abstracting practitioners from executing repetitive and tedious tasks, there is \textbf{less chance of mistakes}~\citeB{bots_agile,5_reasons_for_codebots,why_bots_format_code,microsoft_azure_to_DevSecOps}. This is since (1) humans tend to commit errors if constantly repeating the same task, and (2) software can apply quality standards pre-defined. 

Moreover, such an abstraction can be used by managers to improve the risk management activity~\citeB{bots_agile,5_reasons_for_codebots,why_bots_format_code} and \textbf{facilitate the adoption of new methodologies and technologies} in their team~\citeB{bots_agile,microsoft_azure_to_DevSecOps}. For example, by using a chatbot to improve code review, managers could provide developers with easy tips and good practices without assigning human effort to such a purpose.

\subsubsection{\underline{Technical Benefits}}

%\topar{"in conclusione, i bot possono anche portare benefici nel codice sorgente (da 9 a 10)"}
Lastly, bots and conversational agents can also provide many benefits from a technical and product view. For example, bots can \textbf{improve the maintainability and quality of source code} by detecting technical debt—e.g., code smells and test smells—or by enforcing good programming practices~\citeA{carr2016automatic,wyrich_2019_bot_for_code_refactoring,alizadeh_2019_refbot_software_refactoring_bot,hu_2019_improving_feedback_pull_request_with_bots,erlenhov_2020_bot_characteristics_and_challenges_from_practitioners,okanovi_2020_chatbot_to_support_load_testing,carvalho_2020_C_3PR_fix_static_analysis_violations,gilson_2020_recording_design_decision_on_the_fly,wyrich_2020_acceptance_of_refactoring_bots}\citeB{best_bots_to_improve_your_software_development_process,bots_agile,5_reasons_for_codebots,why_bots_format_code,microsoft_azure_to_DevSecOps}.
Moreover, bots can collect information from different artifacts—e.g., source code, documentation, and reports—to spread to the entire team and \textbf{reduce socio-technical problems}, e.g., turnover. Furthermore—it is always related to the knowledge-sharing side—bots can also be used to collect information about API and new technologies and provide them ``on demand'' to the users. Such communication can occur in any communication channel used by the development team~\citeA{beschastnikh_2017_research_and_bot,labeuf_2017_software_bots,erlenhov_2020_bot_characteristics_and_challenges_from_practitioners,lin_2020_bot_framework_for_service_development_assistance,romero_2020_experiences_building_an_answer_bot,dominic_2020_bot_for_SE_newcomers,gilson_2020_recording_design_decision_on_the_fly}\citeB{best_bots_to_improve_your_software_development_process,microsoft_azure_to_DevSecOps,new_grey_1}.

\section{Discussion, Implications, and Future Steps}
\label{sec:discussion}

The results of our study pointed out several observations worth further discussion. We started from them to identify potential future research lines that the research community can explore. Moreover, since the multivocal literature review also comprises gray literature, practitioners can benefit from our observations.
In the following paragraphs, we report the observations and implications from our review, organized by topic.

%%%%%%%%%%%%%%%%%%%%%%%%%%%%%%%%%%%%%%%%%%%%%%%%%%%%%%%%%%%%%%%%%%%%%%%%
\subsection{Conversational Agents and Human Aspects of Software Development}
Conversational Agents (CAs) are \emph{de facto} software systems in which interaction with users represents the main characteristic, thus implying that making the potential audience more involved during design phases is crucial to project success~\citeA{labeuf_2017_software_bots,srivastava_2019_architecture_applications_with_conversational_components,cerezo_2019_builing_an_expert_recommender_bot,pinheiro_2019_bot_development_challenges_motivations,castro_2019_bot_usability,melo_2020_what_do_devs_expect_from_CAs,erlenhov_2020_bot_characteristics_and_challenges_from_practitioners,abdellatif_2020_challenges_in_chatbot_development_from_stackoverflow,wyrich_2020_acceptance_of_refactoring_bots,wessel_2021_challenges_interacting_with_bot_on_OSS,saadat_2021_bots_modify_workflow_of_github_teams}\citeB{storey2020_botse,bots_challenges_for_development,bots_challenges_for_development_2,bots_NLP_or_NLU,7_chatbot_development_challenges}. Moreover, since most of the reported software bots operate with different stakeholders simultaneously, the interaction patterns between different actors also acquire importance. For this reason, representing collaboration and communication patterns in software communities in a quantitative way could help facilitate the CAs' interaction workflow design. In order to operationalize such collaboration patterns, state of the art defined \emph{community smells}~\cite{tamburri2019software}—i.e., a set of anti-patterns in collaboration activities in software development communities that are precursors of socio-technical problems.
Based on the above consideration, we can provide two future fields of exploration:

\begin{itemize}
    \item[\faHandORight \hspace{0.05cm}] \underline{Community Smells to support bots design.} %Community Smells could be used as input for the bot design phase in order to improve the fitness of such systems with their audience. Researchers should put effort into studying how such an approach could impact the development process.
    Community Smells could be used as input when designing bot to improve the communication and increase the trust that users have with them. Researchers could study how this aspect impacts the development process.
    \item[\faHandORight \hspace{0.05cm}] \underline{Bots-Community Smells.} Community Smells represent collaboration and communication anti-patterns in software communities. Adopting bots for software development purposes could lead to the creation of a community over the community, characterized by the interaction between bots and developers and bots with other bots. This could lead to the emergence of a new way to represent patterns—and anti-patterns—through a particular type of community smells. The research community could (1) quantitatively represent them and (2) provide practitioners with instruments to improve their relationship with conversational systems.
    %Community Smells represent collaboration and communication anti-patterns from developers in software communities. Moreover, adopting bots for software development purposes could create a community over the community, characterized by the interaction between bots and developers and bots with other bots. Then, this could lead to the emergence of new opportunities to represent such patterns—and anti-patterns—through a particular type of community smells, allowing the research community to (1) quantitatively represent them and (2) provide practitioners with instruments to improve their relationship with conversational systems.
\end{itemize}

%%%%%%%%%%%%%%%%%%%%%%%%%%%%%%%%%%%%%%%%%%%%%%%%%%%%%%%%%%%%%%%%%%%%%%%%
\subsection{Conversational Agents for Training and Support}

%\topar{Molti lavori si concentrano sull'usare i conversational agents per supportare (1) l'insegnamento e (2) il training.}
Besides practical uses, the adoption of bots also revolves around knowledge sharing~\citeA{storey_2016_bot_uses_taxonomy,tian2017APIbot,beschastnikh_2017_research_and_bot,paikari_2018_framework_chatbot_and_their_future,fukui_2019_bot_suggesting_questions,kumar_2019_sankie,ni_2019_CrowDevBot_CA_for_software_crowdsourcing,erlenhov_2020_bot_characteristics_and_challenges_from_practitioners,lin_2020_bot_framework_for_service_development_assistance,romero_2020_experiences_building_an_answer_bot,dominic_2020_bot_for_SE_newcomers,chatterjee_2021_automatic_extraction_of_opinion}\citeB{best_bots_to_improve_your_software_development_process,storey2020_botse,bots_and_AI_in_project_management,bots_and_AI_in_project_management_2}. Specifically, practitioners use bots to (1) share information between team members and (2) record decision rationale. Moreover, conversational agents are used to provide information on software modules to new developers in order to facilitate training phases.
Based on our findings, we can provide the following considerations:

\begin{itemize}
    \item[\faHandORight \hspace{0.05cm}] \underline{Bots and Turnover.} Bots can be used to collect expert knowledge and provide it to newcomers~\citeB{best_bots_to_improve_your_software_development_process,storey2020_botse}. In this way, the consequences of turnover~\cite{foucault2015_impact_of_turnover}—i.e., the continuous influx and retreat of human resources in a team— can mitigate the progressive loss of knowledge on a software module. From here, researchers should focus on how (i) bots should collect such knowledge and (ii) how this should be provided to newcomers. 
    %\item[\faHandORight \hspace{0.05cm}] \stefano{Recommendation Systems}
\end{itemize}

%%%%%%%%%%%%%%%%%%%%%%%%%%%%%%%%%%%%%%%%%%%%%%%%%%%%%%%%%%%%%%%%%%%%%%%%
\subsection{Addressing Limitations with Best Practices}
\label{Sec_addressing_limitation}

Our literature survey identified challenges and best practices related to bots and conversational agents~\cite{wessel2021_dont_disturb_me_botChallenges}. We tried to map such challenges to best practices based on the various papers to provide solutions and mitigation strategies supported by evidence in research and practitioners' contexts. We reported our findings in Figure \ref{fig:challenges_best_practices}.

\begin{itemize}
    \item[\faHandORight \hspace{0.05cm}] \underline{Challenges and Best Practices.} Our mapping—based on literature evidence—should be validated through qualitative studies and could be a valuable contribution for practitioners facing bot challenges.
\end{itemize}

\begin{figure}
    \centering
    \includegraphics[width=1\linewidth]{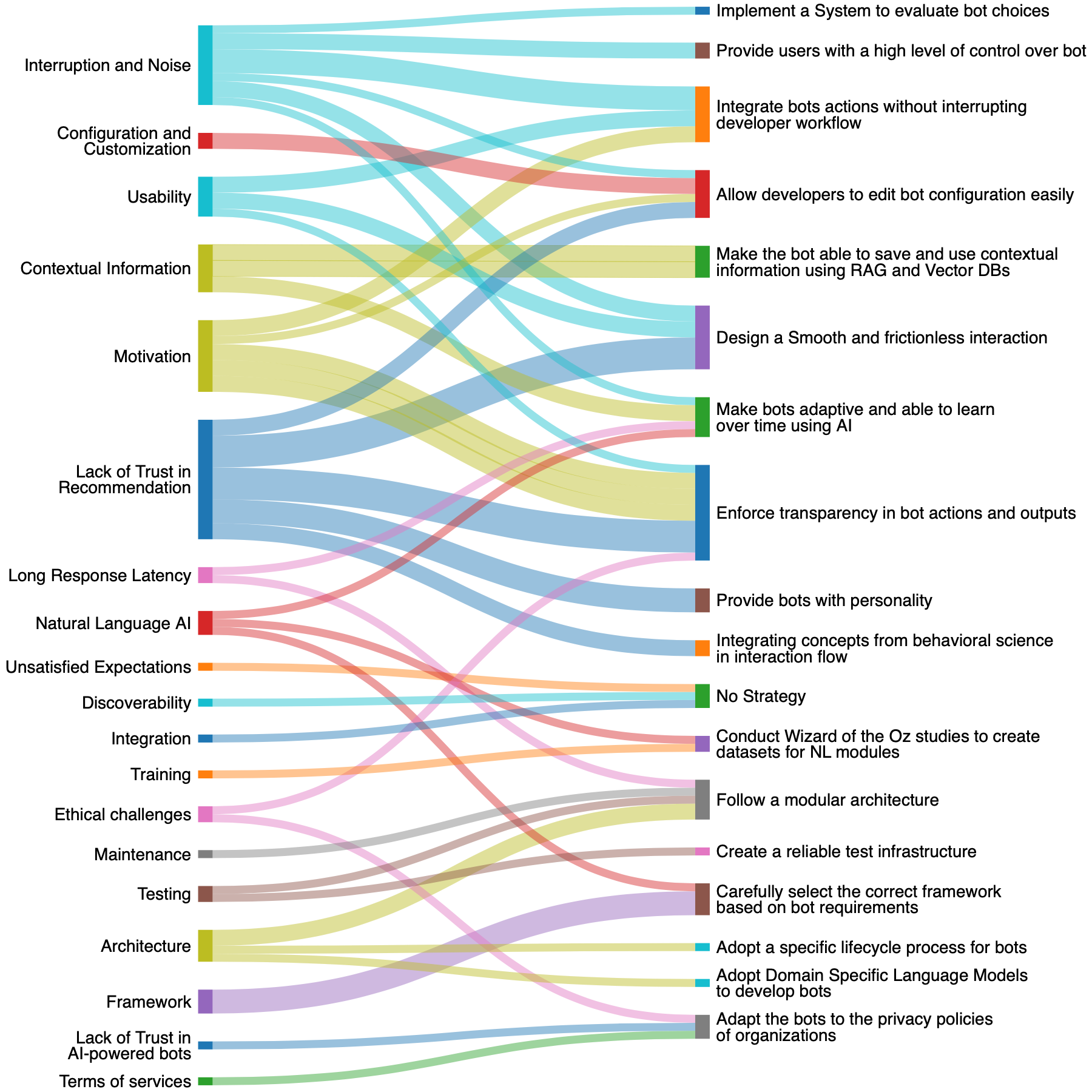}
    \caption{Challenges (on the right) and Associated Best Practices (on the left).}
    \label{fig:challenges_best_practices}
\end{figure}

%%%%%%%%%%%%%%%%%%%%%%%%%%%%%%%%%%%%%%%%%%%%%%%%%%%%%%%%%%%%%%%%%%%%%%%%
\subsection{Conversational Agents for Project Management: How far are we?}

Project Management is the practice of applying knowledge, skills, methods, and tools to meet the requirements of a specific project, achieving quality and performance goals~\cite{PMBOK}.
In the context of software development, project management is a highly complex and hard-to-standardize activity due to the volatile nature of software and the extreme heterogeneity of stakeholders involved—both in terms of background and competencies~\cite{MyticalManMonth,cerdeiral2019_software_project_management_and_maturity_model, hayat2019_influence_of_agile_on_software_project_management}.
Moreover, according to the Standish Group's CHAOS Report~\cite{chaos_report}, approximately 70\% of software projects fail—some of them with catastrophic consequences~\cite{project_failure_2, project_failure_3, friden_FBI_failure_report} and state of the art demonstrated that management aspects significantly impact project outcomes, leading management activities as a priority to consider~\cite{It_projects_fail,projectFailure}.
Software bots—and particularly Conversational Agents (CAs)—could help improve and support management activities:

\begin{itemize}
    \item[\faHandORight \hspace{0.05cm}] \underline{CAs as Managers' Twins.} Project managers (PM) are responsible for ensuring that the project meets the requirements. In order to achieve this objective, managers should have various skills and—most critically—manage various factors simultaneously~\cite{PMBOK}. To support this goal, having a community of bots collecting different metrics and monitoring different activities could support PMs in their daily tasks~\citeA{voria2022cadocs}. Moreover, making a hierarchical bots structure in which pieces of information are filtered and summarized differently at each level of the organization could help managers focus on essential tasks, resulting in increased productivity and mental health quality~\citeA{wessel2022bots_orchestratore}.
    
    \item[\faHandORight \hspace{0.05cm}] \underline{CAs as Meeting Facilitator.} There are tools designed to help practitioners manage information sharing during meetings~\citeA{gilson_2019_NLP_and_collaborativeSE,gilson_2020_recording_design_decision_on_the_fly,kim_2020_bot_to_facilitate_group_discussion}\citeB{storey2020_botse}. Since software engineering is \emph{de facto} a social activity involving different stakeholders with different backgrounds, having such support could greatly benefit the state of practice. However, the reported effectiveness of these tools could be better. Therefore, investing in improving their capabilities and using Generative AI could be a worthy future research topic.
\end{itemize}

%%%%%%%%%%%%%%%%%%%%%%%%%%%%%%%%%%%%%%%%%%%%%%%%%%%%%%%%%%%%%%%%%%%%%%%%

\subsection{Open Issues for bots and conversational agents}

Despite the significant attention reserved to the topic by researchers and practitioners, using bots for software development is still a young and unexplored field. For such a reason, there are a series of open issues and exploration opportunities that the research community should care of:

\begin{itemize}
    \item[\faHandORight \hspace{0.05cm}] \underline{Usability Research for CAs.} Although CAs put the interaction with users at their center, to the best of our knowledge, there needs to be more research on the usability of such systems. Moreover, practitioners should put more effort into defining standards and guidelines to design interaction workflows in conversational systems.
    \item[\faHandORight \hspace{0.05cm}] \underline{Testing of Bots.} Testing bots and conversational agents arose as a complex topic~\citeA{srivastava_2019_architecture_applications_with_conversational_components,pinheiro_2019_bot_development_challenges_motivations,erlenhov_2020_challenges_and_guidelines_for_test_bot_testing,cabot_2021_testing_challenges_for_NLP_intensive_bots}\citeB{bots_challenges_for_development_2}. Most of the testing strategies for these systems fall under the category of system testing, and only a few have been proposed at lower levels of testing (e.g., unit and integration).
    \item[\faHandORight \hspace{0.05cm}] \underline{Behavioral Science and Bots.} Motivating users to adopt bots is a widely documented problem~\citeA{monperrus_2019_automated_bug_fixes,erlenhov_2020_bot_characteristics_and_challenges_from_practitioners,wessel_2021_challenges_interacting_with_bot_on_OSS}\citeB{storey2020_botse,software_bot_explained,7_chatbot_development_challenges}. Various works proposed adopting behavioral science concepts—e.g., nudge and interaction principles—to improve the design phases of these systems~\citeA{brown_2020_recommendation_for_bot_architectures}. For this aim, researchers from both fields—computer science and social science—could join forces to open new research opportunities.
    \item[\faHandORight \hspace{0.05cm}] \underline{Interruption and Noise: an Open Challenge.} Interruption and noise are significant and documented problems for conversational agents and bots~\citeA{labeuf_2017_software_bots,erlenhov_2020_bot_characteristics_and_challenges_from_practitioners,wessel_2021_challenges_interacting_with_bot_on_OSS,saadat_2021_bots_modify_workflow_of_github_teams}\citeB{storey2020_botse}. Despite the work already done to propose mitigation strategies for such phenomena, more should be done to increase the adoption and effectiveness—in terms of accomplished contribution—of bots for development purposes.
    \item[\faHandORight \hspace{0.05cm}] \underline{Generative AI for Software Development.} The recent emergence of large language models in the consumer market has inevitably led software engineers to experiment with integrating these tools into their workflow~\citeA{purwoko2023analysis,scoccia2023exploring}. When such LLMs are used in their chatbot versions (e.g., ChatGPT), they constitute software bots. Despite their impressive capabilities and the potential they could have in all aspects of software development, some clear limitations continue to exist, likely due to the technology's still early stage.
\end{itemize}

\section{Limitations and Threats}
\label{sec:threat}
Threats to the validity of an MLR review are typically related to the inaccuracy of data extraction and incomplete set of studies due to search terms limitation, selection of academic search engines, and researcher bias regarding exclusion/inclusion criteria. In this section, we identified and carefully addressed potential threats to the validity \cite{kitchenham2009_SLR_definition,wohlin2012experimentation} of our study by taking steps to minimize or mitigate them.

\paragraph{\textbf{Internal Validity}}
Threats in this category regard the source and data selection approach, as well as the possibility that the findings are affected by casualty factors without the researcher's knowledge~\cite{wohlin2012experimentation}.
We relied on a previous systematic approach proposed in the literature \cite{kitchenham2009_SLR_definition} and described in Section \ref{sec:methodology}. We defined and reported each step of the process carefully—i.e., search engines, search terms, and inclusion/exclusion criteria—to ensure replicability. These are usually the problematic steps. 
As for the search terms used for studies extraction, we relied not only on the most common words—e.g., bot—but also on similar terms used by different authors for pointing to a similar concept—e.g., conversational agents. We followed a formal search using defined keywords, followed by a manual search using the references of the initial pool in our field of study. As for the search engine, we included official comprehensive academic databases, e.g., Scopus, but also fewer ones like Google Scholar. While for the grey literature, we relied on Google Search Engine—like other Grey Literature and Multivocal Literature Reviews~\cite{islam2019_mlr_example_1,garousi2016_mlr_example_2}. 
Therefore, we believe that we collected all the relevant studies in the field. Finally, as for the inclusion/exclusion criteria, they can usually suffer from researchers' judgment and experience. For this reason, all the paper's authors were involved in checking criteria by applying joint voting for study selection. As a result, only those with high scores were selected for this study. 

\paragraph{\textbf{Construct Validity}}
Threats in this category regard the extent to which the object of study truly represents the theory behind the study and are mainly due to imprecision in performed measurements~\cite{wohlin2012experimentation}.
One threat in this category concerns the lack of empirical evidence in the studies coming from grey literature. Indeed, this literature reports facts based on opinion or experience. Since the grey literature represents the voice of the practitioners, we could assume that they would be speaking from personal experience. This can be a crucial limitation; however, the goal of an MLR is also to show different ways of thinking, i.e., empirical versus practical, since it can have two types of audience: academic and industrial. Moreover, the inclusion of grey literature can help to overcome the scarcity of studies retrieved by automatic searches in the digital databases of scientific literature.

\paragraph{\textbf{Conclusion Validity}}
Threats in this category concern the degree to which the conclusion we reach is credible or believable~\cite{wohlin2012experimentation}.
The first two authors of the paper reviewed each step of the process to ensure the reliability of the study and mitigate bias in data selection and extraction. Each severe disagreement between authors was resolved by consensus among all the authors.
Furthermore, in order to strengthen our results, we followed both the guidelines from Kitchenham et al.~\cite{keele2007_slr_guidelines_Kitchenham}—related to the formal literature selection—and from Garousi et al.~\cite{garousi2019_mlr_guidelines}—associated with the grey literature inclusion.
Moreover, we followed a rigid quality criteria assessment—both for white and grey literature—and we extracted our sources from the most used and known databases.

\paragraph{\textbf{External Validity}}
Threats in this category concern the extent to which our study's results can be generalized~\cite{wohlin2012experimentation}.
First of all, we conducted our MLR on the most known and suggested databases of literature~\cite{garousi2019_mlr_guidelines,keele2007_slr_guidelines_Kitchenham} to increase the quality of extracted sources and generalizability of our findings.
Furthermore, we extracted only primary sources written in English, so the rest in other languages were excluded, thus possibly threatening our generalizability. Nonetheless, (i) the top venues in software engineering accept only papers written in English, and (ii) most of the literature reviews already present in the literature include exclusion criteria based on the English language.

\section{Conclusion}
\label{sec:conclusion}

%\topar{Risultati}
In this work,  we reported on a multivocal literature review to shed light on adopting bots and conversational agents for software engineering purposes. The reason behind the choice of a multivocal review is related to the fact that bots and CAs are extremely linked to the practitioners' field, so we wanted to contribute both to researchers and practitioners. Specifically, we aimed to provide a representation and categorization of the (1) motivation for adopting bots, (2) challenges arising from their integration into software development processes, (3) best practices applicable contrasting the before-mentioned challenges, and (4) benefits derived from the adoption of bots.
Moreover, we also provide a list of discussion points and opportunities for researchers to encourage new investigations on the matter. Indeed, our contributions could be considered a fundamental base built to advance the state of the art and provide an advantage both for researchers and practitioners.

\section*{Acknowledgments}
The authors would like to thank the two anonymous Reviewers and the editor for their insightful comments on our paper.

\bibliographystyleA{unsrt}
\bibliographyA{formal_literature}

\bibliographystyleB{unsrt}
\bibliographyB{grey_literature}

\bibliographystyle{ACM-Reference-Format}
\bibliography{bibliography}

\begin{thebibliography}{10}

\bibitem{wessel_2021_challenges_interacting_with_bot_on_OSS}
Mairieli Wessel, Igor Wiese, Igor Steinmacher, and Marco~A. Gerosa.
\newblock Don't disturb me: Challenges of interacting with software bots on open source software projects, 2021.

\bibitem{zamanirad2017programming}
Shayan Zamanirad, Boualem Benatallah, Moshe~Chai Barukh, Fabio Casati, and Carlos Rodriguez.
\newblock Programming bots by synthesizing natural language expressions into api invocations.
\newblock In {\em 2017 32nd IEEE/ACM International Conference on Automated Software Engineering (ASE)}, pages 832--837, 2017.

\bibitem{srivastava_2019_architecture_applications_with_conversational_components}
Saurabh Srivastava and TV~Prabhakar.
\newblock A reference architecture for applications with conversational components.
\newblock In {\em 2019 IEEE 10th International Conference on Software Engineering and Service Science (ICSESS)}, pages 1--5. IEEE, 2019.

\bibitem{van_2019_programming_repair_bots}
Rijnard van Tonder and Claire Le~Goues.
\newblock Towards s/engineer/bot: Principles for program repair bots.
\newblock In {\em 2019 IEEE/ACM 1st International Workshop on Bots in Software Engineering (BotSE)}, pages 43--47. IEEE, 2019.

\bibitem{monperrus_2019_automated_bug_fixes}
Martin Monperrus.
\newblock Explainable software bot contributions: Case study of automated bug fixes.
\newblock In {\em 2019 IEEE/ACM 1st International Workshop on Bots in Software Engineering (BotSE)}, pages 12--15, May 2019.

\bibitem{gilson_2019_NLP_and_collaborativeSE}
Fabian Gilson and Danny Weyns.
\newblock When natural language processing jumps into collaborative software engineering.
\newblock In {\em 2019 IEEE International Conference on Software Architecture Companion (ICSA-C)}, pages 238--241, March 2019.

\bibitem{ren_2020_collaborative_modelling_bot_vs_online_tools}
Ranci Ren, John~W. Castro, Adri\'{a}n Santos, Sara P\'{e}rez-Soler, Silvia~T. Acu\~{n}a, and Juan de~Lara.
\newblock Collaborative modelling: Chatbots or on-line tools? an experimental study.
\newblock In {\em Proceedings of the Evaluation and Assessment in Software Engineering}, EASE '20, page 260–269, New York, NY, USA, 2020. Association for Computing Machinery.

\bibitem{gilson_2020_recording_design_decision_on_the_fly}
Fabian Gilson, Sam Annand, and Jack Steel.
\newblock Recording software design decisions on the fly.
\newblock In {\em SEED/NLPaSE@ APSEC}, pages 53--66, 2020.

\bibitem{purwoko2023analysis}
Justine~Winata Purwoko, Tegar Abdullah, Budiman Wijaya, Alexander Agung~Santoso Gunawan, and Karen~Etania Saputra.
\newblock Analysis chatgpt potential: Transforming software development with ai chat bots.
\newblock In {\em 2023 International Conference on Networking, Electrical Engineering, Computer Science, and Technology (IConNECT)}, pages 36--41. IEEE, 2023.

\bibitem{saini2022automated}
Rijul Saini, Gunter Mussbacher, Jin~LC Guo, and J{\"o}rg Kienzle.
\newblock Automated, interactive, and traceable domain modelling empowered by artificial intelligence.
\newblock {\em Software and Systems Modeling}, 21(3):1015--1045, 2022.

\bibitem{almonte2021automating}
Lissette Almonte, Sara P{\'e}rez-Soler, Esther Guerra, Iv{\'a}n Cantador, and Juan de~Lara.
\newblock Automating the synthesis of recommender systems for modelling languages.
\newblock In {\em Proceedings of the 14th ACM SIGPLAN International Conference on Software Language Engineering}, pages 22--35, 2021.

\bibitem{qasse2023chat2code}
Ilham Qasse, Shailesh Mishra, Bj{\"o}rn {\th}{\'o}r~J{\'o}nsson, Foutse Khomh, and Mohammad Hamdaqa.
\newblock Chat2code: A chatbot for model specification and code generation, the case of smart contracts.
\newblock In {\em 2023 IEEE International Conference on Software Services Engineering (SSE)}, pages 50--60. IEEE, 2023.

\bibitem{scoccia2023exploring}
Gian~Luca Scoccia.
\newblock Exploring early adopters' perceptions of chatgpt as a code generation tool.
\newblock In {\em 2023 38th IEEE/ACM International Conference on Automated Software Engineering Workshops (ASEW)}, pages 88--93. IEEE, 2023.

\bibitem{perez_2019_CA_flexible_modelling}
Sara Pérez-Soler, Esther Guerra, and Juan de~Lara.
\newblock Flexible modelling using conversational agents.
\newblock In {\em 2019 ACM/IEEE 22nd International Conference on Model Driven Engineering Languages and Systems Companion (MODELS-C)}, pages 478--482, Sep. 2019.

\bibitem{wyrich_2019_bot_for_code_refactoring}
Marvin Wyrich and Justus Bogner.
\newblock Towards an autonomous bot for automatic source code refactoring.
\newblock In {\em 2019 IEEE/ACM 1st International Workshop on Bots in Software Engineering (BotSE)}, pages 24--28. IEEE, 2019.

\bibitem{alizadeh_2019_refbot_software_refactoring_bot}
Vahid Alizadeh, Mohamed~Amine Ouali, Marouane Kessentini, and Meriem Chater.
\newblock Refbot: intelligent software refactoring bot.
\newblock In {\em 2019 34th IEEE/ACM International Conference on Automated Software Engineering (ASE)}, pages 823--834. IEEE, 2019.

\bibitem{hu_2019_improving_feedback_pull_request_with_bots}
Zhewei Hu and Edward~F. Gehringer.
\newblock Improving feedback on github pull requests: A bots approach.
\newblock In {\em 2019 IEEE Frontiers in Education Conference (FIE)}, pages 1--9, Oct 2019.

\bibitem{carvalho_2020_C_3PR_fix_static_analysis_violations}
Antônio Carvalho, Welder Luz, Diego Marcílio, Rodrigo Bonifácio, Gustavo Pinto, and Edna Dias~Canedo.
\newblock C-3pr: A bot for fixing static analysis violations via pull requests.
\newblock In {\em 2020 IEEE 27th International Conference on Software Analysis, Evolution and Reengineering (SANER)}, pages 161--171, Feb 2020.

\bibitem{wyrich_2020_acceptance_of_refactoring_bots}
Marvin Wyrich., Regina Hebig., Stefan Wagner., and Riccardo Scandariato.
\newblock Perception and acceptance of an autonomous refactoring bot.
\newblock In {\em Proceedings of the 12th International Conference on Agents and Artificial Intelligence - Volume 1: ICAART,}, pages 303--310. INSTICC, SciTePress, 2020.

\bibitem{wyrich_2021_bots_interaction_with_pull_requests}
Marvin Wyrich, Raoul Ghit, Tobias Haller, and Christian M{\""u}ller.
\newblock Bots don't mind waiting, do they? comparing the interaction with automatically and manually created pull requests.
\newblock {\em arXiv preprint arXiv:2103.03591}, 2021.

\bibitem{serban_2021_fixes_for_statyc_analysis_warning}
Dragos Serban, Bart Golsteijn, Ralph Holdorp, and Alexander Serebrenik.
\newblock Saw-bot: Proposing fixes for static analysis warnings with github suggestions.
\newblock In {\em 2021 IEEE/ACM Third International Workshop on Bots in Software Engineering (BotSE)}, pages 26--30, June 2021.

\bibitem{storey_2016_bot_uses_taxonomy}
Margaret-Anne Storey and Alexey Zagalsky.
\newblock Disrupting developer productivity one bot at a time.
\newblock In {\em Proceedings of the 2016 24th ACM SIGSOFT International Symposium on Foundations of Software Engineering}, FSE 2016, page 928–931, New York, NY, USA, 2016. Association for Computing Machinery.

\bibitem{urli_2018_repairnator_project}
Simon Urli, Zhongxing Yu, Lionel Seinturier, and Martin Monperrus.
\newblock How to design a program repair bot? insights from the repairnator project.
\newblock In {\em 2018 IEEE/ACM 40th International Conference on Software Engineering: Software Engineering in Practice Track (ICSE-SEIP)}, pages 95--104. IEEE, 2018.

\bibitem{paikari_2018_framework_chatbot_and_their_future}
Elahe Paikari and Andr\'{e} van~der Hoek.
\newblock A framework for understanding chatbots and their future.
\newblock In {\em Proceedings of the 11th International Workshop on Cooperative and Human Aspects of Software Engineering}, CHASE '18, page 13–16, New York, NY, USA, 2018. Association for Computing Machinery.

\bibitem{erlenhov_2020_bot_characteristics_and_challenges_from_practitioners}
Linda Erlenhov, Francisco Gomes de~Oliveira Neto, and Philipp Leitner.
\newblock An empirical study of bots in software development: Characteristics and challenges from a practitioner’s perspective.
\newblock In {\em Proceedings of the 28th ACM Joint Meeting on European Software Engineering Conference and Symposium on the Foundations of Software Engineering}, ESEC/FSE 2020, page 445–455, New York, NY, USA, 2020. Association for Computing Machinery.

\bibitem{erlenhov_2020_challenges_and_guidelines_for_test_bot_testing}
Linda Erlenhov, Francisco~Gomes de~Oliveira~Neto, Martin Chukaleski, and Samer Daknache.
\newblock Challenges and guidelines on designing test cases for test bots.
\newblock In {\em Proceedings of the IEEE/ACM 42nd International Conference on Software Engineering Workshops}, ICSEW'20, page 41–45, New York, NY, USA, 2020. Association for Computing Machinery.

\bibitem{okanovi_2020_chatbot_to_support_load_testing}
Du\v{s}an Okanovi\'{c}, Samuel Beck, Lasse Merz, Christoph Zorn, Leonel Merino, Andr\'{e} van Hoorn, and Fabian Beck.
\newblock Can a chatbot support software engineers with load testing? approach and experiences.
\newblock In {\em Proceedings of the ACM/SPEC International Conference on Performance Engineering}, ICPE '20, page 120–129, New York, NY, USA, 2020. Association for Computing Machinery.

\bibitem{dey_2020_bot_that_commit_code}
Tapajit Dey, Sara Mousavi, Eduardo Ponce, Tanner Fry, Bogdan Vasilescu, Anna Filippova, and Audris Mockus.
\newblock Detecting and characterizing bots that commit code.
\newblock In {\em Proceedings of the 17th International Conference on Mining Software Repositories}, MSR '20, page 209–219, New York, NY, USA, 2020. Association for Computing Machinery.

\bibitem{dey_2020_bot_commits}
Tapajit Dey, Bogdan Vasilescu, and Audris Mockus.
\newblock An exploratory study of bot commits.
\newblock In {\em Proceedings of the IEEE/ACM 42nd International Conference on Software Engineering Workshops}, ICSEW'20, page 61–65, New York, NY, USA, 2020. Association for Computing Machinery.

\bibitem{he2023automating}
Runzhi He, Hao He, Yuxia Zhang, and Minghui Zhou.
\newblock Automating dependency updates in practice: An exploratory study on github dependabot.
\newblock {\em IEEE Transactions on Software Engineering}, 2023.

\bibitem{beschastnikh_2017_research_and_bot}
Ivan Beschastnikh, Mircea~F. Lungu, and Yanyan Zhuang.
\newblock Accelerating software engineering research adoption with analysis bots.
\newblock In {\em 2017 IEEE/ACM 39th International Conference on Software Engineering: New Ideas and Emerging Technologies Results Track (ICSE-NIER)}, pages 35--38, 2017.

\bibitem{wessel_2019_bot_close_abbandoned_pull_request_and_issues}
Mairieli Wessel, Igor Steinmacher, Igor Wiese, and Marco~A Gerosa.
\newblock Should i stale or should i close? an analysis of a bot that closes abandoned issues and pull requests.
\newblock In {\em 2019 IEEE/ACM 1st International Workshop on Bots in Software Engineering (BotSE)}, pages 38--42. IEEE, 2019.

\bibitem{mirsaeedi_2020_mitigating_turnover_with_bot}
Ehsan Mirsaeedi and Peter~C. Rigby.
\newblock Mitigating turnover with code review recommendation: Balancing expertise, workload, and knowledge distribution.
\newblock In {\em Proceedings of the ACM/IEEE 42nd International Conference on Software Engineering}, ICSE '20, page 1183–1195, New York, NY, USA, 2020. Association for Computing Machinery.

\bibitem{voria2022cadocs}
Gianmario Voria, Viviana Pentangelo, Antonio Della~Porta, Stefano Lambiase, Gemma Catolino, Fabio Palomba, and Filomena Ferrucci.
\newblock Community smell detection and refactoring in slack: The cadocs project.
\newblock In {\em 2022 IEEE International Conference on Software Maintenance and Evolution (ICSME)}, pages 469--473. IEEE, 2022.

\bibitem{wang2023optimizingEliteDev}
Zhendong Wang, Yi~Wang, and David Redmiles.
\newblock Optimizing workflow for elite developers: Perspectives on leveraging se bots.
\newblock In {\em 2023 IEEE/ACM 5th International Workshop on Bots in Software Engineering (BotSE)}, pages 23--27. IEEE, 2023.

\bibitem{hefny2021proactive}
Abdelrahman~H Hefny, Georgios~A Dafoulas, and Manal~A Ismail.
\newblock A proactive management assistant chatbot for software engineering teams: Prototype and preliminary evaluation.
\newblock In {\em 2021 3rd Novel Intelligent and Leading Emerging Sciences Conference (NILES)}, pages 295--300. IEEE, 2021.

\bibitem{kim_2020_bot_to_facilitate_group_discussion}
Soomin Kim, Jinsu Eun, Changhoon Oh, Bongwon Suh, and Joonhwan Lee.
\newblock Bot in the bunch: Facilitating group chat discussion by improving efficiency and participation with a chatbot.
\newblock In {\em Proceedings of the 2020 CHI Conference on Human Factors in Computing Systems}, CHI '20, page 1–13, New York, NY, USA, 2020. Association for Computing Machinery.

\bibitem{paikari_2019_chatbot_for_conflict_detection_and_resolution}
Elahe Paikari, JaeEun Choi, SeonKyu Kim, Sooyoung Baek, MyeongSoo Kim, SeungEon Lee, ChaeYeon Han, YoungJae Kim, KaHye Ahn, Chan Cheong, et~al.
\newblock A chatbot for conflict detection and resolution.
\newblock In {\em 2019 IEEE/ACM 1st International Workshop on Bots in Software Engineering (BotSE)}, pages 29--33. IEEE, 2019.

\bibitem{cerezo_2019_builing_an_expert_recommender_bot}
Jhonny Cerezo, Juraj Kubelka, Romain Robbes, and Alexandre Bergel.
\newblock Building an expert recommender chatbot.
\newblock In {\em 2019 IEEE/ACM 1st International Workshop on Bots in Software Engineering (BotSE)}, pages 59--63, May 2019.

\bibitem{basu_2021_bot_for_distribution_of_service_requests}
Arkadip Basu and Kunal Banerjee.
\newblock Designing a bot for efficient distribution of service requests.
\newblock In {\em 2021 IEEE/ACM Third International Workshop on Bots in Software Engineering (BotSE)}, pages 16--20, June 2021.

\bibitem{ni_2019_CrowDevBot_CA_for_software_crowdsourcing}
Zeyu Ni, Beijun Shen, Yuting Chen, Zhangyuan Meng, and Junming Cao.
\newblock Crowdevbot: A task-oriented conversational bot for software crowdsourcing platform (s).
\newblock In {\em SEKE}, pages 410--522, 2019.

\bibitem{wessel_2020_code_review_bot_OSS}
Mairieli Wessel, Alexander Serebrenik, Igor Wiese, Igor Steinmacher, and Marco~A. Gerosa.
\newblock Effects of adopting code review bots on pull requests to oss projects.
\newblock In {\em 2020 IEEE International Conference on Software Maintenance and Evolution (ICSME)}, pages 1--11, Sep. 2020.

\bibitem{wessel2022bots_orchestratore}
Mairieli Wessel, Ahmad Abdellatif, Igor Wiese, Tayana Conte, Emad Shihab, Marco~A Gerosa, and Igor Steinmacher.
\newblock Bots for pull requests: The good, the bad, and the promising.
\newblock In {\em Proceedings of the 44th International Conference on Software Engineering}, pages 274--286, 2022.

\bibitem{wessel2022qualityBotCodeReview}
Mairieli Wessel, Alexander Serebrenik, Igor Wiese, Igor Steinmacher, and Marco~A Gerosa.
\newblock Quality gatekeepers: investigating the effects of code review bots on pull request activities.
\newblock {\em Empirical Software Engineering}, 27(5):108, 2022.

\bibitem{ribeiro2022together}
Eric Ribeiro, Ronan Nascimento, Igor Steinmacher, Laerte Xavier, Marco Gerosa, Hugo de~Paula, and Mairieli Wessel.
\newblock Together or apart? investigating a mediator bot to aggregate bot’s comments on pull requests.
\newblock In {\em 2022 IEEE International Conference on Software Maintenance and Evolution (ICSME)}, pages 434--438. IEEE, 2022.

\bibitem{matthies_2019_bot_and_agile_retrospectives}
Christoph Matthies, Franziska Dobrigkeit, and Guenter Hesse.
\newblock An additional set of (automated) eyes: Chatbots for agile retrospectives.
\newblock In {\em 2019 IEEE/ACM 1st International Workshop on Bots in Software Engineering (BotSE)}, pages 34--37. IEEE, 2019.

\bibitem{zhang2024automatic}
Yuxia Zhang, Zhiqing Qiu, Klaas-Jan Stol, Wenhui Zhu, Jiaxin Zhu, Yingchen Tian, and Hui Liu.
\newblock Automatic commit message generation: A critical review and directions for future work.
\newblock {\em IEEE Transactions on Software Engineering}, 2024.

\bibitem{ahmad2023towardsLLMArchitecture}
Aakash Ahmad, Muhammad Waseem, Peng Liang, Mahdi Fahmideh, Mst~Shamima Aktar, and Tommi Mikkonen.
\newblock Towards human-bot collaborative software architecting with chatgpt.
\newblock In {\em Proceedings of the 27th International Conference on Evaluation and Assessment in Software Engineering}, pages 279--285, 2023.

\bibitem{kumar_2019_sankie}
Rahul Kumar, Chetan Bansal, Chandra Maddila, Nitin Sharma, Shawn Martelock, and Ravi Bhargava.
\newblock Building sankie: An ai platform for devops.
\newblock In {\em 2019 IEEE/ACM 1st International Workshop on Bots in Software Engineering (BotSE)}, pages 48--53, May 2019.

\bibitem{lin_2020_bot_framework_for_service_development_assistance}
Chun-Ting Lin, Shang-Pin Ma, and Yu-Wen Huang.
\newblock Msabot: A chatbot framework for assisting in the development and operation of microservice-based systems.
\newblock In {\em Proceedings of the IEEE/ACM 42nd International Conference on Software Engineering Workshops}, ICSEW'20, page 36–40, New York, NY, USA, 2020. Association for Computing Machinery.

\bibitem{carr2016automatic}
Scott~A Carr, Francesco Logozzo, and Mathias Payer.
\newblock Automatic contract insertion with ccbot.
\newblock {\em IEEE Transactions on Software Engineering}, 43(8):701--714, 2016.

\bibitem{tian2017APIbot}
Yuan Tian, Ferdian Thung, Abhishek Sharma, and David Lo.
\newblock Apibot: Question answering bot for api documentation.
\newblock In {\em 2017 32nd IEEE/ACM International Conference on Automated Software Engineering (ASE)}, pages 153--158, 2017.

\bibitem{fukui_2019_bot_suggesting_questions}
Katsunori Fukui, Tomoki Miyazaki, and Masao Ohira.
\newblock Suggesting questions that match each user’s expertise in community question and answering services.
\newblock In {\em 2019 20th IEEE/ACIS International Conference on Software Engineering, Artificial Intelligence, Networking and Parallel/Distributed Computing (SNPD)}, pages 501--506, July 2019.

\bibitem{brown_2019_bots_for_effective_recommendations}
Chris Brown and Chris Parnin.
\newblock Sorry to bother you: Designing bots for effective recommendations.
\newblock In {\em 2019 IEEE/ACM 1st International Workshop on Bots in Software Engineering (BotSE)}, pages 54--58, May 2019.

\bibitem{khanan_2020_JITBot_just_in_time_defect_prediction_bot}
Chaiyakarn Khanan, Worawit Luewichana, Krissakorn Pruktharathikoon, Jirayus Jiarpakdee, Chakkrit Tantithamthavorn, Morakot Choetkiertikul, Chaiyong Ragkhitwetsagul, and Thanwadee Sunetnanta.
\newblock Jitbot: an explainable just-in-time defect prediction bot.
\newblock In {\em Proceedings of the 35th IEEE/ACM International Conference on Automated Software Engineering}, pages 1336--1339, 2020.

\bibitem{romero_2020_experiences_building_an_answer_bot}
Ricardo Romero, Esteban Parra, and Sonia Haiduc.
\newblock Experiences building an answer bot for gitter.
\newblock In {\em Proceedings of the IEEE/ACM 42nd International Conference on Software Engineering Workshops}, ICSEW'20, page 66–70, New York, NY, USA, 2020. Association for Computing Machinery.

\bibitem{dominic_2020_bot_for_SE_newcomers}
James Dominic, Charles Ritter, and Paige Rodeghero.
\newblock Onboarding bot for newcomers to software engineering.
\newblock In {\em Proceedings of the International Conference on Software and System Processes}, ICSSP '20, page 91–94, New York, NY, USA, 2020. Association for Computing Machinery.

\bibitem{chatterjee_2021_automatic_extraction_of_opinion}
Preetha Chatterjee, Kostadin Damevski, and Lori Pollock.
\newblock Automatic extraction of opinion-based q\&a from online developer chats.
\newblock In {\em 2021 IEEE/ACM 43rd International Conference on Software Engineering (ICSE)}, pages 1260--1272. IEEE, 2021.

\bibitem{sadi_2021_RAPID_bot_for_designing_web_APIs}
Mahsa~H Sadi and Eric Yu.
\newblock Rapid: a knowledge-based assistant for designing web apis.
\newblock {\em Requirements Engineering}, 26(2):185--236, 2021.

\bibitem{robe2022designing}
Peter Robe and Sandeep~Kaur Kuttal.
\newblock Designing pairbuddy—a conversational agent for pair programming.
\newblock {\em ACM Transactions on Computer-Human Interaction (TOCHI)}, 29(4):1--44, 2022.

\bibitem{robe2022pairProgramming_dataset}
Peter Robe, Sandeep~K Kuttal, Jake AuBuchon, and Jacob Hart.
\newblock Pair programming conversations with agents vs. developers: challenges and opportunities for se community.
\newblock In {\em Proceedings of the 30th ACM Joint European Software Engineering Conference and Symposium on the Foundations of Software Engineering}, pages 319--331, 2022.

\bibitem{labeuf_2017_software_bots}
Carlene Lebeuf, Margaret-Anne Storey, and Alexey Zagalsky.
\newblock Software bots.
\newblock {\em IEEE Software}, 35(1):18--23, January 2018.

\bibitem{saadat_2021_bots_modify_workflow_of_github_teams}
Samaneh Saadat, Natalia Colmenares, and Gita Sukthankar.
\newblock Do bots modify the workflow of github teams?
\newblock In {\em 2021 IEEE/ACM Third International Workshop on Bots in Software Engineering (BotSE)}, pages 1--5, June 2021.

\bibitem{pinheiro_2019_bot_development_challenges_motivations}
André~M. Pinheiro, Caio~S. Rabello, Leonardo~B. Furtado, Gustavo Pinto, and Cleidson~R.B. de~Souza.
\newblock Expecting the unexpected: Distilling bot development, challenges, and motivations.
\newblock In {\em 2019 IEEE/ACM 12th International Workshop on Cooperative and Human Aspects of Software Engineering (CHASE)}, pages 51--52, May 2019.

\bibitem{lee_2019_accelerating_bots_development}
Jinho Lee, Inseok Hwang, Thomas~S. Hubregtsen, Anne~E. Gattiker, and Christopher~M. Durham.
\newblock Accelerating conversational agents built with off-the-shelf modularized services.
\newblock {\em IEEE Pervasive Computing}, 18(2):47--57, April 2019.

\bibitem{castro_2019_bot_usability}
John~Wilmar Castro, Ranci Ren, Silvia~T Acu{\~n}a, and Juan~de Lara.
\newblock Usability of chatbots: A systematic mapping study.
\newblock 2019.

\bibitem{melo_2020_what_do_devs_expect_from_CAs}
Glaucia Melo, Edith Law, Paulo Alencar, and Donald Cowan.
\newblock Understanding user understanding: What do developers expect from a cognitive assistant?
\newblock In {\em 2020 IEEE International Conference on Big Data (Big Data)}, pages 3165--3172, Dec 2020.

\bibitem{abdellatif_2020_challenges_in_chatbot_development_from_stackoverflow}
Ahmad Abdellatif, Diego Costa, Khaled Badran, Rabe Abdalkareem, and Emad Shihab.
\newblock Challenges in chatbot development: A study of stack overflow posts.
\newblock In {\em Proceedings of the 17th International Conference on Mining Software Repositories}, MSR '20, page 174–185, New York, NY, USA, 2020. Association for Computing Machinery.

\bibitem{abdellatif2021comparison}
Ahmad Abdellatif, Khaled Badran, Diego~Elias Costa, and Emad Shihab.
\newblock A comparison of natural language understanding platforms for chatbots in software engineering.
\newblock {\em IEEE Transactions on Software Engineering}, 48(8):3087--3102, 2021.

\bibitem{erlenhov2022dependency}
Linda Erlenhov, Francisco~Gomes de~Oliveira~Neto, and Philipp Leitner.
\newblock Dependency management bots in open-source systems—prevalence and adoption.
\newblock {\em PeerJ Computer Science}, 8:e849, 2022.

\bibitem{de2023meet}
Gabriele De~Vito, Stefano Lambiase, Fabio Palomba, and Filomena Ferrucci.
\newblock Meet c4se: Your new collaborator for software engineering tasks.
\newblock In {\em 2023 49th Euromicro Conference on Software Engineering and Advanced Applications (SEAA)}, pages 235--238. IEEE, 2023.

\bibitem{brown_2020_recommendation_for_bot_architectures}
Chris Brown and Chris Parnin.
\newblock Sorry to bother you again: Developer recommendation choice architectures for designing effective bots.
\newblock In {\em Proceedings of the IEEE/ACM 42nd International Conference on Software Engineering Workshops}, ICSEW'20, page 56–60, New York, NY, USA, 2020. Association for Computing Machinery.

\bibitem{gao2022does}
Anze Gao, Yang Zhang, Tao Wang, Sihao Chen, and Jinsheng Deng.
\newblock How does bot affect developer’s sentiment: An empirical study on github issues and prs.
\newblock In {\em 2022 IEEE Smartworld, Ubiquitous Intelligence \& Computing, Scalable Computing \& Communications, Digital Twin, Privacy Computing, Metaverse, Autonomous \& Trusted Vehicles (SmartWorld/UIC/ScalCom/DigitalTwin/PriComp/Meta)}, pages 1856--1861. IEEE, 2022.

\bibitem{gao2022understandingBotsSentiment}
Anze Gao, Sihao Chen, Tao Wang, and Jinsheng Deng.
\newblock Understanding the impact of bots on developers sentiment and project progress.
\newblock In {\em 2022 IEEE 13th International Conference on Software Engineering and Service Science (ICSESS)}, pages 93--96. IEEE, 2022.

\bibitem{perez_2021_creating_and_migrating_chatbot}
Sara P{\'e}rez-Soler, Esther Guerra, and Juan de~Lara.
\newblock Creating and migrating chatbots with conga.
\newblock In {\em 2021 IEEE/ACM 43rd International Conference on Software Engineering: Companion Proceedings (ICSE-Companion)}, pages 37--40. IEEE, 2021.

\bibitem{cabot_2021_testing_challenges_for_NLP_intensive_bots}
Jordi Cabot, Loli Burgueño, Robert Clarisó, Gwendal Daniel, Jorge Perianez-Pascual, and Roberto Rodriguez-Echeverria.
\newblock Testing challenges for nlp-intensive bots.
\newblock In {\em 2021 IEEE/ACM Third International Workshop on Bots in Software Engineering (BotSE)}, pages 31--34, June 2021.

\bibitem{ouaddi2024architecture}
Charaf Ouaddi, Lamya Benaddi, and Abdeslam Jakimi.
\newblock Architecture, tools, and dsls for developing conversational agents: An overview.
\newblock {\em Procedia Computer Science}, 231:293--298, 2024.

\bibitem{hu2023bot}
Zhewei Hu and Edward Gehringer.
\newblock Bot with interactions: Improving github pull-request feedback through two-way communication.
\newblock In {\em 2023 IEEE/ACM 5th International Workshop on Bots in Software Engineering (BotSE)}, pages 28--32. IEEE, 2023.

\bibitem{arteaga2024support}
Emily~Judith Arteaga~Garcia, Jo{\~a}o~Felipe Nicolaci~Pimentel, Zixuan Feng, Marco Gerosa, Igor Steinmacher, and Anita Sarma.
\newblock How to support ml end-user programmers through a conversational agent.
\newblock In {\em Proceedings of the 46th IEEE/ACM International Conference on Software Engineering}, pages 1--12, 2024.

\end{thebibliography}


\begin{thebibliography}{10}

\bibitem{new_grey_2}
Friend or foe: Ai chatbots in software development.
\newblock Available at \url{https://www.synopsys.com/blogs/software-security/ai-chatbots-in-software-development.html} (2024/04/14).

\bibitem{new_grey_6}
Best applications of large language models.
\newblock Available at \url{https://indatalabs.com/blog/large-language-model-apps} (2024/04/14).

\bibitem{best_bots_to_improve_your_software_development_process}
Jordi Cabot.
\newblock Best bots to improve your software development process.
\newblock Available at \url{https://livablesoftware.com/best-bots-software-development/} (2022/07/14).

\bibitem{four_Things_You_Absolutely_Need_to_Know_About_Software_Bots}
Matt Francis.
\newblock 4 things you absolutely need to know about software bots.
\newblock Available at \url{https://workingmouse.com.au/innovation/4-things-you-absolutely-need-to-know-about-software-bots/} (2022/07/14).

\bibitem{software_bot_explained}
Software bot explained.
\newblock Available at \url{https://everything.explained.today/Software_bot/} (2022/07/16).

\bibitem{storey2020_botse}
Margaret-Anne Storey, Alexander Serebrenik, Carolyn~Penstein Ros{\'e}, Thomas Zimmermann, and James~D Herbsleb.
\newblock {BOT}se: Bots in software engineering (dagstuhl seminar 19471).
\newblock In {\em Dagstuhl Reports}, volume~9. Schloss Dagstuhl-Leibniz-Zentrum f{\"u}r Informatik, 2020.

\bibitem{new_grey_1}
A complete guide on the role of chatbots in devops.
\newblock Available at \url{https://www.invensislearning.com/blog/chatbots-in-devops/} (2024/04/14).

\bibitem{bots_and_AI_in_project_management}
The role of bots and {AI} in project management.
\newblock Available at \url{https://blog.proofhub.com/the-role-of-bots-and-ai-in-project-management-87d29b78e304} (2022/07/14).

\bibitem{bots_and_AI_in_project_management_2}
Four ways {AI} \& bots will change project management.
\newblock Available at \url{https://blog.planview.com/ai-bots-project-management/} (2022/07/14).

\bibitem{bots_challenges_for_development}
6 smart solutions to combat chatbot development challenges.
\newblock Available at \url{https://www.goodfirms.co/blog/chatbot-development-challenges} (2022/07/14).

\bibitem{bots_challenges_for_development_2}
4 biggest challenges in chatbot development and how to avoid them.
\newblock Available at \url{https://insights.daffodilsw.com/blog/4-biggest-challenges-in-chatbot-development-and-how-to-avoid-them} (2022/07/14).

\bibitem{bots_NLP_or_NLU}
Chatbots: When to use {NLP} \& when to use {NLU}.
\newblock Available at \url{https://cobusgreyling.medium.com/chatbots-when-to-use-nlp-and-when-to-use-nlu-8eba17c0a4bc} (2022/07/14).

\bibitem{7_chatbot_development_challenges}
7 chatbot development challenges.
\newblock Available at \url{https://www.a3logics.com/blog/chatbot-development-challenges-you-cannot-ignore} (2022/07/14).

\bibitem{new_grey_3}
What is retrieval-augmented generation, and what does it do for generative ai?
\newblock Available at \url{https://github.blog/2024-04-04-what-is-retrieval-augmented-generation-and-what-does-it-do-for-generative-ai/} (2024/04/14).

\bibitem{new_grey_4}
Ai chatbots: Understanding the benefits and limitations.
\newblock Available at \url{https://synoptek.com/insights/it-blogs/data-insights/ai-chatbots-understanding-the-benefits-and-limitations/} (2024/04/14).

\bibitem{new_grey_5}
What’s next for ai in 2024.
\newblock Available at \url{https://www.technologyreview.com/2024/01/04/1086046/whats-next-for-ai-in-2024/} (2024/04/14).

\bibitem{10_best_chatbot_framework}
10 best {AI} chatbot development frameworks comparison.
\newblock Available at \url{https://www.spaceo.ca/blog/top-ai-chatbot-frameworks/} (2022/07/14).

\bibitem{top_framework_part_1}
Top chatbot development frameworks you should know – exploring paid solutions (part 1).
\newblock Available at \url{https://thirdeyedata.io/top-chatbot-development-frameworks-you-should-know-exploring-paid-solutions-part-1/} (2022/07/14).

\bibitem{probot}
Probot.
\newblock Available at \url{https://github.com/probot/probot} (2022/07/14).

\bibitem{8_chatbot_environments}
A comparison of eight chatbot environments.
\newblock Available at \url{https://cobusgreyling.medium.com/updated-a-comparison-of-eight-chatbot-environments-7f57d4e2dc09} (2022/07/14).

\bibitem{understanding_CA_architecture}
Understanding the conversational chatbot architecture.
\newblock Available at \url{https://blog.vsoftconsulting.com/blog/understanding-the-architecture-of-conversational-chatbot} (2022/07/14).

\bibitem{implement_bot_12_steps}
Implement a successful bot strategy in 12 steps.
\newblock Available at \url{https://www.thinkhdi.com/library/supportworld/2020/implement-successful-bot-strategy-12-steps.aspx} (2022/07/14).

\bibitem{10_steps_for_chatbot_strategy}
10 steps to define your chatbot strategy.
\newblock Available at \url{https://www.digiteum.com/10-steps-to-define-your-chatbot-strategy/} (2022/07/14).

\bibitem{8_best_practices_for_bot_development}
8 best practices for bot development.
\newblock Available at \url{https://marutitech.com/8-best-practices-bot-development/} (2022/07/14).

\bibitem{bots_agile}
Bots: What they are and why your agile software development team should use them.
\newblock Available at \url{https://www.klipfolio.com/blog/bots-agile-software-development} (2022/07/14).

\bibitem{5_reasons_for_codebots}
5 reasons developers are choosing codebots.
\newblock Available at \url{https://workingmouse.com.au/app-development/5-reasons-developers-are-choosing-codebots/} (2022/07/14).

\bibitem{microsoft_azure_to_DevSecOps}
Using microsoft azure bots \& {AI} to automate {D}ev{S}ec{O}ps.
\newblock Available at \url{https://www.xgility.com/webinar-recording-using-microsoft-azure-bots-ai-to-automate-devsecops/} (2022/07/14).

\bibitem{why_bots_format_code}
Why robots should format our code for us.
\newblock Available at \url{https://www.freecodecamp.org/news/why-robots-should-format-our-code-159fd06d17f7} (2022/07/14).

\end{thebibliography}



\begin{thebibliography}{50}

%%% the Web address.

\ifx \showCODEN    \undefined \def \showCODEN     #1{\unskip}     \fi
\ifx \showDOI      \undefined \def \showDOI       #1{#1}\fi
\ifx \showISBNx    \undefined \def \showISBNx     #1{\unskip}     \fi
\ifx \showISBNxiii \undefined \def \showISBNxiii  #1{\unskip}     \fi
\ifx \showISSN     \undefined \def \showISSN      #1{\unskip}     \fi
\ifx \showLCCN     \undefined \def \showLCCN      #1{\unskip}     \fi
\ifx \shownote     \undefined \def \shownote      #1{#1}          \fi
\ifx \showarticletitle \undefined \def \showarticletitle #1{#1}   \fi
\ifx \showURL      \undefined \def \showURL       {\relax}        \fi
% invisible to TeX
\providecommand\bibfield[2]{#2}
\providecommand\bibinfo[2]{#2}
\providecommand\natexlab[1]{#1}
\providecommand\showeprint[2][]{arXiv:#2}

\bibitem[Abdellatif et~al\mbox{.}(2022)]%
        {bot_se_research_repository}
\bibfield{author}{\bibinfo{person}{Ahmad Abdellatif}, \bibinfo{person}{Khaled Badran}, {and} \bibinfo{person}{Emad Shihab}.} \bibinfo{year}{2022}\natexlab{}.
\newblock \bibinfo{title}{A Repository of Research Articles on Software Bots}.
\newblock
\newblock
\urldef\tempurl%
\url{http://papers.botse.org}
\showURL{%
\tempurl}


\bibitem[Adams et~al\mbox{.}(2017)]%
        {adams2017_grey_literature_inclusion_checklist}
\bibfield{author}{\bibinfo{person}{Richard~J Adams}, \bibinfo{person}{Palie Smart}, {and} \bibinfo{person}{Anne~Sigismund Huff}.} \bibinfo{year}{2017}\natexlab{}.
\newblock \showarticletitle{Shades of grey: guidelines for working with the grey literature in systematic reviews for management and organizational studies}.
\newblock \bibinfo{journal}{\emph{International Journal of Management Reviews}} \bibinfo{volume}{19}, \bibinfo{number}{4} (\bibinfo{year}{2017}), \bibinfo{pages}{432--454}.
\newblock


\bibitem[Al~Neimat(2005)]%
        {It_projects_fail}
\bibfield{author}{\bibinfo{person}{Taimour Al~Neimat}.} \bibinfo{year}{2005}\natexlab{}.
\newblock \showarticletitle{Why IT projects fail}.
\newblock \bibinfo{journal}{\emph{The project perfect white paper collection}}  \bibinfo{volume}{8} (\bibinfo{year}{2005}).
\newblock


\bibitem[Brooks~Jr(1995)]%
        {MyticalManMonth}
\bibfield{author}{\bibinfo{person}{Frederick~P Brooks~Jr}.} \bibinfo{year}{1995}\natexlab{}.
\newblock \bibinfo{booktitle}{\emph{The mythical man-month: essays on software engineering}}.
\newblock \bibinfo{publisher}{Pearson Education}.
\newblock


\bibitem[Cerdeiral and Santos(2019)]%
        {cerdeiral2019_software_project_management_and_maturity_model}
\bibfield{author}{\bibinfo{person}{Cristina~T Cerdeiral} {and} \bibinfo{person}{Gleison Santos}.} \bibinfo{year}{2019}\natexlab{}.
\newblock \showarticletitle{Software project management in high maturity: A systematic literature mapping}.
\newblock \bibinfo{journal}{\emph{Journal of Systems and Software}}  \bibinfo{volume}{148} (\bibinfo{year}{2019}), \bibinfo{pages}{56--87}.
\newblock


\bibitem[Cerpa and Verner(2009)]%
        {projectFailure}
\bibfield{author}{\bibinfo{person}{Narciso Cerpa} {and} \bibinfo{person}{June~M. Verner}.} \bibinfo{year}{2009}\natexlab{}.
\newblock \showarticletitle{Why Did Your Project Fail?}
\newblock \bibinfo{journal}{\emph{Commun. ACM}} \bibinfo{volume}{52}, \bibinfo{number}{12} (\bibinfo{date}{Dec.} \bibinfo{year}{2009}), \bibinfo{pages}{130–134}.
\newblock
\showISSN{0001-0782}
\urldef\tempurl%
\url{https://doi.org/10.1145/1610252.1610286}
\showDOI{\tempurl}


\bibitem[Cruzes and Dyb{\aa}(2010)]%
        {cruzes2010_synthesizing_evidence_in_SE}
\bibfield{author}{\bibinfo{person}{Daniela~S Cruzes} {and} \bibinfo{person}{Tore Dyb{\aa}}.} \bibinfo{year}{2010}\natexlab{}.
\newblock \showarticletitle{Synthesizing evidence in software engineering research}. In \bibinfo{booktitle}{\emph{Proceedings of the 2010 ACM-IEEE International Symposium on Empirical Software Engineering and Measurement}}. \bibinfo{pages}{1--10}.
\newblock


\bibitem[Del~Carpio and Angarita(2023)]%
        {del2023assistant}
\bibfield{author}{\bibinfo{person}{Alvaro~Fern{\'a}ndez Del~Carpio} {and} \bibinfo{person}{Leonardo~Berm{\'o}n Angarita}.} \bibinfo{year}{2023}\natexlab{}.
\newblock \showarticletitle{Assistant Solutions in Software Engineering: A Systematic Literature Review}. In \bibinfo{booktitle}{\emph{2023 IEEE 14th International Conference on Software Engineering and Service Science (ICSESS)}}. IEEE, \bibinfo{pages}{93--100}.
\newblock


\bibitem[Foucault et~al\mbox{.}(2015)]%
        {foucault2015_impact_of_turnover}
\bibfield{author}{\bibinfo{person}{Matthieu Foucault}, \bibinfo{person}{Marc Palyart}, \bibinfo{person}{Xavier Blanc}, \bibinfo{person}{Gail~C Murphy}, {and} \bibinfo{person}{Jean-R{\'e}my Falleri}.} \bibinfo{year}{2015}\natexlab{}.
\newblock \showarticletitle{Impact of developer turnover on quality in open-source software}. In \bibinfo{booktitle}{\emph{Proceedings of the 2015 10th joint meeting on foundations of software engineering}}. \bibinfo{pages}{829--841}.
\newblock


\bibitem[Frieden(2005)]%
        {friden_FBI_failure_report}
\bibfield{author}{\bibinfo{person}{Terry Frieden}.} \bibinfo{year}{2005}\natexlab{}.
\newblock \showarticletitle{Report: FBI wasted millions on'Virtual Case File'}.
\newblock \bibinfo{journal}{\emph{Retrieved April}}  \bibinfo{volume}{9} (\bibinfo{year}{2005}), \bibinfo{pages}{2005}.
\newblock


\bibitem[Gargani and Donaldson(2011)]%
        {gargani2011works}
\bibfield{author}{\bibinfo{person}{John Gargani} {and} \bibinfo{person}{Stewart~I Donaldson}.} \bibinfo{year}{2011}\natexlab{}.
\newblock \showarticletitle{What works for whom, where, why, for what, and when? Using evaluation evidence to take action in local contexts}.
\newblock \bibinfo{journal}{\emph{New Directions for Evaluation}} \bibinfo{volume}{2011}, \bibinfo{number}{130} (\bibinfo{year}{2011}), \bibinfo{pages}{17--30}.
\newblock


\bibitem[Garousi et~al\mbox{.}(2017)]%
        {garousi2017_mlr_example_1}
\bibfield{author}{\bibinfo{person}{Vahid Garousi}, \bibinfo{person}{Michael Felderer}, {and} \bibinfo{person}{Tuna Hacalo{\u{g}}lu}.} \bibinfo{year}{2017}\natexlab{}.
\newblock \showarticletitle{Software test maturity assessment and test process improvement: A multivocal literature review}.
\newblock \bibinfo{journal}{\emph{Information and Software Technology}}  \bibinfo{volume}{85} (\bibinfo{year}{2017}), \bibinfo{pages}{16--42}.
\newblock


\bibitem[Garousi et~al\mbox{.}(2019)]%
        {garousi2019_mlr_guidelines}
\bibfield{author}{\bibinfo{person}{Vahid Garousi}, \bibinfo{person}{Michael Felderer}, {and} \bibinfo{person}{Mika~V M{\"a}ntyl{\"a}}.} \bibinfo{year}{2019}\natexlab{}.
\newblock \showarticletitle{Guidelines for including grey literature and conducting multivocal literature reviews in software engineering}.
\newblock \bibinfo{journal}{\emph{Information and Software Technology}}  \bibinfo{volume}{106} (\bibinfo{year}{2019}), \bibinfo{pages}{101--121}.
\newblock


\bibitem[Garousi and M{\"a}ntyl{\"a}(2016)]%
        {garousi2016_mlr_example_2}
\bibfield{author}{\bibinfo{person}{Vahid Garousi} {and} \bibinfo{person}{Mika~V M{\"a}ntyl{\"a}}.} \bibinfo{year}{2016}\natexlab{}.
\newblock \showarticletitle{When and what to automate in software testing? A multi-vocal literature review}.
\newblock \bibinfo{journal}{\emph{Information and Software Technology}}  \bibinfo{volume}{76} (\bibinfo{year}{2016}), \bibinfo{pages}{92--117}.
\newblock


\bibitem[Hayat et~al\mbox{.}(2019)]%
        {hayat2019_influence_of_agile_on_software_project_management}
\bibfield{author}{\bibinfo{person}{Faisal Hayat}, \bibinfo{person}{Ammar~Ur Rehman}, \bibinfo{person}{Khawaja~Sarmad Arif}, \bibinfo{person}{Kanwal Wahab}, {and} \bibinfo{person}{Muhammad Abbas}.} \bibinfo{year}{2019}\natexlab{}.
\newblock \showarticletitle{The influence of agile methodology (Scrum) on software project management}. In \bibinfo{booktitle}{\emph{2019 20th IEEE/ACIS International Conference on Software Engineering, Artificial Intelligence, Networking and Parallel/Distributed Computing (SNPD)}}. IEEE, \bibinfo{pages}{145--149}.
\newblock


\bibitem[Hussain and Mkpojiogu(2016)]%
        {project_failure_2}
\bibfield{author}{\bibinfo{person}{Azham Hussain} {and} \bibinfo{person}{Emmanuel~OC Mkpojiogu}.} \bibinfo{year}{2016}\natexlab{}.
\newblock \showarticletitle{Requirements: Towards an understanding on why software projects fail}. In \bibinfo{booktitle}{\emph{AIP Conference Proceedings}}, Vol.~\bibinfo{volume}{1761}. AIP Publishing LLC, \bibinfo{pages}{020046}.
\newblock


\bibitem[Institute(2021)]%
        {PMBOK}
\bibfield{author}{\bibinfo{person}{Project~Management Institute}.} \bibinfo{year}{2021}\natexlab{}.
\newblock \bibinfo{booktitle}{\emph{A Guide to the Project Management Body of Knowledge} (\bibinfo{edition}{7} ed.)}.
\newblock 250 pages.
\newblock
\showISBNx{1628256648}


\bibitem[Islam et~al\mbox{.}(2019)]%
        {islam2019_mlr_example_1}
\bibfield{author}{\bibinfo{person}{Chadni Islam}, \bibinfo{person}{Muhammad~Ali Babar}, {and} \bibinfo{person}{Surya Nepal}.} \bibinfo{year}{2019}\natexlab{}.
\newblock \showarticletitle{A multi-vocal review of security orchestration}.
\newblock \bibinfo{journal}{\emph{ACM Computing Surveys (CSUR)}} \bibinfo{volume}{52}, \bibinfo{number}{2} (\bibinfo{year}{2019}), \bibinfo{pages}{1--45}.
\newblock


\bibitem[Johnson et~al\mbox{.}(2012)]%
        {johnson2012_choice_architecture_tools}
\bibfield{author}{\bibinfo{person}{Eric~J Johnson}, \bibinfo{person}{Suzanne~B Shu}, \bibinfo{person}{Benedict~GC Dellaert}, \bibinfo{person}{Craig Fox}, \bibinfo{person}{Daniel~G Goldstein}, \bibinfo{person}{Gerald H{\"a}ubl}, \bibinfo{person}{Richard~P Larrick}, \bibinfo{person}{John~W Payne}, \bibinfo{person}{Ellen Peters}, \bibinfo{person}{David Schkade}, {et~al\mbox{.}}} \bibinfo{year}{2012}\natexlab{}.
\newblock \showarticletitle{Beyond nudges: Tools of a choice architecture}.
\newblock \bibinfo{journal}{\emph{Marketing letters}} \bibinfo{volume}{23}, \bibinfo{number}{2} (\bibinfo{year}{2012}), \bibinfo{pages}{487--504}.
\newblock


\bibitem[Johnson(2020)]%
        {chaos_report}
\bibfield{author}{\bibinfo{person}{Jim Johnson}.} \bibinfo{year}{2020}\natexlab{}.
\newblock \showarticletitle{CHAOS 2020: Beyond Infinity}.
\newblock \bibinfo{journal}{\emph{Standish Group}} (\bibinfo{year}{2020}).
\newblock


\bibitem[J{\o}rgensen and Mol{\o}kken-{\O}stvold(2006)]%
        {project_failure_3}
\bibfield{author}{\bibinfo{person}{Magne J{\o}rgensen} {and} \bibinfo{person}{Kjetil Mol{\o}kken-{\O}stvold}.} \bibinfo{year}{2006}\natexlab{}.
\newblock \showarticletitle{How large are software cost overruns? A review of the 1994 CHAOS report}.
\newblock \bibinfo{journal}{\emph{Information and Software Technology}} \bibinfo{volume}{48}, \bibinfo{number}{4} (\bibinfo{year}{2006}), \bibinfo{pages}{297--301}.
\newblock


\bibitem[Kageki(2012)]%
        {kageki2012_uncanny_valley}
\bibfield{author}{\bibinfo{person}{Norri Kageki}.} \bibinfo{year}{2012}\natexlab{}.
\newblock \showarticletitle{An uncanny mind: Masahiro Mori on the uncanny valley and beyond}.
\newblock \bibinfo{journal}{\emph{IEEE spectrum}} \bibinfo{volume}{12}, \bibinfo{number}{06} (\bibinfo{year}{2012}), \bibinfo{pages}{2012}.
\newblock


\bibitem[Keele et~al\mbox{.}(2007)]%
        {keele2007_slr_guidelines_Kitchenham}
\bibfield{author}{\bibinfo{person}{Staffs Keele} {et~al\mbox{.}}} \bibinfo{year}{2007}\natexlab{}.
\newblock \bibinfo{booktitle}{\emph{Guidelines for performing systematic literature reviews in software engineering}}.
\newblock \bibinfo{type}{{T}echnical {R}eport}. \bibinfo{institution}{Technical report, Ver. 2.3 EBSE Technical Report. EBSE}.
\newblock


\bibitem[Kitchenham et~al\mbox{.}(2009)]%
        {kitchenham2009_SLR_definition}
\bibfield{author}{\bibinfo{person}{Barbara Kitchenham}, \bibinfo{person}{O~Pearl Brereton}, \bibinfo{person}{David Budgen}, \bibinfo{person}{Mark Turner}, \bibinfo{person}{John Bailey}, {and} \bibinfo{person}{Stephen Linkman}.} \bibinfo{year}{2009}\natexlab{}.
\newblock \showarticletitle{Systematic literature reviews in software engineering--a systematic literature review}.
\newblock \bibinfo{journal}{\emph{Information and software technology}} \bibinfo{volume}{51}, \bibinfo{number}{1} (\bibinfo{year}{2009}), \bibinfo{pages}{7--15}.
\newblock


\bibitem[Kumara et~al\mbox{.}(2021)]%
        {kumara2021_sglr_example_1_infrastructure_code}
\bibfield{author}{\bibinfo{person}{Indika Kumara}, \bibinfo{person}{Mart{\'\i}n Garriga}, \bibinfo{person}{Angel~Urbano Romeu}, \bibinfo{person}{Dario Di~Nucci}, \bibinfo{person}{Fabio Palomba}, \bibinfo{person}{Damian~Andrew Tamburri}, {and} \bibinfo{person}{Willem-Jan van~den Heuvel}.} \bibinfo{year}{2021}\natexlab{}.
\newblock \showarticletitle{The do’s and don’ts of infrastructure code: A systematic gray literature review}.
\newblock \bibinfo{journal}{\emph{Information and Software Technology}}  \bibinfo{volume}{137} (\bibinfo{year}{2021}), \bibinfo{pages}{106593}.
\newblock


\bibitem[Lambiase et~al\mbox{.}(2022)]%
        {online_appendix}
\bibfield{author}{\bibinfo{person}{Stefano Lambiase}, \bibinfo{person}{Gemma Catolino}, \bibinfo{person}{Fabio Palomba}, {and} \bibinfo{person}{Filomena Ferrucci}.} \bibinfo{year}{2022}\natexlab{}.
\newblock \bibinfo{title}{Motivations, Challenges, Best Practices, and Benefits for Bots and Conversational Agents in Software Engineering: A Multivocal Literature Review - Online Appendix}.
\newblock
\newblock
\urldef\tempurl%
\url{https://github.com/StefanoLambiase/Bots-and-CAs-for-SE-MLR-OnlineAppendix}
\showURL{%
\tempurl}


\bibitem[Lebeuf et~al\mbox{.}(2018)]%
        {labeuf_2017_software_bots}
\bibfield{author}{\bibinfo{person}{Carlene Lebeuf}, \bibinfo{person}{Margaret-Anne Storey}, {and} \bibinfo{person}{Alexey Zagalsky}.} \bibinfo{year}{2018}\natexlab{}.
\newblock \showarticletitle{Software Bots}.
\newblock \bibinfo{journal}{\emph{IEEE Software}} \bibinfo{volume}{35}, \bibinfo{number}{1} (\bibinfo{date}{January} \bibinfo{year}{2018}), \bibinfo{pages}{18--23}.
\newblock
\showISSN{1937-4194}
\urldef\tempurl%
\url{https://doi.org/10.1109/MS.2017.4541027}
\showDOI{\tempurl}


\bibitem[Lewandowski et~al\mbox{.}(2021)]%
        {lewandowski2021_bot_SLR}
\bibfield{author}{\bibinfo{person}{Tom Lewandowski}, \bibinfo{person}{Jasmin Delling}, \bibinfo{person}{Christian Grotherr}, {and} \bibinfo{person}{Tilo B{\"o}hmann}.} \bibinfo{year}{2021}\natexlab{}.
\newblock \showarticletitle{State-of-the-Art Analysis of Adopting AI-based Conversational Agents in Organizations: A Systematic Literature Review}. In \bibinfo{booktitle}{\emph{25th Pacific Asia Conference on Information Systems, Dubai, UAE}}. \bibinfo{pages}{1--14}.
\newblock


\bibitem[Luo et~al\mbox{.}(2014)]%
        {luo2014_test_flakiness_definition}
\bibfield{author}{\bibinfo{person}{Q. Luo}, \bibinfo{person}{F. Hariri}, \bibinfo{person}{L. Eloussi}, {and} \bibinfo{person}{D. Marinov}.} \bibinfo{year}{2014}\natexlab{}.
\newblock \showarticletitle{An empirical analysis of flaky tests}. In \bibinfo{booktitle}{\emph{ESEC/FSE 2014}}. \bibinfo{pages}{643--653}.
\newblock


\bibitem[Mahood et~al\mbox{.}(2014)]%
        {mahood2014_searching_for_grey_literature}
\bibfield{author}{\bibinfo{person}{Quenby Mahood}, \bibinfo{person}{Dwayne Van~Eerd}, {and} \bibinfo{person}{Emma Irvin}.} \bibinfo{year}{2014}\natexlab{}.
\newblock \showarticletitle{Searching for grey literature for systematic reviews: challenges and benefits}.
\newblock \bibinfo{journal}{\emph{Research synthesis methods}} \bibinfo{volume}{5}, \bibinfo{number}{3} (\bibinfo{year}{2014}), \bibinfo{pages}{221--234}.
\newblock


\bibitem[M{\"a}ntyl{\"a} and Smolander(2016)]%
        {mantyla2016_mlr_example_1_gamification_of_software_testing}
\bibfield{author}{\bibinfo{person}{Mika~V M{\"a}ntyl{\"a}} {and} \bibinfo{person}{Kari Smolander}.} \bibinfo{year}{2016}\natexlab{}.
\newblock \showarticletitle{Gamification of software testing-an mlr}. In \bibinfo{booktitle}{\emph{International conference on product-focused software process improvement}}. Springer, \bibinfo{pages}{611--614}.
\newblock


\bibitem[Moguel-S{\'a}nchez et~al\mbox{.}(2023)]%
        {moguel2023bots}
\bibfield{author}{\bibinfo{person}{R Moguel-S{\'a}nchez}, \bibinfo{person}{CS~Sergio Mart{\'\i}nez-Palacios}, \bibinfo{person}{JO Ochar{\'a}n-Hern{\'a}ndez}, \bibinfo{person}{X Lim{\'o}n}, {and} \bibinfo{person}{AJ S{\'a}nchez-Garc{\'\i}a}.} \bibinfo{year}{2023}\natexlab{}.
\newblock \showarticletitle{Bots in Software Development: A Systematic Literature Review and Thematic Analysis}.
\newblock \bibinfo{journal}{\emph{Programming and Computer Software}} \bibinfo{volume}{49}, \bibinfo{number}{8} (\bibinfo{year}{2023}), \bibinfo{pages}{712--734}.
\newblock


\bibitem[Mori et~al\mbox{.}(2012)]%
        {mori2012_uncanny_valley}
\bibfield{author}{\bibinfo{person}{Masahiro Mori}, \bibinfo{person}{Karl~F MacDorman}, {and} \bibinfo{person}{Norri Kageki}.} \bibinfo{year}{2012}\natexlab{}.
\newblock \showarticletitle{The uncanny valley}.
\newblock \bibinfo{journal}{\emph{IEEE Robotics \& automation magazine}} \bibinfo{volume}{19}, \bibinfo{number}{2} (\bibinfo{year}{2012}), \bibinfo{pages}{98--100}.
\newblock


\bibitem[Motger et~al\mbox{.}(2022)]%
        {motger2022_software_bot_SLR}
\bibfield{author}{\bibinfo{person}{Quim Motger}, \bibinfo{person}{Xavier Franch}, {and} \bibinfo{person}{Jordi Marco}.} \bibinfo{year}{2022}\natexlab{}.
\newblock \showarticletitle{Software-based dialogue systems: survey, taxonomy, and challenges}.
\newblock \bibinfo{journal}{\emph{Comput. Surveys}} \bibinfo{volume}{55}, \bibinfo{number}{5} (\bibinfo{year}{2022}), \bibinfo{pages}{1--42}.
\newblock


\bibitem[Ogawa and Malen(1991)]%
        {ogawa1991_MLR_definition}
\bibfield{author}{\bibinfo{person}{Rodney~T Ogawa} {and} \bibinfo{person}{Betty Malen}.} \bibinfo{year}{1991}\natexlab{}.
\newblock \showarticletitle{Towards rigor in reviews of multivocal literatures: Applying the exploratory case study method}.
\newblock \bibinfo{journal}{\emph{Review of educational research}} \bibinfo{volume}{61}, \bibinfo{number}{3} (\bibinfo{year}{1991}), \bibinfo{pages}{265--286}.
\newblock


\bibitem[Okonkwo and Ade-Ibijola(2021)]%
        {okonkwo2021_chatbots_in_education_SLR}
\bibfield{author}{\bibinfo{person}{Chinedu~Wilfred Okonkwo} {and} \bibinfo{person}{Abejide Ade-Ibijola}.} \bibinfo{year}{2021}\natexlab{}.
\newblock \showarticletitle{Chatbots applications in education: A systematic review}.
\newblock \bibinfo{journal}{\emph{Computers and Education: Artificial Intelligence}}  \bibinfo{volume}{2} (\bibinfo{year}{2021}), \bibinfo{pages}{100033}.
\newblock


\bibitem[Park et~al\mbox{.}(2022)]%
        {park2022_bot_SLR}
\bibfield{author}{\bibinfo{person}{Dong-Min Park}, \bibinfo{person}{Seong-Soo Jeong}, {and} \bibinfo{person}{Yeong-Seok Seo}.} \bibinfo{year}{2022}\natexlab{}.
\newblock \showarticletitle{Systematic Review on Chatbot Techniques and Applications}.
\newblock \bibinfo{journal}{\emph{Journal of Information Processing Systems}} \bibinfo{volume}{18}, \bibinfo{number}{1} (\bibinfo{year}{2022}), \bibinfo{pages}{26--47}.
\newblock


\bibitem[Patton(1991)]%
        {patton1991_MLR_definition_1}
\bibfield{author}{\bibinfo{person}{Michael~Quinn Patton}.} \bibinfo{year}{1991}\natexlab{}.
\newblock \showarticletitle{Towards utility in reviews of multivocal literatures}.
\newblock \bibinfo{journal}{\emph{Review of Educational Research}} \bibinfo{volume}{61}, \bibinfo{number}{3} (\bibinfo{year}{1991}), \bibinfo{pages}{287--292}.
\newblock


\bibitem[P{\'e}rez-Soler et~al\mbox{.}(2018)]%
        {perez2018_collaborative_modeling_with_chatbot}
\bibfield{author}{\bibinfo{person}{Sara P{\'e}rez-Soler}, \bibinfo{person}{Esther Guerra}, {and} \bibinfo{person}{Juan de Lara}.} \bibinfo{year}{2018}\natexlab{}.
\newblock \showarticletitle{Collaborative modeling and group decision making using chatbots in social networks}.
\newblock \bibinfo{journal}{\emph{IEEE Software}} \bibinfo{volume}{35}, \bibinfo{number}{6} (\bibinfo{year}{2018}), \bibinfo{pages}{48--54}.
\newblock


\bibitem[Petersen and Wohlin(2009)]%
        {petersen2009context}
\bibfield{author}{\bibinfo{person}{Kai Petersen} {and} \bibinfo{person}{Claes Wohlin}.} \bibinfo{year}{2009}\natexlab{}.
\newblock \showarticletitle{Context in industrial software engineering research}. In \bibinfo{booktitle}{\emph{2009 3rd international symposium on empirical software engineering and measurement}}. IEEE, \bibinfo{pages}{401--404}.
\newblock


\bibitem[Settles(2009)]%
        {settles2009_active_learning}
\bibfield{author}{\bibinfo{person}{Burr Settles}.} \bibinfo{year}{2009}\natexlab{}.
\newblock \showarticletitle{Active learning literature survey}.
\newblock  (\bibinfo{year}{2009}).
\newblock


\bibitem[Shevat(2017)]%
        {shevat2017_designing_bots}
\bibfield{author}{\bibinfo{person}{Amir Shevat}.} \bibinfo{year}{2017}\natexlab{}.
\newblock \bibinfo{booktitle}{\emph{Designing bots: Creating conversational experiences}}.
\newblock \bibinfo{publisher}{" O'Reilly Media, Inc."}.
\newblock


\bibitem[Storey and Zagalsky(2016)]%
        {storey2016_disrupting_developer_productivity_botDefinition}
\bibfield{author}{\bibinfo{person}{Margaret-Anne Storey} {and} \bibinfo{person}{Alexey Zagalsky}.} \bibinfo{year}{2016}\natexlab{}.
\newblock \showarticletitle{Disrupting developer productivity one bot at a time}. In \bibinfo{booktitle}{\emph{Proceedings of the 2016 24th ACM SIGSOFT international symposium on foundations of software engineering}}. \bibinfo{pages}{928--931}.
\newblock


\bibitem[Suhaili et~al\mbox{.}(2021)]%
        {suhaili2021_bot_SLR}
\bibfield{author}{\bibinfo{person}{Sinarwati~Mohamad Suhaili}, \bibinfo{person}{Naomie Salim}, {and} \bibinfo{person}{Mohamad~Nazim Jambli}.} \bibinfo{year}{2021}\natexlab{}.
\newblock \showarticletitle{Service chatbots: A systematic review}.
\newblock \bibinfo{journal}{\emph{Expert Systems with Applications}}  \bibinfo{volume}{184} (\bibinfo{year}{2021}), \bibinfo{pages}{115461}.
\newblock


\bibitem[Tamburri(2019)]%
        {tamburri2019software}
\bibfield{author}{\bibinfo{person}{Damian~A Tamburri}.} \bibinfo{year}{2019}\natexlab{}.
\newblock \showarticletitle{Software architecture social debt: Managing the incommunicability factor}.
\newblock \bibinfo{journal}{\emph{IEEE Transactions on Computational Social Systems}} \bibinfo{volume}{6}, \bibinfo{number}{1} (\bibinfo{year}{2019}), \bibinfo{pages}{20--37}.
\newblock


\bibitem[Thaler and Sunstein(2009)]%
        {thaler2009_nudge_theory_definition}
\bibfield{author}{\bibinfo{person}{Richard~H Thaler} {and} \bibinfo{person}{Cass~R Sunstein}.} \bibinfo{year}{2009}\natexlab{}.
\newblock \bibinfo{booktitle}{\emph{Nudge: Improving decisions about health, wealth, and happiness}}.
\newblock \bibinfo{publisher}{Penguin}.
\newblock


\bibitem[Thaler et~al\mbox{.}(2013)]%
        {thaler2013_choice_architecture}
\bibfield{author}{\bibinfo{person}{Richard~H Thaler}, \bibinfo{person}{Cass~R Sunstein}, {and} \bibinfo{person}{John~P Balz}.} \bibinfo{year}{2013}\natexlab{}.
\newblock \bibinfo{booktitle}{\emph{Choice architecture}}. Vol.~\bibinfo{volume}{2013}.
\newblock \bibinfo{publisher}{Princeton University Press Princeton, NJ}.
\newblock


\bibitem[Verma and Rubin(2018)]%
        {verma2018_fairness_definition}
\bibfield{author}{\bibinfo{person}{Sahil Verma} {and} \bibinfo{person}{Julia Rubin}.} \bibinfo{year}{2018}\natexlab{}.
\newblock \showarticletitle{Fairness definitions explained}. In \bibinfo{booktitle}{\emph{2018 ieee/acm international workshop on software fairness (fairware)}}. IEEE, \bibinfo{pages}{1--7}.
\newblock


\bibitem[Wessel et~al\mbox{.}(2021)]%
        {wessel2021_dont_disturb_me_botChallenges}
\bibfield{author}{\bibinfo{person}{Mairieli Wessel}, \bibinfo{person}{Igor Wiese}, \bibinfo{person}{Igor Steinmacher}, {and} \bibinfo{person}{Marco~Aurelio Gerosa}.} \bibinfo{year}{2021}\natexlab{}.
\newblock \showarticletitle{Don't Disturb Me: Challenges of Interacting with Software Bots on Open Source Software Projects}.
\newblock \bibinfo{journal}{\emph{Proceedings of the ACM on Human-Computer Interaction}} \bibinfo{volume}{5}, \bibinfo{number}{CSCW2} (\bibinfo{year}{2021}), \bibinfo{pages}{1--21}.
\newblock


\bibitem[Wohlin et~al\mbox{.}(2012)]%
        {wohlin2012experimentation}
\bibfield{author}{\bibinfo{person}{Claes Wohlin}, \bibinfo{person}{Per Runeson}, \bibinfo{person}{Martin H{\"o}st}, \bibinfo{person}{Magnus~C Ohlsson}, \bibinfo{person}{Bj{\"o}rn Regnell}, {and} \bibinfo{person}{Anders Wessl{\'e}n}.} \bibinfo{year}{2012}\natexlab{}.
\newblock \bibinfo{booktitle}{\emph{Experimentation in software engineering}}.
\newblock \bibinfo{publisher}{Springer Science \& Business Media}.
\newblock


\end{thebibliography}
%%%%%%%%%%%%%%%%%%%%%%
%\input{otherTEx/appendix}

\end{document}